\documentstyle[12pt,epsf,axodraw,twoside]{article}
\title{{\bf Charmless decays $B \to PP, PV$, and effects of new strong and
electroweak penguins in Topcolor-assisted Technicolor model }}
\author{ Zhenjun Xiao$^{(1,2,3)}$
Wenjun Li$^{2}$, Libo Guo$^{(1,4)}$ and Gongru Lu$^{1,2}$\\
{\small 1. CCAST(World Laboratory) P.O. Box 8730, Beijing 100080,
P.R.China} \\
{\small 2. Department of Physics, Henan Normal University,
Xinxiang, 453002 P.R. China.} \thanks{Mailing address} \\
{\small 3. Department of Physics, Peking University,
Beijing, 100871 P.R. China.} \\
{\small 4. Department of Physics, Wuhan University, Wuhan, 430000 P.R.
China.} }
\date{\today}

\begin{document}
\maketitle
\begin{abstract}
Based on the low energy effective Hamiltonian with generalized factorization, we
calculate the new physics contributions to the branching ratios
and CP-violating asymmetries of the two-body charmless hadronic
decays $B \to PP, PV$ from the new strong and electroweak penguin diagrams
in the Topcolor-assisted Technicolor (TC2) model.
The top-pion penguins dominate the new physics corrections, and both new
gluonic and electroweak  penguins contribute effectively to most decay modes.
For tree-dominated decay modes $B \to \pi \pi, \rho \pi, etc,$ the new
physics corrections are less than $10\%$.
For decays $B \to K^{(*)} \pi$,  $K^{(*)} \eta $,  $\pi^0 \eta^{(')}$,
$\eta^{(')}\eta^{(')}$, $K \overline{K}^0$, $\overline{K}^{*0} K$, $etc$,
the new physics enhancements can be rather large ( from $- 70\%$ to $ \sim
200\%$ ) and are insensitive to the variations
of $N_c^{eff}$, $k^2$, $\eta$ and $m_{\tilde{\pi}}$ within the
reasonable ranges.
For decays $B^0 \to \phi \pi$, $\phi \eta^{(')}$,  $K^* \overline{K}^0$ and
$\rho^+ K^0$, $\delta {\cal B}$ is strongly $N_c^{eff}-$dependent: varying from
$-90\%$ to $\sim 1680\%$ in the range of $N_c^{eff}=2-\infty$.
The new physics corrections to the CP-violating asymmetries  ${\cal A}_{CP}$
vary greatly for different B decay channels. For five measured CP
asymmetries of $B \to K \pi, K \eta^{\prime}, \omega \pi$ decays,
$\delta {\cal A}_{CP}$ is only about $20\%$ and will be masked by
large theoretical uncertainties. The new physics  enhancements
to interesting $B \to  K \eta^{\prime}$ decays are significant in size
($\sim 50\%$), insensitive to the variations of input parameters and hence
lead to a plausible interpretation for the unexpectedly large
$B \to K \eta^{\prime} $ decay rates.
The TC2 model predictions for branching ratios and CP-violating asymmteries
of  all fifty seven $B \to PP, PV$ decay modes are  consistent
with the available data within one or two standard deviations.
\end{abstract}


\vspace{.5cm}

\newcommand{\beq}{\begin{eqnarray}}
\newcommand{\eeq}{\end{eqnarray}}

\newcommand{\bsga}{ b \to s \gamma}
\newcommand{\aepa}{a_{\epsilon'}}
\newcommand{\aepb}{a_{\epsilon + \epsilon'}}

\newcommand{\bsl}{  B_{SL}}
\newcommand{\nc}{ n_c}
\newcommand{\as}{ \alpha_s }
\newcommand{\aem}{ \alpha_{em} }

\newcommand{\mw}{ M_W }
\newcommand{\mhp}{ M_{H^{+}}}
\newcommand{\pepm}{ P^{\pm}_8 }
\newcommand{\ppm}{ P^{\pm} }
\newcommand{\pipm}{ \pi^{\pm} }
\newcommand{\piz}{ \pi^0 }

\newcommand{\paa}{\pi_1^{\pm}}
\newcommand{\pbb}{\pi_8^{\pm}}
\newcommand{\pcc}{\tilde{\pi}^{\pm}}
\newcommand{\pcm}{\tilde{\pi}^-}
\newcommand{\pcp}{\tilde{\pi}^+}
\newcommand{\pcz}{\tilde{\pi}^0}
\newcommand{\pdd}{\tilde{H}^{\pm}}

\newcommand{\mpaa}{m_{\pi_1}}
\newcommand{\mpbb}{m_{\pi_8}}
\newcommand{\mpcc}{m_{\tilde{\pi}}}
\newcommand{\mpdd}{m_{\tilde{H}}}
\newcommand{\fpit}{F_{ \tilde{\pi}}}

\def\vma{{_{V-A}}}
\def\vpa{{_{V+A}}}
\def\etap{\eta^{\prime}}
\def\etapp{\eta^{(')}}
\def\acp{{\cal A}_{CP}}

\newcommand{\tab}[1]{Table \ref{#1}}
\newcommand{\fig}[1]{Fig.\ref{#1}}
\newcommand{\real}{{\rm Re}\,}
\newcommand{\non}{\nonumber\\ }
\newcommand{\cimw}{C_{i}(M_W) }
\newcommand{\obar}[1]{\shortstack{{\tiny (\rule[.4ex]{1em}{.1mm})}
\\ [-.7ex] $#1$}}

\newpage
\section{ Introduction } \label{sec:1}

The main goals of B experiments undertaken by CLEO, BarBar, Belle and other
collaborations  are to explore the physics of CP violation, to test the
standard model (SM) at an unexpected level of precision, and to make an
exhaustive search for possible effects of physics beyond the SM
\cite{slac504,belle94}.
Precision measurements of B meson system can provide an insight into very
high energy scales via the indirect loop effects of new physics(NP).
The B system therefore offers a complementary probe to the search for new
physics at the Tevatron, LHC and NLC, and in some cases may yield constraint
which surpass those from direct searches or rule out some kinds of NP
models\cite{slac504}.

In B experiments, new physics beyond the standard model may manifest itself,
for example,  in the following ways\cite{slac504,fj99}:
\begin{itemize}

\item
Decays which are expected to be rare in the standard model are found
to have large branching ratios;

\item
CP-violating asymmetries which are
expected to vanish or be very small in the SM are found to be significantly large
or with a very different pattern with what predicted in the SM;

\item
Mixing in B decays is found to differ significantly from SM predictions;

\end{itemize}

These potential deviations may originate from the virtual effects of new physics
through box and/or penguin diagrams in various new physics models
\cite{misiak97,atwood98,wu99,iltan99,xiao96,xiao20}.

Due to the anticipated importance of  two-body charmless hadronic decays
$B \to h_1 h_2$ ( where $h_1$ and $h_2$ are the light pseudo-scalar (P)
and/or vector(V) mesons ) in understanding the phenomenon of CP violation, great
effort have been made by many authors \cite{bh1h2,du97,ali9804,ali98,chen99}.
It is well known that the low energy effective Hamiltonian is the basic tool
to calculate the branching ratios and $A_{CP}$ of B meson decays. The
short-distance QCD corrected Lagrangian at NLO level is available now,
but we do not know how to calculate hadronic matrix element from first
principles. One conventionally resort to the factorization approximation
\cite{bsw87}. However, we also know that non-factorizable contribution
really exists and can not be neglected numerically for most hadronic B
decay channels. To remedy factorization hypothesis, some authors
\cite{cheng98,ali9804,ali98} introduced a
phenomenological parameter $N_{eff}$ (i.e. the effective number of color)
to model the non-factorizable contribution to hadronic matrix element,
which is commonly called generalized factorization.

On the other hand, as pointed by Buras and Silverstrini \cite{bs98},
such generalization suffers from the problems of gauge  and infrared
dependence since the constant matrix $\hat{r}_V$ appeared
in the expressions of $C_i^{eff}$ depends on both the gauge chosen
and the external momenta. Very recently, Cheng {\it et al.} \cite{cheng99a}
studied and resolved above controversies on the gauge dependence and
infrared singularity of $C_i^{eff}$ by using the perturbative QCD
factorization theorem. Based on  this progress, Chen {\it et al.}
\cite{chen99}  calculated the charmless hadronic two-body
decays of $B_u$ and $B_d$ mesons within the framework of generalized
factorization, in which the effective Wilson coefficients $C_i^{eff}$ are
gauge invariant, infrared safe, and renormalization-scale and -scheme
independent.

On the experimental side, the observation of thirteen $B\to PP, PV$ decays
by CLEO, BaBar and Belle collaborations
\cite{cleo99,cleo9912,cleo2000,cleoacp,cleosum,babar2000,belle2000}
signaled the beginning of the golden age of B physics. For $B \to K \pi,
\pi \pi$ decays, the data are well accounted for  in the effective
Hamiltonian\cite{buchalla96a,buras98} with the generalized factorization
approach\cite{bsw87,ali9804,chen99}. For $B \to K \etap$ decays,
however, the unexpectedly large decay rate
${\cal B}(B \to K \etap)=(80 ^{+10}_{-9} \pm 7)\times 10^{-6}$
\cite{cleo9912} still has no completely satisfactory explanation
and has aroused considerable controversy\cite{lipkin2000}.

In this paper, we will present our systematic calculation of branching ratios
and CP-violating asymmetries for two-body charmless hadronic decays $B \to P P$,
$P V $ (with charged $B_u$, neutral $B_d$ mesons ) in the framework of
Topcolor-assisted technicolor (TC2) model \cite{hill95}
by employing the effective Hamiltonian with the generalized factorization.
Since the scale of new strong interactions is expected around $ 1$ TeV, the
tree-level new physics contributions are strongly suppressed and will be
neglected. We therefore will focus on the loop effects of new physics on
two-body charmless hadronic B meson decays. We will
evaluate analytically all new strong and electroweak penguin diagrams
induced by exchanges of charged top-pions $\pcc$ and technipions $\paa$ and
$\pbb$ in the quark level processes $b \to s V^*$ with $V=\gamma,
gluon,\; Z$, and then combine the new physics contributions with
their SM counterparts, find the effective Wilson coefficients
and finally calculate the new physics contributions to the
branching ratios and
CP-violating asymmetries for all fifty seven decay modes under
consideration. We will concentrate on the new physics effects on charmless
$ B \to PP, PV$ decays and compare the theoretical predictions in TC2 model
with the SM predictions as well as the experimental measurements.
For the phenomenologically interesting $B \to  K \eta^{\prime}$ decays, we
found that the new physics  enhancements are significant in size,
$\sim 50\%$, insensitive to the variations of input parameters and hence
lead to a plausible interpretation for the large $B \to K \eta^{\prime} $
decay rates.

This paper is organized as follows. In Sec.2, we describe the basic
structures  of the TC2 model and examine the allowed parameter space of
the TC2 model from  currently available data. In Sec.3, we give a brief
review about the effective Hamiltonian, and then evaluate analytically the
new penguin diagrams and find the effective Wilson coefficients
$C_i^{eff}$ and effective numbers $a_i$ with the inclusion of new physics
contributions. In Sec.4 and 5, we calculate and show the numerical results
of branching  ratios  and CP-violating asymmetries for all fifty seven
$B \to PP, PV$ decay modes, respectively. We concentrate on modes with
well-measured branching ratio and sizable yields. The conclusions and
discussions are included in the final section.

\section{ TC2 model and experimental constraint} \label{sec:2}

Apart from some differences in group structure and/or particle contents,
all TC2 models \cite{hill95,lane96} have the following common
features:
(a) strong Topcolor interactions, broken near 1 TeV, induce a large
top condensate and all but a few GeV of the top quark mass, but contribute
little to electroweak symmetry breaking; (b) Technicolor
\cite{weinberg76}  interactions are
responsible for electroweak symmetry breaking, and Extended Technicolor
(ETC) \cite{etc79} interactions generate
the hard masses of all quarks and leptons, except that of the top quarks;
(c) there exist top-pions $\pcc$ and $\pcz$ with a decay constant
$\fpit \approx 50$ GeV.  In this paper we will chose the well-motivated
and most frequently studied TC2 model proposed by Hill \cite{hill95}
as the typical TC2 model to
calculate the contributions to the charmless hadronic B decays in question
from the relatively light unit-charged pseudo-scalars.
It is straightforward to extend the studies in this paper to
other TC2 models.

In the TC2 model\cite{hill95}, after integrating out the heavy coloron and
$Z'$, the effective four-fermion interactions have the
form \cite{buchalla96b}
\beq
{\cal L}_{eff} =\frac{4 \pi}{M_V^2} \left \{
 \left ( \kappa + \frac{2 \kappa_1}{27} \right )
 \overline{\psi}_L t_R \overline{t}_R \psi_L
+  \left ( \kappa - \frac{ \kappa_1}{27} \right )
\overline{\psi}_L b_R \overline{b}_R \psi_L \right \}, \label{eff1}
\eeq
where  $\kappa= (g_3^2/4\pi)\cot ^2\theta$ and
$\kappa_1= (g_1^2/4\pi)\cot ^2\theta'$, and $M_V$ is the mass of
coloron $V^\alpha$ and $Z'$. The effective interactions of (\ref{eff1}) can
be written in terms of two
auxiliary scalar doublets $\phi_1$ and $\phi_2$. Their couplings to quarks
are given by \cite{kominis95}
\beq
{\cal L}_{eff} = \lambda_1 \overline{\psi}_L \phi_1 \overline{t}_R
+ \lambda_2 \overline{\psi}_L \phi_2 \overline{b}_R, \label{eff2}
\eeq
where $\lambda_1^2 = 4\pi (\kappa + 2\kappa_1/27)$ and
$\lambda_2^2 = 4\pi (\kappa - \kappa_1/27)$. At energies below the Topcolor
scale $\Lambda \sim 1$ TeV the auxiliary fields acquire kinetic terms,
becoming physical degrees of freedom. The properly renormalized
$\phi_1$ and $\phi_2$ doublets take the form
\beq
\phi_1 = \left ( \begin{array}{cc}
\fpit + \frac{1}{\sqrt{2}}(h_t + i \pcz) \\ \pcm
\end{array} \right ), \ \
\phi_2 = \left ( \begin{array}{cc}
\tilde{H}^+\\ \frac{1}{\sqrt{2}}(\tilde{H}^0 + i \tilde{A}^0)
\end{array} \right ), \label{phi12}
\eeq
where $\pcc$ and $\pcz$ are the top-pions, $\tilde{H}^{\pm,0}$ and
$\tilde{A}^0$ are the b-pions, $h_t$ is the top-Higgs,
and $\fpit \approx 50 GeV$ is the top-pion decay constant.

From eq.(\ref{eff2}), the couplings of top-pions
to t- and b-quark can be written as \cite{hill95}:
\beq
\frac{m_t^*}{ \fpit } \left[ i\; \bar{t} t \tilde{\pi}^0 +
  i\; \overline{t}_R b_L \tilde{\pi}^+
+ i\; \frac{m_b^*}{m_t^*}  \overline{t}_L b_R \tilde{\pi}^+ + h.c.
\right], \eeq
where $m_t^* = (1-\epsilon) m_t$ and $m_b^* \approx 1 GeV$ denote the masses
of top and bottom quarks generated by topcolor interactions.

For the mass of top-pions, the current $1-\sigma$ lower mass bound from the
Tevatron data is $\mpcc \geq 150 GeV$\cite{lane96}, while the theoretical
expectation is $\mpcc \approx (150 - 300 GeV)$\cite{hill95}. For the mass
of b-pions, the current theoretical estimation is $m_{\tilde{H}^0} \approx
m_{\tilde{A}^0} \approx (100 - 350)  GeV$ and $\mpdd = m_{\tilde{H}^0}^2
+ 2 m_t^2$ \cite{burdman97}. For the technipions $\paa$ and
$\pbb$, the theoretical estimations are $\mpaa
\geq 50 GeV$ and $\mpbb \approx 200GeV$\cite{eichten86,epj981}.
The effective Yukawa couplings of ordinary technipions $\paa$ and $\pbb$ to
fermion pairs, as well as  the gauge couplings of unit-charged scalars to gauge
bosons $\gamma, Z^0$ and $gluon$ are basically model-independent,  can be
found in refs.\cite{eichten86,epj981,ellis81}.

At low energy, potentially large flavor-changing neutral currents (FCNC) arise
when the quark fields are rotated from their weak eigenbasis to their mass
eigenbasis, realized by the matrices $U_{L,R}$ for the up-type quarks,
and by $D_{L,R}$ for the down-type quarks. When we make the replacements,
for example,
\beq
b_L \to  D_L^{bd} d_L  +   D_L^{bs} s_L + D_L^{bb} b_L, \\
b_R \to D_R^{bd} d_R + D_R^{bs} s_R + D_R^{bb} b_R,
\eeq
the FCNC interactions will be induced. In TC2 model, the corresponding
flavor changing effective Yukawa  couplings  are
\beq
\frac{m_t^*}{\fpit} \left[
i\; \tilde{\pi}^+ ( D_L^{bs}\bar{t}_R  s_L +  D_L^{bd}\bar{t}_R d_L) +
i\; \tilde{H}^+ ( D_R^{bs} \bar{t}_L s_R +   D_R^{bd}\bar{t}_L d_R)
+ h.c. \right ].
\eeq

For the mixing matrices in the TC2 model, authors usually use the
``square-root ansatz":  to take the
square root of the standard model CKM matrix ($V_{CKM}=U_L^+ D_L$)
as an indication of the size of realistic mixings. It should be denoted that
the square root ansatz must be modified because of the strong constraint from
the data of $B^0 - \overline{B^0}$ mixing \cite{kominis95,lane96b,epj983}.
In TC2 model, the neutral scalars $\tilde{H}^0$ and $\tilde{A}^0$ can induce
a contribution to the $B_q^0-\overline{B_q^0}$ ($q=d, s$) mass difference
\cite{buchalla96b,kominis95}
\beq
\frac{\Delta M_{B_q}}{M_{B_q}}
= \frac{7}{12}\frac{m_t^2}{\fpit^2 m^2_{\tilde{H}^0}}
\delta_{bq}B_{B_q} F_{B_q}^2 ,
\label{deltabd}
\eeq
where $M_{B_q}$ is the mass of $B_q$ meson, $F_{B_q}$ is the $B_q$-meson
decay constant, $B_{B_q}$ is the renormalization group invariant parameter,
and $\delta_{bq} \approx |D_L^{bq}D_R^{bq}|$.
For $B_d$ meson, using the data of $\Delta M_{B_d}=(3.05 \pm 0.12)\times
10^{-10} MeV$\cite{pdg98} and setting $\fpit =50GeV$,
$\sqrt{B_{B_d}}F_{B_d}=200 MeV$,  one has the bound $\delta_{bd} \leq 0.76
\times 10^{-7}$ for $m_{\tilde{H}^0} \leq 600 GeV $.
This is an important and strong bound on the product of mixing elements
$D_{L,R}^{bd}$. As pointed in \cite{buchalla96b},
if one naively uses the square-root ansatz for {\em both}  $D_L$ and $D_R$,
this bound  is violated by about 2 orders of magnitudes.
The constraint on both $D_L$ and $D_R$ from the data of $\bsga$ decay is
weaker than that from the $B^0-\bar{B}^0$ mixings\cite{buchalla96b}.
By taking into account above experimental constraints, we naturally set that
$D_R^{ij}=0$ for $i\neq j$. Under this assumption, only the charged
technipions $\paa, \pbb$ and the charged top-pions $\pcc$ contribute to the
inclusive charmless decays $b \to s \bar{q} q,\; d \bar{q} q$ with $q\in \{ u,d,s\}$
through the strong and electroweak penguin diagrams.

In the numerical calculations, we will use the ``square-root ansatz" for
$D_L^{bd}$ and $D_L^{bs}$, i.e, setting $D_L^{bd}=V_{td}/2$ and
$D_L^{bs}=V_{ts}/2$, respectively.
We also fix the following parameters of the TC2 model in the numerical calculation
\footnote{From explicit numerical calculations in next section, we know that the
new physics contributions from technipions $\paa$ and $\pbb$ are much smaller
than those from top-pion $\pcc$ within the reasonable parameter space. We therefore
fix  $\mpaa=100$GeV and $\mpbb=200$GeV for the sake of simplicity. }:
\beq
\mpaa=100GeV,\; \mpbb=200GeV, \; \fpit=50 GeV, \; F_{\pi}=120 GeV,\;
\epsilon=0.05, \label{eq:tc2fix}
\eeq
where $F_{\pi}$ and $\fpit$ are the decay constants for technipions and
top-pions, respectively. For $\mpcc$, we consider the range of
$\mpcc=200 \pm 100 $ GeV to check the mass dependence of branching ratios
and CP-violating asymmetries of charmless B decays.

\section{ Effective Hamiltonian and Wilson coefficients } \label{sec:3}

We here present the well-known effective Hamiltonian for the two-body charmless
decays $B \to h_1 h_2$. For more details about the effective Hamiltonian with
generalized factorization for B decays one can see for example
refs.\cite{ali9804,chen99,buchalla96a,buras98}.

\subsection{Operators and Wilson coefficients in SM}

The standard theoretical frame to calculate the inclusive three-body decays
$b \to s \bar{q} q $  \footnote{For $b \to d \bar{q} q$ decays, one simply make the
replacement $s \to d$.} is based on the effective Hamiltonian
\cite{buras98,ali9804},
\beq
{\cal H}_{eff}(\Delta B=1) = \frac{G_F}{\sqrt{2}} \left \{
\sum_{j=1}^2 C_j \left ( V_{ub}V_{us}^* Q_j^u  + V_{cb}V_{cs}^* Q_j^c \right )
- V_{tb}V_{ts}^* \left [ \sum_{j=3}^{10}  C_j Q_j  + C_{g} Q_{g} \right ]
\right \}. \label{heff2}
\eeq
Here the operator basis reads:
\beq
Q_1&=& (\bar{s}q)_{V-A} (\bar{q}b)_{V-A},\ \
Q_2=   (\bar{s}_\alpha q_{\beta})_{V-A} (\bar{q}_{\beta}b_{\alpha})_{V-A},
\label{q1-q2}
\eeq
with $q=u$ and  $q=c$, and
\beq
Q_3&=& (\bar{s}b)_{V-A} \sum_{q'} (\bar{q'}q')_{V-A}, \ \
Q_4 = (\bar{s}_\alpha b_{\beta})_{V-A} \sum_{q'} (\bar{q'}_{\beta}q'_{\alpha})_{V-A}, \label{q3-q4} \\
Q_5&=& (\bar{s}b)_{V-A} \sum_{q'}
(\bar{q'}q')_{V+A}, \ \
Q_6= (\bar{s}_\alpha b_{\beta})_{V-A} \sum_{q'}
(\bar{q'}_{\beta}q'_{\alpha})_{V+A}, \label{q5-q6} \\
Q_7&=& \frac{3}{2}(\bar{s}b)_{V-A} \sum_{q'} e_{q'} (\bar{q'}q')_{V+A}, \ \
Q_8= \frac{3}{2} (\bar{s}_\alpha b_{\beta})_{V-A} \sum_{q'} e_{q'}
(\bar{q'}_{\beta}q'_{\alpha})_{V+A}, \label{q7-q8} \\
Q_9&=& \frac{3}{2}(\bar{s}b)_{V-A} \sum_{q'} e_{q'} (\bar{q'}q')_{V-A}, \ \
Q_{10} = \frac{3}{2} (\bar{s}_\alpha b_{\beta})_{V-A} \sum_{q'} e_{q'}
(\bar{q'}_{\beta}q'_{\alpha})_{V-A}, \label{q9-q10} \\
Q_{g}&=& \frac{g_s}{8\pi^2}m_b \bar{s}_\alpha \sigma^{\mu \nu}
(1+ \gamma_5)T^a_{\alpha \beta} b_{\beta} G^a_{\mu \nu},
\label{q8}
\eeq
where $\alpha$ and $\beta$ are the $SU(3)$ color indices, $T^a_{\alpha \beta}$
( $a=1,...,8$) are the Gell-Mann matrices. The sum over $q'$ runs over the
quark fields that are active at the scale $\mu=O(m_b)$, i.e., $q'\in \{u,d,s,c,b\}$.
The operator $Q_1$ and $Q_2$ are current-current operators, $Q_3 - Q_6$
are QCD penguin operators induced by gluonic penguin diagrams, and the
operators $Q_7 - Q_{10}$ are generated by electroweak
penguins and box diagrams. The overall factor $2/3$ is introduced
for convenience, and the charge $e_{q'}$ is the charge of the quark $q'$ with
$q'=u,d,s,c,b$. The operator $Q_{g}$ is the chromo-magnetic dipole operator generated from the
magnetic gluon penguin. Following ref.\cite{ali9804}, we also neglect
the effects of the electromagnetic penguin operator $Q_{7\gamma}$, and do not
consider the effect of the weak annihilation and exchange diagrams.

Within the SM and at scale $\mw$, the Wilson coefficients $C_1(M_W), \cdots,
C_{10}(\mw)$ and $C_{g}(\mw)$ have been given for example in
\cite{buchalla96a,buras98}. They read in the naive dimensional
regularization (NDR)  scheme
\beq
C_1(\mw) &=& 1 - \frac{11}{6} \; \frac{\as(\mw)}{4\pi}
               - \frac{35}{18} \; \frac{\aem}{4\pi} \, , \non
C_2(\mw) &=&     \frac{11}{2} \; \frac{\as(\mw)}{4\pi} \, , \non
C_3(\mw) &=& -\frac{\as(\mw)}{24\pi} \left [ E_0(x_t) -\frac{2}{3} \right ]
             +\frac{\aem}{6\pi} \frac{1}{\sin^2\theta_W}
             \left[ 2 B_0(x_t) + C_0(x_t) \right] \, ,\non
C_4(\mw) &=& \frac{\as(\mw)}{8\pi} \left [ E_0(x_t) -\frac{2}{3} \right ] \, ,\non
C_5(\mw) &=& -\frac{\as(\mw)}{24\pi} \left [E_0(x_t) -\frac{2}{3} \right ] \, , \non
C_6(\mw) &=& \frac{\as(\mw)}{8\pi} \left [E_0(x_t) -\frac{2}{3} \right ] \, ,\non
C_7(\mw) &=& \frac{\aem}{6\pi} \left[ 4 C_0(x_t) + D_0(x_t) -\frac{4}{9}\right]\, , \non
C_8(\mw) &=& 0 \, , \non
C_9(\mw) &=& \frac{\aem}{6\pi} \left[ 4 C_0(x_t) + D_0(x_t)-\frac{4}{9} +
             \frac{1}{\sin^2\theta_W} (10 B_0(x_t) - 4 C_0(x_t)) \right] \, ,\non
C_{10}(\mw) &=& 0 \, , \label{eq:cimwsm}\\
C_{g}(\mw) &=& -\frac{E'_0(x_t)}{2}\, , \label{eq:c8gmwsm}
\eeq
where $x_t=m_t^2/M_W^2$, the functions $B_0(x)$, $C_0(x)$, $D_0(x)$, $E_0(x)$ and
$E'_0(x)$ are the familiar Inami-Lim functions \cite{inami81},
\beq
B_0(x) &=& \frac{1}{4} \left[ \frac{x}{1-x} + \frac{x \ln x}{(x-1)^2}
\right]\, , \label{eq:b0xt} \\
C_0(x) &=& \frac{x}{8} \left[ \frac{x-6}{x-1} + \frac{3 x + 2}{(x-1)^2}
\ln x \right]\, , \label{eq:c0xt} \\
D_0(x) &=& -\frac{4}{9} \ln x + \frac{-19 x^3 + 25 x^2}{36 (x-1)^3} +
         \frac{x^2 (5 x^2 - 2 x - 6)}{18 (x-1)^4} \ln x \, , \label{eq:d0xt} \\
E_0(x)&=& \frac{18x -11 x^2 - x^3}{12(1-x)^3}
-\frac{4 -16x +9 x^2}{6 (1-x)^4} \ln[x], \label{eq:e0xt}  \\
E'_0(x)&=& \left [ \frac{2x + 5x^2 - x^3}{4(1-x)^3}
    + \frac{3x^2}{2(1-x)^4}\log[x] \right ].\label{eq:e0pxt}
\end{eqnarray}
Here function $B_0(x)$ results from the evaluation of the box diagrams with
leaving lepton pair $\nu \bar{\nu}$ or $l^+ l^-$\cite{buras98}, function
$C_0(x)$ from the $Z^0$-penguin, function $D_0(x)$ and $E_0(x)$ from the
photon penguin and the gluon penguin diagram respectively,
and finally function $E'_0(x)$ arise from the magnetic gluon penguin.

By using QCD renormalization group equations\cite{buchalla96a,buras98}, it is
straightforward to run Wilson coefficients $C_i(\mw)$ from the scale $\mu =0( \mw)$
down to the lower scale $\mu =O(m_b)$. Working consistently  to the
next-to-leading order ( NLO ) precision,
the Wilson coefficients $C_i$ for $i=1,\ldots,10$ are needed in NLO precision,
while it is sufficient to use the leading logarithmic value for $C_{g}$.
At NLO level, the Wilson coefficients are usually renormalization scheme(RS)
dependent. In the NDR scheme, by using the input parameters as given in
Appendix A and setting $\mu=2.5$ GeV,  we find:
\beq
 C_1 & = & 1.1245,\ \   C_2 = - 0.2662,\;  C_3 = 0.0186,\; C_4 = -0.0458,\non
 C_5 &=& 0.0113, \; C_6  =  -0.0587, \; C_7 = -5.5\times 10^{-4}, \;
 C_8  = 6.8\times 10^{-4},\non
 C_9 & =&  -0.0095,\;  C_{10} =  0.0026,\;  C_g^{eff}  =  -0.1527.
 \label{eq:cimbsm}
\eeq
Here, $C_g^{eff} = C_{g}+ C_5$. These NLO Wilson coefficients are renormalization
scale and scheme dependent, but such dependence will be cancelled by the
corresponding dependence in the matrix elements of the operators in ${\cal
H}_{eff}$, as shown explicitly in \cite{buras98,fleischer93}.

\subsection{New strong and electroweak penguins in TC2 model}

For the charmless hadronic decays of B meson under consideration, the new physics
will manifest itself by modifying the corresponding Inami-Lim functions $C_0(x),
D_0(x), E_0(x)$
and $E'_0(x)$ which determine the coefficients $C_3(\mw), \ldots, C_{10}(\mw)$ and
$C_{g}(\mw)$, as illustrated in Eqs.(\ref{eq:cimwsm},\ref{eq:c8gmwsm}).
These modifications, in turn, will change for example the standard
model predictions for the branching ratios and CP-violating asymmetries
for decays $B \to h_1 h_2$.

\begin{figure}
\begin{center}
\begin{picture}(400,440)(0,0)
\put(-60,-140) {\epsfxsize200mm\epsfbox{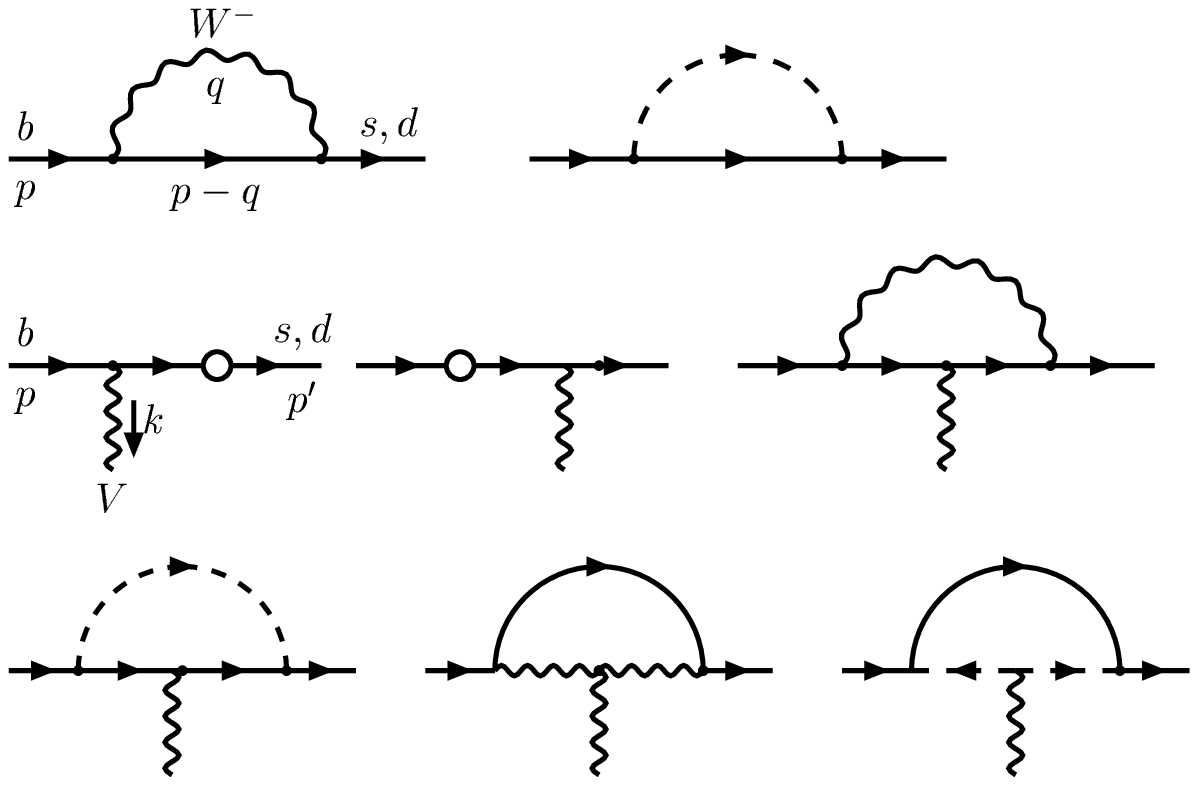}}
\end{picture}
\end{center}
\vspace{-80pt}
\caption{Typical self-energy and penguin diagrams for the quark level decays
$b \to (s, d) V^*$ ($V=\gamma, Z^0, g$),  with $W^{\pm}$ (internal wave lines)
and charged Pseudo-scalar exchanges (internal dash lines) in the SM and TC2 model.
The internal quarks are the upper type quark $u, c$ and $t$.}
\label{fig:fig1}
\end{figure}

The new strong and electroweak penguin diagrams can be obtained from the
corresponding
penguin diagrams in the SM by replacing the internal $W^{\pm}$ lines with the
unit-charged scalar ($\paa, \pbb$ and $\pcc$ ) lines, as shown in \fig{fig:fig1}.
In the analytical calculations of those penguin diagrams, we will use
dimensional regularization to regulate all the ultraviolet divergences in the
virtual loop corrections and adopt the $\overline{MS}$ renormalization
scheme.  It is easy to show that all ultraviolet divergences are canceled for
each kind of charged scalars, respectively.

Following the same procedure of refs.\cite{epj983,inami81}, we calculate
analytically the new $Z^0$-penguin diagrams induced by the exchanges
of charged scalars
$\paa, \pbb$ and $\pcc$, we find the new $C_0$ function which describe the NP
contributions to the Wilson coefficients through the new $Z^0$-penguin diagrams,
\beq
C_0^{TC2} &=&\frac{1}{\sqrt{2} G_F \mw^2 }\left [
\frac{\mpcc^2}{4\fpit^2} T_0(y_t)
+ \frac{\mpaa^2}{3 F_\pi^2 } T_0(z_t)
+ \frac{8 \mpbb^2}{3 F_\pi^2 } T_0(\xi_t) \right], \label{eq:c0tc2}
\eeq
with
\beq
T_0(x)&=&-\frac{x^2}{8 (1-x)} -\frac{x^2}{8 (1-x)^2} \log[x], \label{eq:t0x}
\eeq
where $y_t=m_t^{*2}/\mpcc^2$ with $m_t^*=(1-\epsilon) m_t$,
$z_t=( \epsilon m_t)^2/\mpaa^2$,$\xi_t=( \epsilon m_t)^2/\mpbb^2$.

By evaluating the new $\gamma$-penguin diagrams induced by the exchanges
of three kinds of charged pseudo-scalars ($\pcc, \paa, \pbb$), we find that,
\beq
D_0^{TC2} &=& \left \{ \frac{1}{4 \sqrt{2} G_F \fpit^2 } F_0(y_t)
 + \frac{1}{3 \sqrt{2} G_F F_\pi^2 } \left [  F_0(z_t) +
 8 F_0(\xi_t) \right ]  \right\}, \label{eq:d0tc2}
\eeq
with
\beq
F_0(x)&=& \frac{47-79 x + 38 x^2}{108 (1-x)^3}
     + \frac{3 -6 x^2+4 x^3}{18 (1-x)^4}\log[x]. \label{eq:f0tc2}
\eeq

By evaluating the new $gluon$-penguin diagrams induced by the exchanges
of three kinds of charged pseudo-scalars ($\pcc, \paa, \pbb$) as being
done in \cite{xiao96,xiao20},  we find that,
\beq
E_0^{TC2} &=& \left \{ \frac{1}{4\sqrt{2} G_F \fpit^2 } I_0(y_t)
 + \frac{1}{3 \sqrt{2} G_F F_\pi^2 } \left [  I_0(z_t) +
 8 I_0(\xi_t) + 9 N_0(\xi_t)\right ]  \right\}, \label{eq:e0tc2}\\
{E'}_0^{TC2}&=& \left \{ \frac{1}{8 \sqrt{2} G_F \fpit^2 } K_0(y_t)
 + \frac{1}{6 \sqrt{2} G_F F_\pi^2 } \left [  K_0(z_t) +
 8 K_0(\xi_t) + 9 L_0(\xi_t)\right ]  \right\}, \label{eq:e0ptc2}
\eeq
with
\beq
I_0(x)&=&  \frac{ 7-29 x + 16 x^2}{36 (1-x)^3}
    -\frac{3 x^2-2 x^3}{6 (1-x)^4} \log[x], \label{eq:i0xtc2} \\
K_0(x)&=& -\frac{5-19 x + 20 x^2}{6 (1-x)^3} +
    \frac{x^2-2 x^3}{(1-x)^4}\log[x], \label{eq:k0xtc2}\\
L_0(x)&=& -\frac{4-5x - 5x^2}{6(1-x)^3}  -
    \frac{x-2x^2}{(1-x)^4}\log[x], \label{eq:l0xtc2}\\
N_0(x)&=&  \frac{11-7x + 2x^2}{36(1-x)^3} + \frac{1}{6(1-x)^4}
    \log[x]. \label{eq:n0xtc2}
\eeq

Using the input parameters as given in Appendix A and Eq.(\ref{eq:tc2fix}),
and assuming $\mpcc=200$GeV,  we find numerically that
\beq
\{C_0, D_0, E_0, {E'}_0 \}^{TC2}|_{\mu=\mw}= \{1.27, 0.27, 0.66, -1.58\}
\eeq
if only the new contributions from top-pion penguins are included, while
\beq
\{ C_0, D_0, E_0, {E'}_0 \}^{TC2}|_{\mu=\mw}= \{0.0002, 0.03, 0.04, -0.14\}
\eeq
if only the new contributions from technipion penguins are included. It is
evident that it is the charged top-pion $\pcc$ that strongly dominate the
NP contributions, while the technipions play  a  minor rule only. We therefore
fix the masses of $\paa$ and $\pbb$ in the following numerical calculations.

Using the input parameters as given in Appendix A and Eq.(\ref{eq:tc2fix}) and
assuming $\mpcc=200$ GeV, we find that
\beq
\{ C_0, D_0, E_0, E'_0\}^{SM}|_{\mu=\mw}&=&\{0.81,-0.48,0.27,0.19\},
\label{eq:c0d0sm}\\
\{ C_0, D_0, E_0, E'_0 \}^{TC2}|_{\mu=\mw}&=&\{1.27,0.30,0.71,-1.72
\}.\label{eq:cdeeptc2}
\eeq
It is easy to see that the new physics parts of the functions under study are
comparable in size with their SM counterparts. The SM
predictions, consequently, can be changed significantly through interference.
For $C_0$ and $E_0$ functions, they will interfere constructively.
For $D_0$ and $E'_0$ functions, on contrary, they will interfere
destructively. One also should note that the magnitude of ${E'}_0^{TC2}$ is
much larger than its SM counterpart, and hence  ${E'}_0^{TC2}$ will dominate
in the interference. We will combine the two
pats of the corresponding functions to define the functions as follows,
\beq
C_0(M_W)&=& C_0(M_W)^{SM} + C_0(M_W)^{TC2}, \non
D_0(M_W)&=& D_0(M_W)^{SM} + D_0(M_W)^{TC2}, \non
E_0(M_W)&=& E_0(M_W)^{SM} + E_0(M_W)^{TC2},\non
E'_0(M_W)&=&{E'}_0(M_W)^{SM} + {E'}_0(M_W)^{TC2}, \label{eq:cdeep}
\eeq
where the functions $D_0(M_W)^{SM}$, $E_0(M_W)^{SM}$, $C_0(M_W)^{SM}$
and $E'_0(M_W)^{SM}$ have been given in Eqs.(\ref{eq:c0xt},\ref{eq:d0xt},
\ref{eq:e0xt},\ref{eq:e0pxt}), respectively. While the functions $C_0(M_W)^{TC2}$,
$D_0(M_W)^{TC2}$, $E_0(M_W)^{TC2}$ and $E'_0(M_W)^{TC2}$ have also been defined in
Eqs.(\ref{eq:c0tc2},\ref{eq:d0tc2},\ref{eq:e0tc2},\ref{eq:e0ptc2}), respectively.

Since the heavy charged pseudo-scalars appeared in TC2 model have been integrated out
at the scale $\mw$, the QCD running of the Wilson coefficients $C_i(\mw)$ down to the
scale $\mu=O(m_b)$ after including the NP contributions will be the same
as in the SM.
In the NDR scheme, by using the input parameters as given in Appendix A and
Eq.(\ref{eq:tc2fix}), and setting $\mpcc=200$ GeV and $\mu=2.5$ GeV,  we find that:
\beq
 C_1 & = & 1.1245,\; C_2 = - 0.2662,\;  C_3  =  0.0195,\; C_4 = -0.0441,\non
 C_5 &=& 0.0111, \;  C_6  =  -0.0535, \; C_7  =  0.0026, \; C_8  = 0.0018,\non
 C_9 &=& -0.0175,\; C_{10} = 0.0049, \; C_g^{eff}  =  0.3735, \label{eq:cimbtc2}
\eeq
where $C_g^{eff} = C_{g}+ C_5$. By Comparing the Wilson coefficients in
Eq.(\ref{eq:cimbtc2}) with those given in Eq.(\ref{eq:cimbsm}), we find that
$C_{1,2}$ remain unchanged, $C_{3,4,5,6}$ changed moderately, $C_{7,8,9,10}$
and $C_g^{eff}$ changed significantly because of the inclusion of new physics
contributions.

\subsection{Effective Wilson coefficients}

Using the generalized factorization approach for nonleptonic B meson
decays, the renormalization scale- and
scheme-independent effective Wilson coefficients $C_i^{eff}$
($i=1,\ldots,10$) have been obtained in \cite{cheng98,ali98,ali9804} by
adding to $C_i(\mu)$ the
contributions from vertex-type quark matrix elements, denoted by
anomalous dimensinal matrix $\gamma_V$ and constant matrix
$\hat{r}_V$ as given for example in \cite{ali9804}.
Very recently, Cheng {\it et al.} \cite{cheng99a} studied and resolved
the so-called gauge and infrared problems \cite{bs98} of generalized
factorization approach. They found that the gauge invariance is maintained
under radiative corrections by working in the physical
on-mass-shell scheme, while the infrared divergence in radiative
corrections should be isolated using the dimensional
regularization  and the resultant infrared poles are absorbed into
the universal meson wave functions \cite{cheng99a}.

In the NDR scheme and for $SU(3)_C$, the effective
Wilson coefficients $C_i^{eff}$ can be written as \cite{ali9804,chen99},
\beq
C_1^{eff} &=& C_1 +
\frac{\alpha_s}{4\pi} \, \left( \hat{r}_V^T +
 \gamma_{V}^T \log \frac{m_b}{\mu}\right)_{1j} \, C_j ~,\non
C_2^{eff} &=& C_2 +
\frac{\alpha_s}{4\pi} \, \left( \hat{r}_V^T +
 \gamma_V^T \log \frac{m_b}{\mu} \right)_{2j} \, C_j  ~,\non
C_3^{eff} &=& C_3  + \frac{\alpha_s}{4\pi} \, \left(  \hat{r}_V^T +
\gamma_V^T \log \frac{m_b}{\mu}\right)_{3j} \, C_j
-\frac{1}{6} \frac{\alpha_s}{4\pi} \, (C_t + C_p + C_g) ~,\non
C_4^{eff} &=& C_4 +\frac{\alpha_s}{4\pi} \, \left( \hat{r}_V^T +
\gamma_V^T \log \frac{m_b}{\mu}\right)_{4j} \, C_j
+\frac{1}{2} \frac{\alpha_s}{4\pi} \, (C_t + C_p + C_g)  ~, \non
C_5^{eff} &=& C_5 +\frac{\alpha_s}{4\pi} \, \left(  \hat{r}_V^T +
\gamma_V^T \log \frac{m_b}{\mu}\right)_{5j} \, C_j
-\frac{1}{6} \frac{\alpha_s}{4\pi} \, (C_t + C_p + C_g)  ~, \non
C_6^{eff} &=& C_6  +\frac{\alpha_s}{4\pi} \, \left(  \hat{r}_V^T +
\gamma_V^T \log \frac{m_b}{\mu} \right)_{6j} \, C_j
+\frac{1}{2} \frac{\alpha_s}{4\pi} \, (C_t + C_p + C_g) ~,  \non
C_7^{eff} &=& C_7 + \frac{\alpha_s}{4\pi} \, \left(  \hat{r}_V^T +
\gamma_V^T \log \frac{m_b}{\mu} \right)_{7j} \, C_j
+  \frac{\alpha_{ew}}{8\pi} C_e ~, \non
C_8^{eff} &=& C_8 + \frac{\alpha_s}{4\pi} \, \left(  \hat{r}_V^T +
\gamma_V^T \log \frac{m_b}{\mu} \right)_{8j} \, C_j ~, \non
C_9^{eff} &=& C_9 + \frac{\alpha_s}{4\pi} \, \left(  \hat{r}_V^T +
\gamma_V^T \log \frac{m_b}{\mu} \right)_{9j} \, C_j
+  \frac{\alpha_{ew}}{8\pi} C_e ~, \non
C_{10}^{eff} &=& C_{10} + \frac{\alpha_s}{4\pi} \, \left(  \hat{r}_V^T +
\gamma_V^T \log \frac{m_b}{\mu} \right)_{10j} \, C_j ~, \label{eq:wceff}
\eeq
where the matrices  $\hat{r}_V$ and $\gamma_V$ contain the process-independent
contributions from the vertex diagrams. Like ref.\cite{chen99}, we
here include vertex corrections to $C_7 -C_{10}$\footnote{Numerically, such
corrections are negligibly small.}. The anomalous dimension matrix
$\gamma_V$ has been given explicitly, for example, in Eq.(2.17) of
\cite{chen99}. Note that the correct value of
the element $(\hat{r}_{NDR})_{66}$ and  $(\hat{r}_{NDR})_{88}$ should be  17
instead of 1 as pointed in \cite{cheng00a}, $\hat{r}_V$ in the NDR scheme
takes the form
\begin{equation}
\hat{r}_V^{NDR} = \left(
\begin{array}{ccccc ccccc}
3 & -9 & 0 & 0 & 0 & 0& 0 & 0 & 0 & 0 \\
-9 & 3 & 0 & 0 & 0 & 0& 0 & 0 & 0 & 0 \\
0 & 0  & 3 &-9 & 0 & 0& 0 & 0 & 0 & 0 \\
0 & 0  &-9 & 3 & 0 & 0& 0 & 0 & 0 & 0 \\
0 & 0  & 0 & 0 & -1& 3& 0 & 0 & 0 & 0 \\
0 & 0  & 0 & 0 & -3&17& 0 & 0 & 0 & 0 \\
0 & 0  & 0 & 0 & 0& 0&-1 & 3 & 0 & 0 \\
0 & 0  & 0 & 0 & 0& 0&-3 &17 & 0 & 0 \\
0 & 0  & 0 & 0 & 0& 0& 0 & 0 & 3 &-9 \\
0 & 0  & 0 & 0 & 0& 0& 0 & 0 &-9 & 3 \\
\end{array} \right) \quad .
\end{equation}

The function $C_t$, $C_p$, and $C_g$ in Eq.(\ref{eq:wceff})
describe the contributions arising from the penguin diagrams of the current-current
$Q_{1,2}$ and the QCD operators $Q_3$-$Q_6$, and the tree-level diagram of the
magnetic dipole operator $Q_{g}$, respectively. We here also follow the
procedure of ref.\cite{ali98} to include the contribution of magnetic gluon
penguin operator $Q_g$. The effective Wilson coefficients in
Eq.(\ref{eq:wceff}) are now
renormalization-scheme and -scale independent and do not suffer from gauge
and infrared problems.
The functions $C_t$, $C_p$, and $C_g$ are given in  the NDR scheme by
\cite{ali9804,chen99}\footnote{The constant term $2/3$ in front of
$C_4 + C_6$ in $C_p$ was missed in \cite{ali9804}, but recovered
firstly in \cite{chen99}.}
\beq
C_t &=& \left [ \frac{2}{3} + {\lambda_u\over\lambda_t}G(m_u)
    +{\lambda_c\over\lambda_t} G(m_c) \right ] C_1,\label{cct} \\
C_p &=& \left [\frac{4}{3} - G(m_q) - G(m_b) \right ] C_3
     + \sum_{i=u,d,s,c,b} \left [ \frac{2}{3}- G(m_i) \right ] (C_4+C_6),  \label{ccp}\\
C_e &=& {8\over9}\left [ \frac{2}{3} +  {\lambda_u\over\lambda_t} G(m_u)
    + {\lambda_c\over\lambda_t} G(m_c) \right ] (C_1+3C_2),\label{cce}\\
C_g &=& -{2m_b\over \sqrt{< k^2>}}C^{\rm eff}_{g}, \label{ccg}
\eeq
with $\lambda_{q'}\equiv V_{q'b}V_{q'q}^*$. The function $G(m,k,\mu)$
is of the form\cite{abel98}
\beq
G(m,k,\mu)&=& \frac{10}{9}-\frac{2}{3}\ln[\frac{m^2}{\mu^2}] + \frac{2\mu^2}{3 m^2}
-\frac{2(1 + 2 z )}{3z } g(z)~,
\eeq
where $z=k^2/(4m^2)$, and
\beq
g(z) = \left \{\begin{array}{ll}
\sqrt{\frac{1-z}{z}} \arctan[\frac{z}{1-z}], & z < 1, \\
\sqrt{\frac{1-z}{4z}}\left [
\ln[\frac{\sqrt{z} + \sqrt{z-1}}{\sqrt{z}-\sqrt{z-1}}] - i \pi \right ], &
z >1, \\
\end{array} \right. \label{eq:gz}
\eeq
where $k^2$ is the momentum squared transferred by the gluon, photon or $Z$
to the $q^\prime \overline{q^\prime}$ pair in inclusive three-body decays
$b \to q q^\prime \overline{q^\prime}$, and $m$ is the mass of internal
up-type quark in the penguin diagrams. For $k^2>4m^2$, an imaginary part
of $g(z)$ will appear because of the generation
of a strong phase at the $\bar{u}u$ and $\bar{c}c$ threshold
\cite{abel98,bss79,hou91}.

For the two-body exclusive B meson decays any information on $k^2$ is
lost in the factorization assumption, and it is not clear what "relevant" $k^2$
should be taken in numerical calculation.  Based on  simple estimates
involving two-body kinematics \cite{deshpande90} or  the investigations in
first paper of ref.\cite{bh1h2}, one usually  use the "physical" range for
$k^2$ \cite{deshpande90,hou91,fleischer93,ali9804,chen99},
\beq
\frac{m_b^2}{4}\stackrel{<}{\sim} k^2
\stackrel{<}{\sim}\frac{m_b^2}{2}. \label{eq:k2}
\eeq
Following refs.\cite{ali9804,chen99}, we use $k^2=m_b^2/2$ in the numerical
calculation and will consider the $k^2$-dependence of branching ratios and
CP-violating asymmetries of charmless  B meson decays.
In fact, branching ratios considered here are not sensitive to
the value of $k^2$ within the reasonable range of $k^2$, but the CP-violating
asymmetries are sensitive to the variation of $k^2$.

\section{ Branching ratios of $B \to PP, PV$ decays} \label{sec:br}

In numerical calculations, we focus on the new physics effects on the branching
ratios and CP-violating asymmetries for $B \to PP, PV$ decays.
For the standard model part, we will follow the procedure of
refs.\cite{ali9804} and compare our SM results with those given
in \cite{ali9804,chen99}.
Two sets of form factors at the zero momentum transfer from the  BSW model
\cite{bsw87}, as well as Lattice QCD and Light-cone QCD sum rules (LQQSR)
will be used, respectively. Explicit values
of these form factors can be found in \cite{ali9804} and have also been
listed in Appendix B.

Following \cite{ali9804}, the fifty seven decay channels under study in this
paper are also  classified into five different classes (for more details
about classification, see \cite{ali9804}) as listed in the tables.
The first three and last two classes are tree-dominated and
penguin-dominated decays, respectively.

\begin{itemize}

\item Class-I:  including four decay modes, $B^0 \to \pi^-\pi^+,
\rho^{\pm}\pi^{\mp}$ and $B^0 \to \rho^- K^+$, the large and $N_c^{eff}$
stable coefficient $a_1$ play the major role.

\item
Class-II: including ten decay modes, for example $B^0 \to \pi^0 \pi^0$, and
the relevant coefficient for these decays is $a_2$ which shows a strong
$N_c^{eff}$-dependence.

\item
Class-III: including nine decay modes involving the interference of class-I
and class-II decays, such as the decays  $B^+ \to \pi \etap$.

\item
Class-IV: including twenty two  $B \to PP, PV$ decay modes such as
$B \to K \etapp$ decays. The amplitudes of these decays involve one (or more) of
the dominant penguin coefficients $a_{4,6,9}$ with constructive interference
among them. The Class-IV decays are $N_c^{eff}$ stable.

\item
Class-V:  including twelve $B \to PP, PV$ decay modes, such as
$B \to \pi^0 \etapp$ and $B \to \phi K$ decays. Since the amplitudes of
these decays involve large and delicate cancellations due to
interference between strong $N_c^{eff}$-dependent coefficients $a_{3,5,7,10}$
and the dominant penguin coefficients $a_{4,6,9}$,
these decays are generally not stable against $N_c^{eff}$.

\end{itemize}

\subsection{Decay amplitudes in BSW model}

With the factorization ansatz \cite{bsw87,ali9804,chen99}, the
three-hadron matrix elements or the decay amplitude $<XY|H_{eff}|B>$
can be factorized into a sum of products of two current matrix elements
$<X|J_1^\mu|0>$ and $<Y|J_{2\mu}|B>$ ( or $<Y|J_1^\mu|0>$ and
$<X|J_{2\mu}|B>$). The former matrix
elements are simply given by the corresponding decay constants $f_X$ and
$g_X$\cite{bijnens92}
\beq
<0|J_\mu|X(0^-)> = i f_X k_\mu, \; <0|J_\mu|X(1^-)> = M_X g_X \epsilon_\mu,
\label{eq:e1}
\eeq
where $f_X$ ($g_X$ ) is the decay constant of pesudoscalar (vector) meson,
$\epsilon_\mu$ is the polarization vector of the vector meson. For the second
matrix element $<Y|J_{2\mu}|B>$, the expression in terms of Lorentz-scalar
form factors\cite{bsw87,bijnens92}, are of the form
\beq
<X(0^-)|J_{\mu}|B>&=& \left[ \left( k_B + k_X \right)_{\mu}-
       \frac{M_B^2-M_X^2}{k^2} k_{\mu} \right ] F_1^{B\to X}(k^2)\non
          && + \frac{M_B^2-M_X^2}{k^2} k_{\mu} F_0^{B\to
          X}(k^2),\label{eq:bp}\\
<X(1^-)|J_{\mu}|B>&=& \frac{2}{M_B+M_X} \epsilon_{\mu \nu
             \rho \sigma}\epsilon^{*\nu} k_B^{\rho} k_X^{\sigma}V^{B\to X}(k^2)
             + i \epsilon^* \cdot k\, \frac{2 M_X}{k^2}\, k_{\mu} A_0(k^2) \non
        && + i \left( M_B+ M_X \right)\left[ \epsilon_{\mu}^*-
          \frac{\epsilon^*\cdot {k}}{k^2} k_{\mu}\right]A_1(k^2)\non
       && -i \frac{\epsilon^* \cdot k}{M_B+M_X} \left [(k_B+k_X)_{\mu}-\frac
        {M_B^2-M_X^2}{k^2} k_{\mu} \right ] A_2(k^2),\label{eq:bv}
\eeq
where $ k^{\mu}=k_B^{\mu}-k_X^{\mu}$ and $M_B$, $M_X$, $M_Y$ are the
masses of meson B, X and Y, respectively. The explicit expressions of form factors
$F_{0,1}(k^2), V(k^2)$ and $A_{0,1,2}(k^2)$ have been given in Appendix B.

In the generalized factorization ansatz \cite{ali9804,chen99}, the
effective Wilson coefficients
$C_i^{eff}$ will appear in the decay amplitudes in the combinations,
\beq
a_{2i-1}\equiv C_{2i-1}^{eff} +\frac{{C}_{2i}^{eff}}{N_c^{eff}}, \ \
a_{2i}\equiv C_{2i}^{eff}     +\frac{{C}_{2i-1}^{eff}}{N_c^{eff}}, \ \ \
(i=1,\ldots,5) \label{eq:ai}
\eeq
where the effective number of colors $N_c^{eff}$ is treated as a
free parameter varying in the range of $2 \leq N_c^{eff} \leq \infty$,
in order to get a primary
estimation about the size of non-factorizable contribution to the hadronic
matrix elements. It is evident that  the reliability of generalized
factorization approach has been improved since the effective Wilson
coefficients $C_i^{eff}$ appeared in Eq.(\ref{eq:ai}) are now gauge
invariant and infrared safe. Although  $N_c^{eff}$ can in principle
vary from channel to channel, but in the energetic two-body hadronic B
meson decays, it is expected to be process insensitive as supported by
the data \cite{chen99}.  As argued in ref.\cite{cheng98},
$N_c^{eff}(LL)$ induced by the $(V-A)(V-A)$ operators can be rather
different from $N_c^{eff}(LR)$ generated by  $(V-A)(V+A)$ operators.
Since we here focus on the calculation of new physics
effects on the studied B meson decays induced by the new
penguin diagrams in the TC2 model, we will simply assume that
$N_c^{eff}(LL)\equiv N_c^{eff}(LR)=N_c^{eff}$ and consider the variation
of $N_c^{eff}$ in the range of $2 \leq N_c^{eff} \leq \infty$.
For more details about the cases of
$N_c^{eff}(LL)\neq N_c^{eff}(LR)$, one can see for example ref.\cite{chen99}.
We here will also not  consider the possible effects of final state
interaction (FSI) and the contributions from annihilation channels
although they may play a significant rule for some $B \to PV, VV$ decays.

The effective coefficients $a_i$
are displayed in the \tab{ai:bd} and \tab{ai:bs} for the transitions $b\to d$
( $\bar{b} \to \bar{d}$ ) and $b\to s$ ($\bar{b} \to \bar{s}$ ), respectively.
Theoretical predictions of $a_i$ are made by using the input parameters
as given in Appendix A and Eq.{\ref{eq:tc2fix}}, and assuming
$ k^2=m_b^2/2 $ and $\mpcc=200 GeV$. For coefficients $a_3, \ldots, a_{10}$,
the first and second entries in tables (\ref{ai:bd},\ref{ai:bs}) refer to
the values of $a_i$ in the SM and TC2 model respectively.

The new physics effects on the B decays under study will be included by using
the modified effective coefficients $a_i$ ($i=3,\dots,10$) as given in the
second entries of \tab{ai:bd} and \tab{ai:bs}. In the numerical calculations
the input parameters as given in Appendix A, B and Eq.(\ref{eq:tc2fix}) will
be used implicitly.

\begin{table}[tb]
\begin{center}
\caption{Numerical values of $a_i$ for the transitions $b \to d$ [$\bar{b}
\to \bar{d}$ ].  The first and second entries for $a_3, \ldots, a_{10}$ refer
to the values of $a_i$ in the SM and TC2 model respectively. The entries for $a_3,
\ldots, a_{10}$ should be multiplied with $10^{-4}$. } \label{ai:bd}
\vspace{0.2cm}
\begin{tabular} {|l|c|c|c|} \hline \hline
        & $N_c^{eff}=2$                 & $N_c^{eff}=3$ & $N_c^{eff}=\infty$  \\ \hline
 $a_1$  &$0.995 \;[0.995]$           &$1.061\;[1.061]$           & $1.192  \;[1.192]$           \\
 $a_2$  &$0.201 \;[0.201]$           &$0.026\;[0.026]$           & $-0.395 \;[-0.395]$         \\
 $a_3$  &$-16-7i\;[-25-23i]$        &$77    \;[77]$              & $261+13i\;[280+47i]$       \\
        &$-26-8i\;[-35-24i]$        &$90    \;[90]$              & $322+15i\;[340+49i]$       \\
 $a_4$  &$-423-33i\;[-470-117i]$    &$-467-35i\;[-517-125i]$    & $-554-39i\;[-610-141i]$    \\
        &$-534-38i\;[-580-122i]$    &$-588-40i\;[-638-130i]$    & $-695-45i\;[-751-146i]$   \\
 $a_5$  &$-192-7i \;[-202-23i]$      &$-71    \;[-71]$          & $171+ 13i\;[190 + 47i]$    \\
        &$-195-8i \;[-205-24i]$      &$-57    \;[-57]$          & $218+ 15i\;[237 + 49i]$    \\
 $a_6$  &$-642-33i\;[-689-117i]$    &$-671-35i\;[-721-125i]$    & $-728-39i\;[-784-141i]$    \\
        &$-718-38i\;[-764-122i]$    &$-754-40i\;[-804-130i]$    & $-827-45i\;[-884-146i]$  \\ \hline
 $a_7$  &$8.1-0.9i\;[7.7-1.7i]$     &$6.9-0.9i\;[6.4-1.7i]$     & $4.3-0.9i\;[3.9-1.7i]$     \\
        &$34-0.9i \;[34-1.7i]$      &$31-0.9i\;[30-1.7i]$       & $24.3-0.9i\;[23.9-1.7i]$   \\
 $a_8$  &$9.7-0.5i  \;[9.5-0.8i]$   &$9.0-0.3i  \;[8.8-0.6i]$   & $7.5     \;[7.5]$               \\
        &$32-0.5i   \;[31-0.8i]$    &$28-0.3i   \;[27-0.6i]$    & $19.4    \;[19.4]$              \\
 $a_9$  &$-83.7-0.9i\;[-84.1-1.7i]$ &$-90-0.9i\;[-90-1.7i]$ &$-102-0.9i\;[-102-1.7i]$  \\
        &$-153-0.9i \;[-153-1.7i]$  &$-164-0.9i\;[-165-1.7i]$& $-187-0.9i\;[-188-1.7i]$ \\
$a_{10}$&$-14.4-0.5i\;[-14.6-0.8i]$ &$-2.6-0.3i  \;[-2.5-0.6i]$ & $37     \;[37]$              \\
        &$-25-0.5i  \;[-25-0.8i]$   &$-6.6-0.3i\;[-6.5-0.6i]$  & $69       \;[69]$              \\
\hline \hline
\end{tabular}\end{center}
\end{table}

\begin{table}[tb]
\begin{center}
\caption{Same as \tab{ai:bd} but for $b \to s$ [$\bar{b} \to \bar{s}$ ]
transitions. }
\label{ai:bs}
\vspace{0.2cm}
\begin{tabular} {|l|c|c|c|} \hline \hline
       & $N_c^{eff}=2$              & $N_c^{eff}=3$ & $N_c^{eff}=\infty$  \\ \hline
 $a_1$ &$0.995  \;[0.995]$            &$1.061\;[1.061]$            &$1.192  \;[1.192]$               \\
 $a_2$ &$0.201  \;[0.201]$            &$0.026\;[0.026]$            &$-0.395 \;[-0.395]$             \\
 $a_3$ &$-21-14i\;[-19-14i]$        &$77   \;[77]$                  &$272+29i\;[269+29i]$           \\
       &$-31-15i\;[-30-15i]$        &$90   \;[90]$                  &$332+31i\;[329+31i]$            \\
 $a_4$ &$-449-72i\;[-442-72i]$      &$-494-77i\;[-487-77i]$      &$-585-86i\;[-576-86i]$        \\
       &$-560-77i\;[-553-77i]$      &$-615-82i\;[-608-82i]$      &$-725-92i\;[-717-92i]$        \\
 $a_5$ &$-198-14i\;[-196-14i]$      &$-71     \;[-71]$                &$181+ 29i\;[179 + 29i]$        \\
       &$-200-15i\;[-199-15i]$      &$-57     \;[-57]$                &$229+ 31i\;[226 + 31i]$         \\
 $a_6$ &$-667-72i \;[-661-72i]$      &$-698-77i\;[-691-77i]$      &$-758-86i\;[-750-86i]$        \\
       &$-744-77i \;[-737-77i]$      &$-782-82i\;[-774-82i]$      &$-858-92i\;[-850-92i]$       \\ \hline
 $a_7$ &$7.9-1.3i \;[7.9-1.3i]$      &$6.6-1.3i\;[6.7-1.3i]$      &$4.1-1.3i\;[4.2-1.3i]$         \\
       &$34-1.3i\;[34-1.3i]$         &$31-1.3i \;[31-1.3i]$       &$24-1.3i \;[24-1.3i]$        \\
 $a_8$ &$9.6-0.6i  \;[9.6-0.6i]$     &$8.9-0.4i  \;[8.9-0.4i]$    &$7.5       \;[7.5]$                   \\
       &$32-0.6i   \;[32-0.6i]$      &$28-0.4i   \;[28-0.4i]$     &$19.4      \;[19.4]$                  \\
 $a_9$ &$-84-1.3i\;[-84-1.3i]$       &$-90-1.3i\;[-90-1.3i]$      &$-102-1.3i\;[-102-1.3i]$     \\
       &$-153-1.3i \;[-153-1.3i]$    &$-165-1.3i \;[-164-1.3i]$    &$-188-1.3i \;[-187-1.3i]$    \\
$a_{10}$&$-14.5-0.6i\;[-14.5-0.6i]$  &$-2.2-0.4i \;[-2.6-0.4i]$    &$37\;[37]$                 \\
        &$-25-0.6i  \;[-25-0.6i]$    &$-6.6-0.4i \;[-6.6-0.4i]$    &$69\;[69]$                  \\
\hline \hline
\end{tabular}\end{center}
\end{table}

From \tab{ai:bd} and \tab{ai:bs}, one can find several interesting features of
coefficients $a_i$ because of the inclusion of NP effects: (a)
the NP correction to the real part of effective coefficients
is around $20\%$ for $a_{3,4,5,6}$, and can be as large as a factor of 4
for coefficients $a_{7,8,9,10}$; (b)  the NP
correction to the imaginary part of  $a_i$ is negligibly small;
(c) the coefficient $a_1$ and $a_2$ remain unchanged since we have neglected
the very small tree-level NP contributions.

\subsection{Branching ratios of  $B \to PP$ decays}

Using above formulas, it is straightforward to find the decay amplitudes of
$B \to PP, PV$. As an example, we present here the decay
amplitude $M(B^- \to \pi^- \pi^0)=<\pi^-\pi^0|H_{eff}|B_u^->$,
\beq
M(B^- \to \pi^- \pi^0)&=& \frac{G_F}{2}
\left \{ V_{ub}V_{ud}^* \left ( a_1 M^{\pi^-\pi^0}_{uud} + a_2 M^{\pi^-\pi^0}_{duu}
\right )\right. \non
 &&  \left. - V_{tb} V_{td}^* \left [ \left ( a_4+a_{10}+ (a_6+a_8) R_1 \right )
 M^{\pi^-\pi^0}_{duu} \right. \right. \non
&& \left. \left. - \left ( a_4 + \frac{3}{2}(  a_7 -a_9 )
 -\frac{a_{10}}{2}  + (a_6-\frac{a_8}{2})R_2 \right )
 M^{\pi^-\pi^0}_{uud},\right ]\right\} \label{eq:bpipi0}
\eeq
with
\beq
R_1 &=&\frac{2 m_{\pi^-}^2}{(m_b-m_u) (m_u + m_d)},\label{eq:r1}\\
R_2 &=&\frac{m_{\pi^0}^2}{m_d (m_b -m_d)}, \label{eq:r2}\\
M^{\pi^-\pi^0}_{uud} &=& -i(m_B^2-m_{\pi^-}^2)f_{\pi}F^{B\pi}_0(m_{\pi^0}^2),\\
M^{\pi^-\pi^0}_{duu} &=& -i(m_B^2-m_{\pi^0}^2)f_{\pi}F^{B\pi}_0(m_{\pi^-}^2),
\eeq
where $f_{\pi}$ is the decay constant of $\pi$ meson. The form factor
$F_0^{B\pi}(m^2)$ can be found in Appendix B.
Under the approximation of setting $m_u =m_d$ and $m_{\pi^0}=m_{\pi^-}$, the decay
amplitude $M(B^- \to \pi^- \pi^0)$ in Eq.(\ref{eq:bpipi0}) will be reduced to the
same form as the one given in Eq.(80) of \cite{ali9804}:
\beq
M(B^- \to \pi^- \pi^0)&=& -i\, \frac{G_F}{2} f_{\pi} F^{B\pi}_0(m_{\pi}^2)
(m_B^2-m_{\pi}^2) \left \{ V_{ub}V_{ud}^* \left ( a_1  + a_2 \right ) \right. \non
 &&  \left. - V_{tb} V_{td}^* \times \frac{3}{2}
 \left ( - a_7 + a_9 + a_{10}  + a_8 R_2 \right ) \right\} \label{eq:mpp3}
\eeq
In the following numerical calculations, we use the decay amplitudes as given
in Appendix A of ref.\cite{ali9804} directly without further discussions about
the details.

In the B rest frame, the branching ratios of two-body B meson decays can
be written as
\beq
{\cal  B}(B \to X Y )=  \frac{1}{\Gamma_{tot}} \frac{|p|}{8\pi M_B^2}
|M(B\to XY)|^2\label{eq:brbpp}
\eeq
for $B \to P P$ decays, and
\beq
{\cal  B}(B \to X Y )=  \frac{1}{\Gamma_{tot}} \frac{|p|^3}{8\pi M_V^2}
|M(B\to X Y )/(\epsilon \cdot p_{B})|^2\label{eq:brbpv}
\eeq
for $B \to P V$ decays. Here $\Gamma_{tot}(B_u^-)=3.989 \times 10^{-13}$ GeV
and $\Gamma_{tot}(B_d^0)=4.219 \times 10^{-13}$GeV obtained by using
$\tau(B_u^-)=1.65 ps$ and $\tau(B_d^0)=1.56 ps$ \cite{pdg98}, $p_B$ is the
four-momentum of the B meson, $M_V$ and $\epsilon$ is the mass and polarization
vector of the produced light vector meson respectively,  and
\beq
|p| =\frac{1}{2M_B}\sqrt{[M_B^2 -(M_X + M_Y)^2] [ M_B^2 -(M_X-M_Y)^2 ]}
\label{eq:pxy}
\eeq
is the magnitude of momentum of particle X and Y in the B rest frame.

In Tables \ref{bpp1}-\ref{bpv2}, we present the numerical results of the
branching ratios for the twenty $B \to P P $ decays and thirty seven
$B \to PV$ decays in the framework of the SM and TC2 model.
The theoretical predictions are made by using the central values
of input parameters as given in
Eq.(\ref{eq:tc2fix}) and Appendix A and B, and assuming $\mpcc=200$GeV and
$N_c^{eff}=2, 3, \infty$ in  the generalized factorization approach.
The $k^2$-dependence
of the branching ratios is weak in the range of $k^2=m_b^2/2\pm 2\; GeV^2$
and hence the numerical results are given by fixing $k^2=m_b^2/2$. The
currently available CLEO data\cite{cleo99,cleo9912,cleo2000} are
listed in the last column. The branching
ratios collected in the tables are the averages of the branching ratios of $B$ and
anti-$B$ decays. The ratio $\delta {\cal  B}$  describes the new physics
corrections on the SM predictions of corresponding branching ratios
and is defined as
\beq
\delta {\cal  B} (B \to XY) = \frac{{\cal  B}(B \to XY)^{TC2}
-{\cal  B}(B \to XY)^{SM}}{{\cal  B}(B \to XY)^{SM}}~. \label{eq:dbr}
\eeq

By comparing the numerical results with the CLEO data,
the following general features of $B \to PP$ decays  can be understood:
\begin{itemize}

\item
The SM predictions for five measured $B^0 \to \pi^+ \pi^-$ and $B \to K \pi$
decay modes are consistent with the CLEO data. But for the measured $B \to
K \etap$ decays, the observed branching ratio are clearly much larger
than the SM predictions \cite{du97,ali9804,chen99}. All other estimated
branching ratios in \tab{bpp1} and \tab{bpp2} are consistent with the
new CLEO upper limits.

\item
The uncertainties of the SM predictions for the branching ratios of $B \to PP$
decays induced by varying $k^2$ is $\sim 10\%$ within the range of
$k^2 = m_b^2/2 \pm 2 GeV^2$.

\item
For most class-II, IV and V decay channels, such as $B \to \eta \etapp$,
$B \to K \pi$,$B \to K \etap$, $ etc,$ the NP enhancements to the decay
rates can be rather large:  from $~20\%$ to $~70\%$ $w.r.t$
the SM predictions.

\item
For most $B \to P P $ decay channels, the magnitude of NP effects is
insensitive to the variations of  $\mpcc$ and $N_c^{eff}$.

\item
The central values of the branching ratios obtained by using the LQQSR form
factors will be generally increased by about $15\%$ when compared with the
results using the BSW form factors, as can be seen from \tab{bpp1} and
\tab{bpp2}. No matter the BSW  or the LQQSR form factors was used, the
magnitude and whole pattern of the new physics corrections to the decay
rates in study remain  basically unchanged.

\item
Both new gluonic and electroweak penguin diagrams  contribute effectively
to most decay modes.

\end{itemize}

\begin{table}[tb]
\begin{center}
\caption{$B\to PP$  branching ratios (in units of $10^{-6}$) using the
BSW form factors, with $k^2=m_b^2/2$, $A=0.81$, $\lambda=0.2205$,
$\rho=0.12$, $\eta=0.34$, $N_c^{eff}=2,\; 3,\; \infty$ and assuming
$\mpcc=200$ GeV, in the SM and TC2 model by employing generalized
factorization approach. The last column contains measured branching ratios
and upper limits ($90\% C.L.$) [19,20] }
\label{bpp1}
\vspace{0.2cm}
\begin{tabular} {l|l|c|c|c|c|c|c|c|c|c|l} \hline \hline
 & & \multicolumn{3}{c|}{SM }&
\multicolumn{3}{c|}{TC2}& \multicolumn{3}{c|}{$\delta {\cal  B} \; [\%]$} &   \\
\cline{3-11}
Channel & Class & $2$& $3$ & $\infty$
& $2$& $3$ & $\infty$&$2$&$3$& $\infty$ &Data  \\ \hline
$B^0 \to \pi^+ \pi^-$        & I  & $9.10$ &$10.3$&$13.0$&$9.27$ &$10.5$&$13.2$&$1.9$&$1.8$&$1.6$& $ 4.3^{+1.6}_{-1.5} \pm 0.5$ \\
$B^0 \to \pi^0 \pi^0$        & II & $0.28$ &$0.15$&$0.92$&$0.28$ &$0.16$&$0.94$&$1.0$&$6.3$&$2.8$& $<9.3$ \\
$ B^+ \to \pi^+ \pi^0$       & III& $6.41$ &$5.06$&$2.85$&$6.41$ &$5.07$&$2.85$&$0.1$&$0.1$&$0.1$& $ < 12.7 $  \\ \hline
$ B^0 \to \eta \eta$         & II & $0.14$ &$0.10$&$0.29$&$0.20$ &$0.17$&$0.38$&$40 $&$64 $&$30 $& $<18$  \\
$ B^0 \to \eta \eta^\prime$  & II & $0.14$ &$0.08$&$0.38$&$0.19$ &$0.13$&$0.45$&$30 $&$67 $&$19 $& $<27$ \\
$ B^0 \to \eta' \eta^\prime$ & II & $0.05$ &$0.01$&$0.13$&$0.04$ &$0.02$&$0.14$&$13 $&$73 $&$7.8$& $< 47 $  \\ \hline
$ B^+ \to \pi^+ \eta$        & III& $3.51$ &$2.78$&$1.75$&$3.85$ &$3.17$&$2.25$&$10 $&$14 $&$28 $& $<5.7$ \\
$ B^+ \to \pi^+ \eta^\prime$ & III& $2.49$ &$1.88$&$1.02$&$2.59$ &$1.99$&$1.16$&$3.8$&$5.8$&$13 $& $<12$ \\
$ B^0 \to \pi^0 \eta$        &  V & $0.26$ &$0.29$&$0.39$&$0.36$ &$0.42$&$0.57$&$42 $&$44 $&$46 $& $<2.9 $  \\
$ B^0 \to \pi^0 \eta^\prime$ &  V & $0.06$ &$0.08$&$0.14$&$0.08$ &$0.10$&$0.18$&$37 $&$35 $&$26 $& $<5.7$ \\ \hline
$B^+ \to K^+ \pi^0$          & IV & $12.0$ &$13.5$&$16.7$&$19.6$ &$21.8$&$26.5$&$63 $&$61 $&$59 $& $ 11.6^{+3.0 +1.4}_{-2.7 -1.3}$  \\
$B^0 \to K^+ \pi^-$          & IV & $17.8$ &$19.8$&$24.0$&$24.4$ &$26.9$&$32.2$&$37 $&$36 $&$35 $& $17.2^{+2.5}_{-2.4} \pm 1.2$ \\
$B^+ \to K^0 \pi^+$          & IV & $19.9$ &$23.2$&$30.6$&$27.7$ &$32.7$&$44.0$&$39 $&$41 $&$44 $& $18.2^{+4.6}_{-4.0}\pm 1.6 $\\
$B^0 \to K^0 \pi^0$          & IV & $7.27$ &$8.31$&$10.7$&$7.95$ &$9.36$&$12.6$&$9.3$&$13 $&$18 $& $ 14.6 ^{+5.9\; +2.4}_{-5.1\; -3.3}$ \\ \hline
$ B^+ \to  K^+ \eta$         & IV & $3.91$ &$4.56$&$6.07$&$4.09$ &$5.08$&$7.45$&$4.6$&$11 $&$23 $& $ <6.9$  \\
$ B^+ \to  K^+ \eta^\prime$  & IV & $22.6$ &$28.5$&$42.4$&$33.8$ &$41.6$&$59.5$&$50 $&$46 $&$40 $& $ 80^{+10}_{-9}\pm 7 $ \\
$B^0 \to K^0 \eta$           & IV & $3.22$ &$3.63$&$4.58$&$3.33$ &$3.90$&$5.23$&$3.6$&$7.5$&$14 $& $ <9.3 $ \\
$B^0 \to K^0 \eta^\prime$    & IV & $21.9$ &$28.2$&$43.0$&$32.9$ &$41.3$&$61.2$&$50 $&$47 $&$43 $& $ 89^{+18}_{-16}\pm 9 $  \\ \hline
$ B^+ \to K^+ \bar K^0$      & IV & $1.16$ &$1.35$&$1.78$&$1.61$ &$1.90$&$2.55$&$38 $&$40 $&$43 $& $<5.1 $ \\
$ B^0 \to K^0\bar K^0$       & IV & $1.10$ &$1.28$&$1.68$&$1.52$ &$1.80$&$2.41$&$38 $&$40 $&$43 $& $ <17$  \\
\hline \hline
\end{tabular}\end{center}
\end{table}

\begin{table}[tb]
\begin{center}
\caption{Same as \tab{bpp1}, but using the LQQSR form factors.}
\label{bpp2}
\vspace{0.2cm}
\begin{tabular} {l|l|c|c|c|c|c|c|c|c|c|l} \hline \hline
 & & \multicolumn{3}{c|}{SM }&
\multicolumn{3}{c|}{TC2}& \multicolumn{3}{c|}{$\delta {\cal  B}\; [\%]$} &   \\
\cline{3-11}
Channel & Class & $2$& $3$ & $\infty$
& $2$& $3$ & $\infty$&$2$&$3$& $\infty$ &Data  \\ \hline
$B^0 \to \pi^+ \pi^-$        & I  & $10.8$ &$12.3$&$15.5$&$11.0$ &$12.5$&$15.8$&$1.9$&$1.8$&$1.6$& $4.3^{+1.6}_{-1.5} \pm 0.5$ \\
$B^0 \to \pi^0 \pi^0$        & II & $0.33$ &$0.18$&$1.09$&$0.33$ &$0.19$&$1.12$&$1.0$&$6.3$&$2.8$& $<9.3$ \\
$ B^+ \to \pi^+ \pi^0$       & III& $7.62$ &$6.02$&$3.39$&$7.63$ &$6.03$&$3.39$&$0.1$&$0.1$&$0.1$& $<12.7 $  \\ \hline
$ B^0 \to \eta \eta$         & II & $0.17$ &$0.13$&$0.36$&$0.24$ &$0.21$&$0.47$&$40 $&$64 $&$30 $& $<18$  \\
$ B^0 \to \eta \eta^\prime$  & II & $0.17$ &$0.09$&$0.45$&$0.22$ &$0.15$&$0.53$&$30 $&$67 $&$19 $& $<27$ \\
$ B^0 \to \eta' \eta^\prime$ & II & $0.05$ &$0.01$&$0.15$&$0.05$ &$0.02$&$0.16$&$13 $&$73 $&$7.8$& $<47 $  \\ \hline
$ B^+ \to \pi^+ \eta$        & III& $4.25$ &$3.37$&$2.13$&$4.66$ &$3.83$&$2.73$&$9.6$&$14 $&$28 $& $<5.7$ \\
$ B^+ \to \pi^+ \eta^\prime$ & III& $2.90$ &$2.17$&$1.17$&$3.01$ &$2.30$&$1.33$&$3.8$&$5.8$&$14 $& $<12$ \\
$ B^0 \to \pi^0 \eta$        &  V & $0.31$ &$0.35$&$0.47$&$0.43$ &$0.50$&$0.69$&$42 $&$44 $&$46 $& $<2.9 $  \\
$ B^0 \to \pi^0 \eta^\prime$ &  V & $0.07$ &$0.09$&$0.17$&$0.09$ &$0.12$&$0.21$&$37 $&$36 $&$26 $& $<5.7$ \\ \hline
$B^+ \to K^+ \pi^0$          & IV & $14.3$ &$16.0$&$19.8$&$23.2$ &$25.8$&$31.4$&$63 $&$61 $&$58 $& $11.6^{+3.0 +1.4}_{-2.7 -1.3}$  \\
$B^0 \to K^+ \pi^-$          & IV & $21.2$ &$23.5$&$28.5$&$29.0$ &$32.0$&$38.4$&$37 $&$36 $&$35 $& $17.2^{+2.5}_{-2.4} \pm 1.2$ \\
$B^+ \to K^0 \pi^+$          & IV & $23.7$ &$27.7$&$36.4$&$33.0$ &$38.9$&$52.3$&$39 $&$41 $&$43 $& $18.2^{+4.6}_{-4.0}\pm 1.6 $\\
$B^0 \to K^0 \pi^0$          & IV & $8.68$ &$9.92$&$12.7$&$9.51$ &$11.2$&$15.1$&$9.6$&$13 $&$18 $& $14.6^{+5.9\; +2.4}_{-5.1\; -3.3}$ \\ \hline
$ B^+ \to  K^+ \eta$         & IV & $4.37$ &$5.10$&$6.80$&$4.54$ &$5.66$&$8.33$&$3.9$&$11 $&$22 $& $<6.9$  \\
$ B^+ \to  K^+ \eta^\prime$  & IV & $26.2$ &$33.1$&$49.2$&$39.2$ &$48.2$&$69.1$&$50 $&$46 $&$40 $& $80^{+10}_{-9}\pm 7 $ \\
$B^0 \to K^0 \eta$           & IV & $3.57$ &$4.02$&$5.07$&$3.67$ &$4.30$&$5.76$&$2.8$&$6.8$&$14 $& $<9.3 $ \\
$B^0 \to K^0 \eta^\prime$    & IV & $25.5$ &$32.7$&$49.9$&$38.1$ &$48.0$&$71.1$&$50 $&$47 $&$43 $& $89^{+18}_{-16}\pm 9 $  \\ \hline
$ B^+ \to K^+ \bar K^0$      & IV & $1.35$ &$1.58$&$2.07$&$1.87$ &$2.21$&$2.96$&$38 $&$40 $&$43 $& $<5.1 $ \\
$ B^0 \to K^0\bar K^0$       & IV & $1.28$ &$1.49$&$1.96$&$1.77$ &$2.09$&$2.80$&$38 $&$40 $&$43 $& $<17$  \\
\hline \hline
\end{tabular}\end{center}
\end{table}

\begin{table}[tb]
\begin{center}
\caption{Same as \tab{bpp1}, but for branching ratios of $B \to PP$ decays
with new physics contributions from charged-scalar gluonic penguins only. }
\label{bpp3}
\vspace{0.2cm}
\begin{tabular} {l|l|c|c|c|c|c|c|l} \hline \hline
 & & \multicolumn{3}{c|}{TC2: QCD only } &
  \multicolumn{3}{c|}{$\delta {\cal  B} \; [\%]$} &   \\ \cline{3-8}
Channel & Class & $2$& $3$ & $\infty$&$2$&$3$& $\infty$ &Data  \\ \hline
$B^0 \to \pi^+ \pi^-$        & I  & $9.27$ &$10.5$&$13.3$&$1.90$&$1.86$&$1.90$& $ 4.3^{+1.6}_{-1.5} \pm 0.5$ \\
$B^0 \to \pi^0 \pi^0$        & II & $0.34$ &$0.22$&$1.01$&$23.3$&$47.9$&$9.85$& $<9.3$ \\
$ B^+ \to \pi^+ \pi^0$       & III& $6.41$ &$5.06$&$2.85$&$0.0 $&$0.0 $&$0.0 $& $ < 12.7 $  \\ \hline
$ B^0 \to \eta \eta$         & II & $0.17$ &$0.14$&$0.34$&$23.8$&$37.1$&$16.5$& $<18$  \\
$ B^0 \to \eta \eta^\prime$  & II & $0.19$ &$0.13$&$0.45$&$32.1$&$68.7$&$17.7$& $<27$ \\
$ B^0 \to \eta' \eta^\prime$ & II & $0.05$ &$0.02$&$0.15$&$29.3$&$162 $&$16.9$& $< 47 $  \\ \hline
$ B^+ \to \pi^+ \eta$        & III& $3.75$ &$3.05$&$2.10$&$6.86$&$9.86$&$19.7$& $<5.7$ \\
$ B^+ \to \pi^+ \eta^\prime$ & III& $2.65$ &$2.05$&$1.25$&$6.11$&$9.38$&$21.8$& $<12$ \\
$ B^0 \to \pi^0 \eta$        &  V & $0.36$ &$0.41$&$0.54$&$40.2$&$40.5$&$37.8$& $<2.9 $  \\
$ B^0 \to \pi^0 \eta^\prime$ &  V & $0.12$ &$0.15$&$0.24$&$107 $&$97.2$&$65.2$& $<5.7$ \\ \hline
$B^+ \to K^+ \pi^0$          & IV & $16.0$ &$18.0$&$22.4$&$33.0$&$33.5$&$34.1$& $ 11.6^{+3.0 +1.4}_{-2.7 -1.3}$  \\
$B^0 \to K^+ \pi^-$          & IV & $24.5$ &$27.3$&$33.3$&$37.7$&$38.2$&$39.1$& $17.2^{+2.5}_{-2.4} \pm 1.2$ \\
$B^+ \to K^0 \pi^+$          & IV & $27.3$ &$31.7$&$41.5$&$37.0$&$36.5$&$35.7$& $18.2^{+4.6}_{-4.0}\pm 1.6 $\\
$B^0 \to K^0 \pi^0$          & IV & $10.4$ &$11.8$&$15.1$&$42.5$&$42.4$&$42.0$& $ 14.6 ^{+5.9\; +2.4}_{-5.1\; -3.3}$ \\ \hline
$ B^+ \to  K^+ \eta$         & IV & $5.02$ &$5.90$&$7.95$&$28.4$&$29.4$&$30.8$& $ <6.9$  \\
$ B^+ \to  K^+ \eta^\prime$  & IV & $37.4$ &$45.2$&$63.2$&$65.6$&$58.8$&$49.0$& $ 80^{+10}_{-9}\pm 7 $ \\
$B^0 \to K^0 \eta$           & IV & $4.25$ &$4.85$&$6.22$&$32.1$&$33.6$&$35.8$& $ <9.3 $ \\
$B^0 \to K^0 \eta^\prime$    & IV & $36.1$ &$44.2$&$63.1$&$64.4$&$57.1$&$46.9$& $ 89^{+18}_{-16}\pm 9 $  \\ \hline
$ B^+ \to K^+ \bar K^0$      & IV & $1.59$ &$1.84$&$2.41$&$36.4$&$35.9$&$35.2$& $<5.1 $ \\
$ B^0 \to K^0\bar K^0$       & IV & $1.50$ &$1.74$&$2.27$&$36.4$&$35.9$&$35.2$& $ <17$  \\
\hline \hline
\end{tabular}\end{center}
\end{table}

\subsubsection{$B \to \pi \pi, \; K \pi$ decays }

There are so far seven measured $B \to PP$ decay modes
\cite{cleo9912,cleo2000,babar2000,belle2000}:
\beq
{\cal B}(B \to \pi^+ \pi^-)&=& \left \{\begin{array}{ll}
( 4.3 ^{+1.6}_{-1.5} \pm 0.5 )\times 10^{-6} & {\rm [CLEO]}, \\
( 9.3 ^{+2.8\; +1.2 }_{-2.1\; -1.4} )\times 10^{-6} & {\rm [BaBar]},  \\
\end{array} \right. \label{eq:brexp01} \\
{\cal B}(B \to K^+ \pi^0)&=& \left \{\begin{array}{ll}
( 11.6 ^{+3.0\; +1.4}_{-2.7\; -1.3} )\times 10^{-6} & {\rm [CLEO]}, \\
( 18.8 ^{+5.5}_{-4.9} \pm 2.3 )\times 10^{-6} & {\rm [Belle]},  \\
\end{array} \right. \label{eq:brexp11} \\
{\cal B}(B \to K^+ \pi^-)&=& \left \{\begin{array}{ll}
( 17.2 ^{+2.5}_{-2.4} \pm 1.2)\times 10^{-6} & {\rm [CLEO]}, \\
( 12.5 ^{+3.0\; +1.3}_{-2.6\; -1.7} \pm 2.3 )\times 10^{-6} & {\rm [BaBar]},  \\
( 17.4 ^{+5.1}_{-4.6} \pm 3.4)\times 10^{-6} & {\rm [Belle]}, \\
\end{array} \right. \label{eq:brexp12} \\
{\cal B}(B \to K^0 \pi^+)&=&
( 18.2 ^{+4.6}_{-4.0} \pm 1.6)\times 10^{-6}\ \   {\rm [CLEO]},
\label{eq:brexp13}\\
{\cal B}(B \to K^0 \pi^0)&=& \left \{\begin{array}{ll}
( 14.6 ^{+5.9\; +2.4}_{-5.1\; -3.3} )\times 10^{-6} & {\rm [CLEO]}, \\
( 21  ^{+9.3\; +2.5}_{-7.8\; -2.3} )\times 10^{-6} & {\rm [Belle]},  \\
\end{array} \right. \label{eq:brexp14} \\
{\cal B}(B \to K^+ \etap )&=& \left \{\begin{array}{ll}
( 80 ^{+10}_{-9} \pm 7 )\times 10^{-6} & {\rm [CLEO]}, \\
( 62 \pm 18 \pm 8 )\times 10^{-6} & {\rm [BaBar]},  \\
\end{array} \right. \label{eq:brexp16} \\
{\cal B}(B \to K^0 \etap )&=&
( 89 ^{+18}_{-16} \pm 9 )\times 10^{-6} \ \  {\rm [CLEO]},
\label{eq:brexp18}
\eeq
The measurements of CLEO, BaBar and Belle are in good agreement
within errors.

As a Class-I decay channel, the $B^0 \to \pi^+  \pi^-$ decay are dominated by the
$b \to u$ tree diagram. This mode together with $B^0 \to \pi^0 \pi^0$ and
$B^+ \to \pi^+ \pi^0$ decays play an important role in determination of angle
$\alpha$. For all three $B \to \pi \pi$ decay modes, the new penguin enhancement
is very small, $\leq 6.3\%$ for $N_c^{eff}=2 - \infty$, as listed in
tables 3 and 4. The theoretical predictions in the SM and TC2 model
are consistent with the CLEO data.

For $B^0 \to \etapp \etapp $ decays, the NP enhancement is varying in the
range of $10\%$ to $70\%$. For $B^+ \to \pi^+ \etapp $ decays, the NP
enhancement is around $ 10\%$ and depend on $N_c^{eff}$ moderately.
For $B^0 \to \pi^0 \etapp $ decays,
the NP enhancement is large, $30\% -60\%$, and insensitive to the variation
of $N_c^{eff}$.

In the SM, the four Class-IV decays $B \to K \pi$ are dominated by the
$b\to s g$ gluonic penguin diagram, with additional contributions from
$b \to u$ tree and electroweak penguin diagrams. Measurements of
$B\to K \pi$ decays are particularly important to measure the angle $\gamma$.
In the TC2 model, the new  penguin diagrams will interfere with their SM
counterparts and change the SM predictions for the branching ratios and
CP-violating asymmetries.

It is well known that the effective Hamiltonian calculations of charmless hadronic
B meson decays contain many uncertainties including form factors,
light quark masses, CKM matrix elements, QCD scale and external
momentum $k^2$. As a simple illustration of the theoretical uncertainties,
we calculate and show the branching ratios of four $B \to K \pi$ decay modes
by using $F_0^{B\pi}(0)=0.20$  ( preferred by the CLEO measurement of
$B \to \pi^+ \pi^-$ mode \cite{9912372} ) instead of the ordinary BSW
value $F_0^{B\pi}(0)=0.33$ ( all other input parameters remain unchanged )
and by varying $\eta$, $k^2$ and $\mpcc$ in the ranges of
$\eta=0.34 \pm 0.08$, $k^2=m_b^2/2 \pm 2 GeV^2$, $\mpcc=200\pm 100$ GeV,
and setting $N_c^{eff}=2,3,\infty$:
\beq
{\cal B}(B \to K^+ \pi^0)&=& \left \{\begin{array}{ll}
( 5.8 \pm 0.1 ^{+1.6\;  +1.4}_{-0.4\; -0.7} )\times 10^{-6} & {\rm in\ \ SM}, \\
( 10.1 \pm 0.1 ^{+1.0\; +2.3 +2.2}_{-0.6\; -1.0\; -1.2} )\times 10^{-6} &
{\rm in \ \ TC2},  \\
\end{array} \right. \label{eq:brpp11} \\
{\cal B}(B \to K^+ \pi^-)&=& \left \{\begin{array}{ll}
( 7.3 \pm 0.1 ^{+0.7\; +1.5}_{-0.5 \; -0.8} )\times 10^{-6} & {\rm in \ \ SM}, \\
( 9.9 \pm 0.1 ^{+1.2\;  +1.9\; 0.7\;}_{-0.9\; -0.9\; -0.6 } )\times 10^{-6}
& {\rm in \ \ TC2 },  \\ \end{array} \right. \label{eq:brpp12} \\
{\cal B}(B \to K^0 \pi^+)&=& \left \{\begin{array}{ll}
( 8.5 \pm 0.0 ^{+0.8 \; +2.7 }_{-1.5\; -1.2} )\times 10^{-6} & {\rm in \ \ SM}, \\
( 12.0 \pm 0.0 ^{+1.5\;  +4.2\; +1.2}_{-1.0\;  -1.8\; -0.8 } )\times 10^{-6} & {\rm in \ \ TC2},  \\
\end{array} \right. \label{eq:brpp13} \\
{\cal B}(B \to K^0 \pi^0)&=& \left \{\begin{array}{ll}
( 2.5 \pm 0.0 ^{+0.3\; +0.6}_{-0.2\; -0.3} )\times 10^{-6} & {\rm in \ \ SM}, \\
( 2.3 \pm 0.0 ^{+0.4\; +0.8\;  +0.1}_{-0.3\; -0.3\; -0.4} )\times 10^{-6} & {\rm in \ \ TC2},  \\
\end{array} \right. \label{eq:brpp14}
\eeq
where the first, second and third error correspond to the uncertainty
$\delta \eta =\pm 0.08$, $\delta q^2 =\pm 2$ and $2\leq N_c^{eff} \leq
\infty$ respectively, while the fourth error refers to $\delta \mpcc=
\pm 100$ GeV. By comparing the ratios in
tables (\ref{bpp1}, \ref{bpp2}) and  in Eqs.(\ref{eq:brpp11}-\ref{eq:brpp14}),
it is easy to see that the central values of the branching ratios
${\cal B}(B \to K \pi)$ are greatly reduced by using $F_0^{B\pi}(0)=0.20$
instead of $0.33$, the new physics enhancements therefore become essential to
make the theoretical predictions being consistent with data.

\begin{figure}[t] 
\vspace{-60pt}
\begin{minipage}[t]{0.95\textwidth}
\centerline{\epsfxsize=\textwidth \epsffile{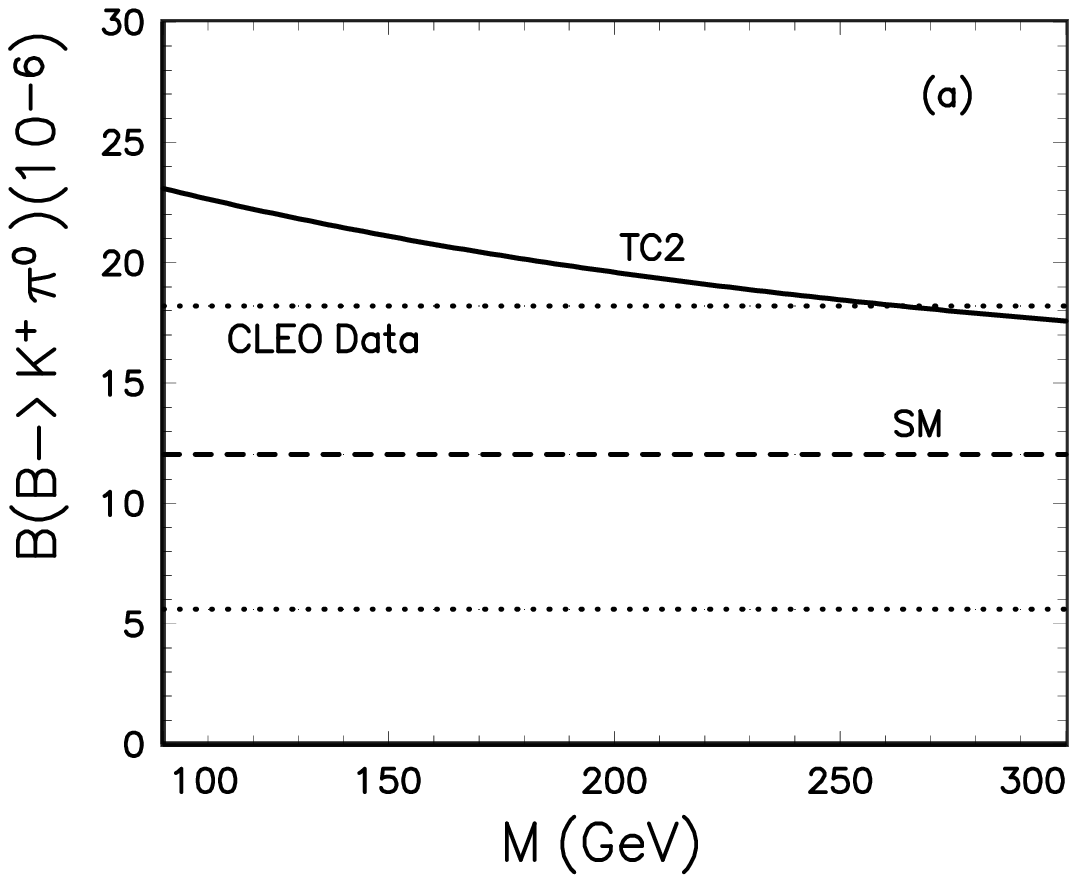}}
\vspace{-60pt}
\centerline{\epsfxsize=\textwidth \epsffile{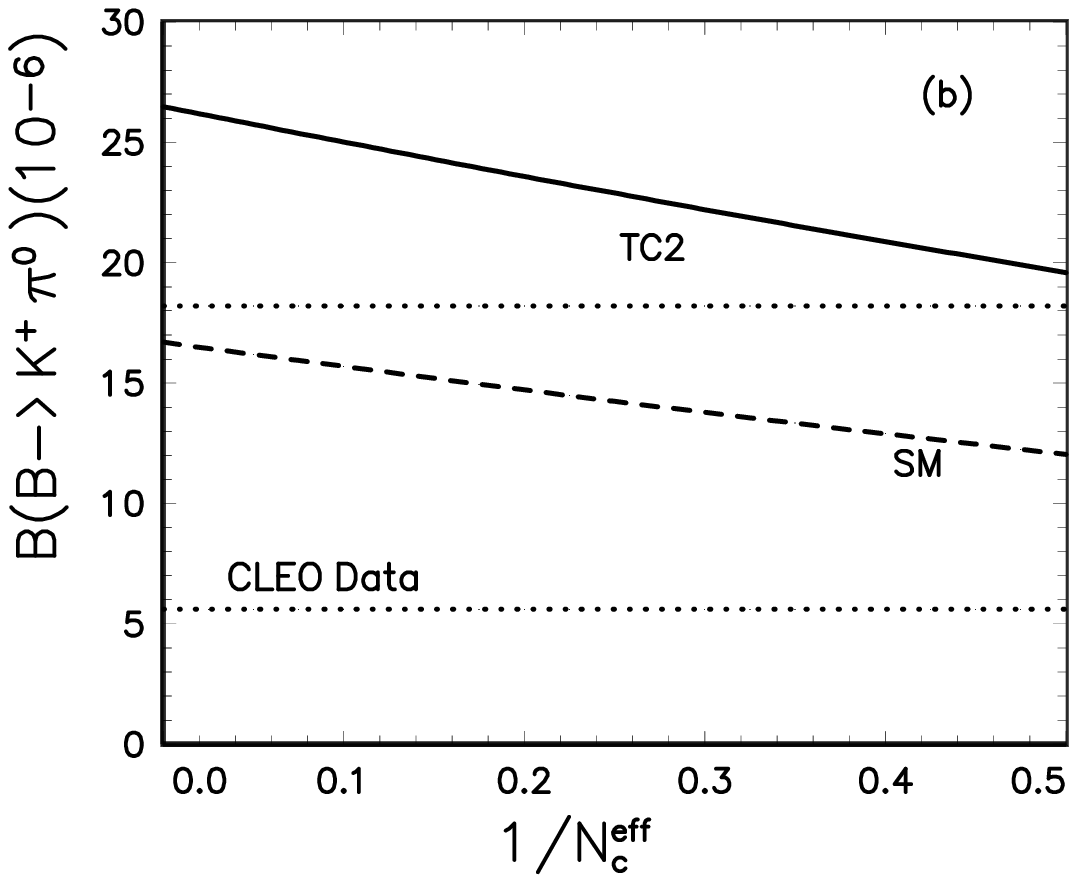}}
\caption{Plots of branching ratios of $B \to K^+ \pi^0$ decay versus
$\mpcc$ and $1/N_c^{eff}$ in the SM and TC2 model. For (a) and (b),
we set $N_c^{eff}=2$ and $\mpcc=200$GeV, respectively.
The short-dashed line  and solid curve show the branching ratio in
the SM and TC2 model, respectively. The dots band corresponds to
the CLEO data with $2\sigma$ errors: ${\cal B}(B \to K^+ \pi^0)=
(11.6 ^{+6.6}_{-6.0})\times 10^{-6}$.}
\label{fig:fig2}
\end{minipage}
\end{figure}

Fig.\ref{fig:fig2} shows the mass and $N_c^{eff}$-dependence of the
ratios ${\cal B}(B \to K^+ \pi^0$ ) in the SM and TC2 model using the input
parameters as given in Appendix A and B and employing the BSW form factors.
In Fig.(2a), we set $N_c^{eff}=2$ and assume that $\mpcc=100 - 300$GeV.
In Fig.(2b), we set $\mpcc=200$GeV and assume that $\xi=1/N_c^{eff}=0 - 0.5$.
In Fig.(\ref{fig:fig2}) the short-dashed line and solid curve show the
branching ratio  of $B^0 \to K^+  \pi^0$ decay in the SM and TC2 model,
respectively. The  band between two dots lines corresponds to the
CLEO data with $2\sigma$ errors:  ${\cal B}(B \to K^+
\pi^0)=(11.6^{+6.6}_{-6.0})\times 10^{-6}$.

Same as Fig.\ref{fig:fig2}, Figs.(3,4,5) show the mass and
$N_c^{eff}$-dependence of the branching ratios of decay
$B \to K^+ \pi^0, K^+ \pi^-, K^0 \pi^+$ and $K^0 \pi^0$, respectively.
In these three figures, the short-dashed lines and solid curves show the
branching ratios for relevant decay modes in the SM and TC2 model.
The band again refers to the corresponding CLEO data with $2\sigma$
errors, respectively:
${\cal B}(B \to K^+\pi^-)=(17.2^{+5.6}_{-5.4})\times 10^{-6}$,
${\cal B}(B \to K^0\pi^+)=(18.2^{+9.8}_{-8.6})\times 10^{-6}$ and
${\cal B}(B \to K^0\pi^0)=(14.6^{+12.8}_{-12.2})\times 10^{-6}$.
The large theoretical uncertainties are not shown in all four figures.

Although the new physics enhancements to the branching ratios of
$B \to K^+ \pi$ and $K^0 \pi^+$ decays are relatively large as
illustrated in
Figs.(2,3,4), the theoretical predictions for $B \to K \pi$ decays
in the TC2 model are still consistent with the CLEO measurements within
$2\sigma$ errors after taking into account existed large theoretical
uncertainties. If one uses $F_0^{B\pi}(0)\approx 0.20$ instead of $0.33$,
the new physics effects will play an important role to boost the theoretical
predictions for branching ratios of $B \to K \pi$ decays.

\subsubsection{$B \to K \etapp $ decays and new physics effects}

In the SM, the Class-IV decays $B \to K \etapp$ are expected to proceed
primarily through $b\to s$ penguin diagrams and $b \to u$ tree diagram. In
TC2 model, the new gluonic and electroweak penguins will also contribute
through interference with their SM counterparts.
The CLEO data of $B\to K \etapp$ decays with recent
measurements of $B\to \pi \pi$, $K \pi, etc,$ provide important constraints on the
theoretical picture for these charmless B meson decays.

For $B^+ \to K^+ \eta$ and $B^0 \to K^0 \eta$ decay modes, the new
physics enhancement is less than $10\%$ for $N_c^{eff} \sim 3$. The
theoretical predictions in both the SM and TC2 model are consistent with
the new CLEO upper limits:
${\cal  B}(B \to K^+ \eta) < 6.9\times 10^{-6}$ and
${\cal  B}(B \to K^0 \eta) < 9.3\times 10^{-6}$\cite{cleo9912}.

For $B \to K \etap$ decay modes, the situation is very interesting now.
Unexpectedly large $B \to K \etap$ rates  were firstly reported by CLEO
in 1997\cite{cleo98}, and confirmed very recently\cite{cleo9912,cleo-t0020}.
The $K \etap$ signal is large, stable and has small
errors ($\sim 14\%$). Those measured ratios as given in
Eqs.(\ref{eq:brexp16},\ref{eq:brexp18}) are clearly
much larger than the SM predictions (the contribution from the decay
$b \to s (c\bar{c} ) \to s (\eta, \etap)$ have been included
\cite{ali98,ali9804} ) as given in tables (\ref{bpp1},\ref{bpp2}) and
illustrated  by the short-dashed line in Figs.(\ref{fig:fig6},\ref{fig:fig7})
where only the central values of theoretical predictions are shown.
Furthermore, Lipkin's sum rule \cite{lipkin98}
\beq
{\cal B}(K^+ \etap) + {\cal B}( K^+ \eta) =
{\cal B}(K^+ \pi^0) + {\cal B}( K^0 \pi^+)
\eeq
is also strongly violated ($\sim 4\sigma$) \cite{cleo-t0020}:
$82.2 ^{+12.5}_{-11.6} =  29.8 ^{+5.7}_{-5.2}$.
At present, it is indeed difficult to explain the
observed large rate for $B \to K \etap $ in the framework of SM
\cite{cleo9912,cleo-t0020}. This fact
strongly suggest the requirement for additional contributions unique to
the $\etap $ meson in the framework of the SM, or from new physics
beyond the SM \cite{cleo9912}.

By varying $\eta$, $k^2$ and $\mpcc$ in the ranges of
$\eta=0.34 \pm 0.08$, $k^2=m_b^2/2 \pm 2 GeV^2$, $\mpcc=200\pm 100$ GeV,
and setting $N_c^{eff}=2,3,\infty$, we find that
\beq
{\cal B}(B \to K^+ \etap )&=& \left \{\begin{array}{ll}
( 26.5 \pm 0.1 ^{+2.7\;  +13.9}_{-2.2\; -6.9} )\times 10^{-6} & {\rm in\ \ SM}, \\
( 41.6 \pm 0.1 ^{+6.2\; +17.9\; +3.3 }_{-4.3\; -7.8\; -2.7} )\times 10^{-6} &
{\rm in \ \  TC2},  \\
\end{array} \right. \label{eq:brpp16} \\
{\cal B}(B \to K^0 \etap )&=& \left \{\begin{array}{ll}
( 28.2 \pm 0.1   ^{+3.1\; +14.8}_{-2.1\; -6.3} )\times 10^{-6}
& {\rm in \ \ SM}, \\
( 41.3 \pm 0.1   ^{+6.1\; +19.9\; +3.6 }_{-4.1\; -8.4\; -2.7} )
\times 10^{-6} & {\rm in \ \ TC2},  \\
\end{array} \right. \label{eq:brpp18}
\eeq
where the first to the fourth error corresponds to the uncertainty
$\delta \eta =\pm 0.08$, $\delta q^2 =\pm 2$ and $2\leq N_c^{eff} \leq
\infty$ and $\delta \mpcc= \pm 100$ GeV, respectively. If we use the
LQQSR form factors instead of the BSW form factors, the central
values of $BR(B \to K \etapp)$ will be increased by about $15\%$.
The NP enhancements to $B \to  K \etap$
decays are significant numerically, $\sim 50\%$ for $\mpcc=200$ GeV.

Taking into account all uncertainties considered here, the theoretical
predictions for the magnitude of ${\bf B}(B \to K \etap)$ in the SM and
TC2 model are
\beq
{\cal B}(B \to K^+ \etap )&=& \left \{\begin{array}{ll}
( 20 - 53 )\times 10^{-6} & {\rm in \ \ SM}, \\
( 30 - 74 )\times 10^{-6} & {\rm in \ \ TC2},  \\
\end{array} \right. \label{eq:brpp16th} \\
{\cal B}(B \to K^0 \etap )&=& \left \{\begin{array}{ll}
( 19 - 52 )\times 10^{-6} & {\rm in \ \ SM}, \\
( 28 - 76 )\times 10^{-6} & {\rm in \ \ TC2},  \\
\end{array} \right. \label{eq:brpp18th}
\eeq
It is evident that the theoretical predictions for ratios
${\cal B}(B \to K \etap)$ become now consistent with the CLEO data
due to the NP enhancements. This is a plausible new physics
interpretation for the large $B \to K \etap$ decay rates.

Figs.(\ref{fig:fig6},\ref{fig:fig7}) show the mass and $N_c^{eff}$
dependence of the ratios ${\it B}(B \to K \etap)$ in the SM and TC2
model using the input parameters as given in Appendix A and B and
employing the BSW form factors.
The short-dashed and solid curves in Figs.(6,7) show
the central values of theoretical predictions. The band corresponds
to the CLEO measurements with $2\sigma$ errors.

\subsection{Branching ratios of  $B \to PV$ decays}

In tables (\ref{bpvsm}-\ref{bpv2}) we present the branching ratios
for the thirty seven $B \to PV$ decay modes involving $b\to d$ and  $b \to s$
transitions in the SM and TC2 model by using the BSW and LQQSR form factors
and by employing generalized factorization approach. Theoretical
predictions are made by using the same input parameters as those for the
$B \to PP$ decays in last subsection. The measured branching ratios
from CLEO \cite{cleo99,cleo9912,cleosum} for six $B \to PV$ decay modes,
$B \to \rho^{\pm} \pi^{\mp}, \rho^0 \pi^+$, $\omega \pi^+$, $K^{*+} \eta$,
$K^{*0} \eta$, $K^{*+}\pi^-$, have been given in last column of
\tab{bpvsm}. BaBar and Belle also reported their measurements for $
{\cal B}(B^0 \to \rho^-\pi^+)$ \cite{babar2000} and
${\cal B}(B \to K^{\pm} \phi)$ \cite{belle2000}:
\beq
{\cal B}(B^0 \to \rho^{\pm} \pi^{\mp} )&=& \left \{\begin{array}{ll}
( 49 \pm 13 ^{+6}_{-5} )\times 10^{-6} & {\rm [BaBar]}, \\
( 27.6 ^{+8.4}_{-7.4} \pm 4.2 )\times 10^{-6} & {\rm [CLEO]},  \\
\end{array} \right. \label{eq:brpv1} \\
{\cal B}(B^{\pm} \to K^{\pm} \phi ) &=&
( 17.2 ^{+6.7}_{-5.4} \pm 1.8 )\times 10^{-6} \ \  {\rm [Belle]}.
\eeq
The pattern ${\cal B}(\eta K) < {\cal B}(\eta K^*) < {\cal B}(\etap K) $
and ${\cal B}(\etap K^*) < {\cal B}(\etap K) $ is found by CLEO
\cite{cleo9912}.

\begin{table}[htp]
\begin{center}
\caption{$B\to PV$ branching ratios (in units of $10^{-6}$) using the BSW
[ LQQSR ] form factors in the SM. The last column shows the CLEO and
Belle measurements and upper limits ($90\%$ C.L.) [19,20,25].}
\label{bpvsm}
\begin{tabular} {l|l|c|c|c|l} \hline \hline
Channel & Class & $N_c^{eff}=2$& $N_c^{eff}=3$ & $N_c^{eff}=\infty$ &Data  \\ \hline
\parbox[c]{2.5cm}{$B^0 \to \rho^+  \pi^- $ \\
                $B^0 \to \rho^-  \pi^+ $}  &\parbox[l]{1cm}{ I\\I} &
\parbox[c]{2.5cm}{\centering$21.1\;[25.1]$\\
                          $5.7\;[6.5]$ }&
\parbox[c]{2.5cm}{\centering$24.0 \;[28.5]$ \\
                          $6.48 \;[7.4]$ }&
\parbox[c]{2.5cm}{\centering$30.3\;[36.0]$\\
                          $8.19 \;[9.4]$} & $ \}\; 27.6^{+8.4}_{-7.4}\pm 4.2$  \\
$ B^0 \to \rho^0 \pi^0$       &II &$0.49  \;[0.58 ] $ &$0.06 \; [0.07]$&$2.05 \; [2.41 ]$&$< 5.1$ \\
$ B^+ \to \rho^0 \pi^+ $      &III&$5.72  \;[6.63 ] $ &$3.46 \; [3.97]$&$0.71 \; [0.78 ]$&$10.4^{+3.3}_{-3.4}\pm 2.1$ \\
$ B^+ \to \rho^+ \pi^0$       &III&$13.5  \;[16.0 ] $ &$12.6 \; [15.0]$&$10.9 \; [13.1 ]$&$<43$\\ \hline
$ B^0 \to \rho^0 \eta $       &II &$0.01  \;[0.02 ] $ &$0.02 \; [0.02]$&$0.06 \; [0.08 ]$&$<10$  \\
$ B^0 \to \rho^0 \eta^\prime$ &II &$0.02  \;[0.01 ] $ &$0.002\; [0.003]$&$0.03 \; [0.03]$&$<12$     \\
$ B^+ \to \rho^+ \eta $       &III&$5.44  \;[6.57 ] $ &$4.75 \; [5.79]$&$3.54 \; [4.38 ]$&$<15$     \\
$ B^+ \to \rho^+ \eta ^\prime$&III&$4.35  \;[5.02 ] $ &$3.81 \; [4.40]$&$2.85 \; [3.29 ]$&$<33$     \\ \hline
$ B^0 \to \omega \pi^0$       &II &$0.29  \;[0.35 ] $ &$0.08 \; [0.09]$&$0.15 \; [0.19 ]$&$<5.5$    \\
$ B^+ \to \omega \pi^+ $      &III&$6.32  \;[7.35 ] $ &$3.75 \; [4.31]$&$0.78 \; [0.85 ]$&$11.3^{+3.3}_{-2.9}\pm 1.4$ \\
$ B^0 \to \omega \eta $       &II &$0.32  \;[0.38 ] $ &$0.03 \; [0.04]$&$0.82 \; [0.98 ]$&$<12 $   \\
$ B^0 \to \omega \eta^\prime$ &II &$0.20  \;[0.23 ] $ &$0.001\; [0.002]$&$0.68 \; [0.79 ]$&$<60$    \\ \hline
$ B^0 \to \phi \pi^0$         &V  &$0.03  \;[0.04 ] $ &$0.002\; [0.002]$&$0.23 \; [0.27 ]$&$<5.4$   \\
$ B^+ \to \phi \pi^+ $        &V  &$0.06  \;[0.08 ] $ &$0.004\; [0.005]$&$0.49 \; [0.58 ]$&$<4$     \\
$ B^0 \to \phi \eta $         &V  &$0.01  \;[0.01 ] $ &$0.001\; [0.001]$&$0.09 \; [0.10 ]$&$<9 $    \\
$ B^0 \to  \phi \eta^\prime $ &V  &$0.01  \;[0.01 ] $ &$0.001\; [0.001]$&$0.07 \; [0.08 ]$&$<31$   \\ \hline
$ B^+ \to  \bar K^{*0} K^+ $  &IV &$0.42  \;[0.49 ] $ &$0.53 \; [0.61]$&$0.78 \; [0.90 ]$&$<5.3$    \\
$ B^0 \to \bar{K}^{*0} K^0$   &IV &$0.40  \;[0.46 ] $ &$0.50 \; [0.58]$&$0.73 \; [0.85 ]$&$-$   \\
$ B^+ \to K^{*+}  \bar K^0$ &V    &$0.004 \;[0.006 ] $ &$0.002\; [0.003]$&$0.001 \;[0.001]$&$-$    \\
$ B^0 \to K^{*0} \bar{ K}^0$&IV   &$0.004 \;[0.006 ] $ &$0.002\; [0.003]$&$0.001 \;[0.001]$&$< 12$  \\ \hline
$ B^0 \to \rho^0 K^0$       &IV   &$0.52  \;[0.60 ] $ &$0.53 \; [0.62]$&$0.72  \;[0.83] $&$< 27$ \\
$ B^+ \to \rho^0 K^+ $      &IV   &$0.39  \;[0.46  ]$ &$0.31 \; [0.36]$&$0.31  \;[0.36] $&$< 17$   \\
$ B^0 \to  \rho^- K^{+} $ &I      &$0.54  \;[0.62  ]$ &$0.59 \; [0.68]$&$0.70 \; [0.81]$ &$ <25$      \\
$ B^+ \to \rho^+ K^0$     &IV     &$0.11  \;[0.12  ]$ &$0.05 \; [0.05]$&$0.005\; [0.006]$ &$< 48$      \\ \hline
$ B^+ \to K^{*+} \eta$    &IV     &$2.43  \;[3.12  ]$ &$2.39 \; [3.04]$&$2.32 \; [2.89]$ &$26.4^{+9.6}_{-8.2}\pm 3.3$  \\
$ B^+ \to K^{*+} \etap$  &III     &$0.66  \;[1.14  ]$ &$0.36 \; [0.61]$&$0.24 \; [0.23]$ &$<35$                        \\
$ B^0 \to K^{*0} \eta$    &IV     &$2.32  \;[2.98  ]$ &$2.54 \; [3.23]$&$3.06 \; [3.82]$ &$13.8^{+5.5}_{-4.6} \pm 1.6$ \\
$ B^0 \to K^{*0} \etap$   &V      &$0.33  \;[0.69  ]$ &$0.09 \; [0.23]$&$0.31 \; [0.26]$ &$<20$                        \\ \hline
$ B^0 \to K^{*+} \pi^-$   &IV     &$8.59  \;[10.2  ]$ &$9.67 \; [11.5]$&$12.0 \; [14.3]$ &$22 ^{+8\, +4}_{-6\, -5}$    \\
$ B^0 \to K^{*0} \pi^0$   &IV     &$2.44  \;[2.77  ]$ &$3.02 \; [3.43]$&$4.42 \; [5.01]$ &$<3.6$                       \\
$ B^+ \to K^{*+ }\pi^0$   &IV     &$4.95  \;[6.09  ]$ &$5.55 \; [6.84]$&$6.91 \; [8.52]$ &$<31$                        \\
$ B^+ \to  K^{*0} \pi^+ $ &IV     &$7.35  \;[8.75  ]$ &$9.23 \; [11.0]$&$13.6 \; [16.2]$ &$< 16$                       \\ \hline
$ B^+ \to \phi K^+ $      &V      &$22.1  \;[25.7  ]$ &$11.5 \; [13.4]$&$0.60 \; [0.70]$ &$< 17.2^{+6.7}_{-5.4}\pm 1.8$ \\
$ B^0 \to \phi K^0 $      &V      &$20.9  \;[24.3  ]$ &$10.9 \; [12.6]$&$0.57 \; [0.66]$ &$< 28$                       \\ \hline
$ B^0 \to \omega K^0 $    &V      &$3.31  \;[3.86  ]$ &$0.002\; [0.003]$&$13.3 \; [15.4]$ &$< 21$                       \\
$ B^+  \to \omega K^+ $   &V      &$3.53  \;[4.11  ]$ &$0.25 \; [0.28]$&$16.5 \; [19.2]$ &$< 7.9$                      \\
\hline
\end{tabular}\end{center}
\end{table}

\begin{table}[htp]
\begin{center}
\caption{$B\to PV$ branching ratios (in units of $10^{-6}$)
using the BSW [LQQSR] form factors in the TC2 model. }
\label{bpv1}
\begin{tabular} {l|l|c|c|c|c|c|c} \hline \hline
 & & \multicolumn{3}{c|}{ TC2 }&
 \multicolumn{3}{c}{$\delta {\cal  B}\; [\%]$}  \\ \hline
Channel & Class & $2$& $3$ & $\infty$ & $2$&$3$& $\infty$  \\ \hline
$ B^0 \to \rho^+  \pi^- $     &I  &$21.2  \;[25.3]$&$24.1  \;[28.7]$&$30.4  \;[36.2]$&$ 0.71$&$ 0.63$&$  0.50$  \\
$ B^0 \to \rho^-  \pi^+ $     &I  &$5.70  \;[6.54]$&$6.48  \;[7.44]$&$8.19  \;[9.40]$&$-0.06$&$-0.05$&$ -0.03$  \\
$ B^0 \to \rho^0 \pi^0$       &II &$0.49  \;[0.58]$&$0.06  \;[0.07]$&$2.06  \;[2.43]$&$ 0.05$&$ 5.90$&$  0.53$  \\
$ B^+ \to \rho^0 \pi^+ $      &III&$5.72  \;[6.63]$&$3.46  \;[3.98]$&$0.73  \;[0.81]$&$-0.02$&$ 0.19$&$  3.62$  \\
$ B^+ \to \rho^+ \pi^0$       &III&$13.6  \;[16.2]$&$12.7  \;[15.1]$&$11.0  \;[13.2]$&$ 0.83$&$ 0.86$&$  0.92$  \\ \hline
$ B^0 \to \rho^0 \eta $       &II &$0.03  \;[0.04]$&$0.03  \;[0.04]$&$0.09  \;[0.11]$&$ 96.7$&$ 116 $&$  49.1$  \\
$ B^0 \to \rho^0 \eta^\prime$ &II &$0.01  \;[0.01]$&$0.002  \;[0.002]$&$0.03  \;[0.03]$&$-21.5$&$-30.5$&$  2.39$   \\
$ B^+ \to \rho^+ \eta $       &III&$5.49  \;[6.63]$&$4.82  \;[5.86]$&$3.62  \;[4.48]$&$ 0.96$&$ 1.29$&$  2.26$  \\
$ B^+ \to \rho^+ \eta ^\prime$&III&$4.35  \;[5.01]$&$3.81  \;[4.39]$&$2.85  \;[3.29]$&$-0.19$&$-0.08$&$  0.15$ \\ \hline
$ B^0 \to \omega \pi^0$       &II &$0.33  \;[0.39]$&$0.08  \;[0.10]$&$0.17  \;[0.21]$&$ 12.3$&$ 3.23$&$  13.3$  \\
$ B^+ \to \omega \pi^+ $      &III&$6.58  \;[7.65]$&$3.84  \;[4.43]$&$0.78  \;[0.85]$&$ 4.05$&$ 2.63$&$ -0.10$  \\
$ B^0 \to \omega \eta $       &II &$0.38  \;[0.45]$&$0.06  \;[0.07]$&$0.82  \;[0.98]$&$ 19.0$&$ 107 $&$ -0.43$  \\
$ B^0 \to \omega \eta^\prime$ &II &$0.21  \;[0.24]$&$0.002  \;[0.002]$&$0.69  \;[0.79]$&$ 6.00$&$ 63.8$&$  0.94$  \\ \hline
$ B^0 \to \phi \pi^0$         &V  &$0.03  \;[0.03]$&$0.009  \;[0.01]$&$0.37  \;[0.44]$&$-9.1$&$ 351 $&$  62.8$  \\
$ B^+ \to \phi \pi^+ $        &V  &$0.06  \;[0.07]$&$0.02  \;[0.02]$&$0.79  \;[0.94]$&$-9.1$&$ 351 $&$  62.8$  \\
$ B^0 \to \phi \eta $         &V  &$0.01  \;[0.01]$&$0.003  \;[0.004]$&$0.14  \;[0.17]$&$-9.1$&$ 351 $&$  62.8$  \\
$ B^0 \to  \phi \eta^\prime $ &V  &$0.01  \;[0.01]$&$0.003  \;[0.003]$&$0.11  \;[0.13]$&$-9.1$&$ 351 $&$  62.8$  \\ \hline
$ B^+ \to  \bar K^{*0} K^+ $  &IV &$0.64  \;[0.74]$&$0.81  \;[0.95]$&$1.22  \;[1.43]$&$ 52.1$&$ 54.4$&$  57.7$  \\
$ B^0 \to \bar{K}^{*0} K^0$   &IV &$0.60  \;[0.70]$&$0.77  \;[0.89]$&$1.16  \;[1.35]$&$ 52.1$&$ 54.4$&$  57.7$  \\
$ B^+ \to K^{*+}  \bar K^0$   &V  &$0.00  \;[0.002]$&$0.005  \;[0.001]$&$0.01  \;[0.01]$&$-71.8$&$-72.9$&$  886$  \\
$ B^0 \to K^{*0} \bar{ K}^0$  &IV &$0.001  \;[0.002]$&$0.005  \;[0.001]$&$0.01  \;[0.01]$&$-71.8$&$-72.9$&$  885$  \\ \hline
$ B^0 \to \rho^0 K^0$         &IV &$0.85  \;[0.99]$&$0.92  \;[1.07]$&$1.21  \;[1.41]$&$ 64.5$&$ 72.0$&$  69.3$  \\
$ B^+ \to \rho^0 K^+ $        &IV &$0.86  \;[1.00]$&$0.92  \;[1.07]$&$1.24  \;[1.44]$&$ 118$&$ 198 $&$  299$  \\
$ B^0 \to  \rho^- K^{+} $     &I  &$0.38  \;[0.44]$&$0.45  \;[0.51]$&$0.60  \;[0.69]$&$-29.6$&$-24.5$&$ -14.3$  \\
$ B^+ \to \rho^+ K^0$         &IV &$0.04  \;[0.04]$&$0.005  \;[0.006]$&$0.09  \;[0.11]$&$-66.0$&$-89.7$&$  1686.$  \\ \hline
$ B^+ \to K^{*+} \eta$        &IV &$4.02  \;[5.18]$&$3.81  \;[4.85]$&$3.39  \;[4.23]$&$ 65.4$&$ 59.0$&$  46.1$  \\
$ B^+ \to K^{*+} \etap$       &III&$0.32  \;[0.55]$&$0.24  \;[0.31]$&$0.50  \;[0.44]$&$-50.8$&$-32.8$&$  108$   \\
$ B^0 \to K^{*0} \eta$        &IV &$3.74  \;[4.82]$&$4.07  \;[5.18]$&$4.79  \;[5.97]$&$ 61.4$&$ 59.8$&$  56.6$   \\
$ B^0 \to K^{*0} \etap$       &V  &$0.10  \;[0.24]$&$0.10  \;[0.08]$&$0.93  \;[0.87]$&$-71.3$&$ 2.18$&$  196$   \\ \hline
$ B^0 \to K^{*+} \pi^-$       &IV &$13.6  \;[16.2]$&$14.7  \;[17.5]$&$17.1  \;[20.4]$&$ 58.1$&$ 52.3$&$  42.7$   \\
$ B^0 \to K^{*0} \pi^0$       &IV &$2.74  \;[2.97]$&$3.58  \;[3.89]$&$5.63  \;[6.17]$&$ 12.5$&$ 18.5$&$  27.4$   \\
$ B^+ \to K^{*+ }\pi^0$       &IV &$9.38  \;[11.7]$&$10.2  \;[12.8]$&$12.0  \;[15.1]$&$ 89.4$&$ 84.0$&$  74.3$   \\
$ B^+ \to  K^{*0} \pi^+ $     &IV &$11.2  \;[13.4]$&$14.3  \;[17.1]$&$21.6  \;[25.7]$&$ 53.0$&$ 55.2$&$  58.5$   \\ \hline
$ B^+ \to \phi K^+ $          &V  &$29.4  \;[34.3]$&$15.3  \;[17.9]$&$0.82  \;[0.95]$&$ 33.3$&$ 33.5$&$  35.2$   \\
$ B^0 \to \phi K^0 $          &V  &$27.8  \;[32.4]$&$14.5  \;[16.9]$&$0.77  \;[0.90]$&$ 33.3$&$ 33.5$&$  35.2$   \\ \hline
$ B^0 \to \omega K^0 $        &V  &$4.49  \;[5.23]$&$0.003  \;[0.003]$&$18.0  \;[21.0]$&$ 35.6$&$ 12.7$&$  35.9$   \\
$ B^+  \to \omega K^+ $       &V  &$5.18  \;[6.04]$&$0.24  \;[0.27]$&$22.8  \;[26.5]$&$ 46.9$&$-4.63$&$  38.3$   \\
\hline
\end{tabular}\end{center}
\end{table}

\begin{table}[htp]
\begin{center}
\caption{$B\to PV$ branching ratios (in units of $10^{-6}$) using the BSW form
factors in TC2 model with new contributions induced by charged-Higgs
gluonic penguin diagrams only. }
\label{bpv2}
\begin{tabular} {l|l|c|c|c|c|c|c} \hline \hline
& & \multicolumn{3}{c|}{TC2: QCD only } &
\multicolumn{3}{c}{$\delta {\cal  B} \; [\%]$}   \\ \cline{3-8}
Channel & Class & $2$& $3$ & $\infty$&$2$&$3$& $\infty$   \\ \hline
$ B^0 \to \rho^+  \pi^- $     &I  & $21.2$ &$24.1$ &$30.4$&$ 0.64$&$ 0.62$ &$ 0.59$   \\
$ B^0 \to \rho^-  \pi^+ $     &I  & $5.70$ &$6.48$ &$8.19$&$-0.06$&$-0.05$ &$-0.04$   \\
$ B^0 \to \rho^0 \pi^0$       &II & $0.54$ &$0.11$ &$2.12$&$ 9.68$&$ 95.0$ &$ 3.12$ \\
$ B^+ \to \rho^0 \pi^+ $      &III& $5.77$ &$3.52$ &$0.78$&$ 0.94$&$ 1.75$ &$ 10.6$ \\
$ B^+ \to \rho^+ \pi^0$       &III& $13.6$ &$12.7$ &$11.0$&$ 0.32$&$ 0.38$ &$ 0.54$   \\ \hline
$ B^0 \to \rho^0 \eta $       &II & $0.03$ &$0.03$ &$0.08$&$ 105 $&$ 115 $ &$ 41.1$  \\
$ B^0 \to \rho^0 \eta^\prime$ &II & $0.004$ &$0.003$ &$0.04$&$-51.1$&$-9.35$ &$ 25.5$  \\
$ B^+ \to \rho^+ \eta $       &III& $5.47$ &$4.79$ &$3.59$&$ 0.61$&$ 0.82$ &$ 1.44$ \\
$ B^+ \to \rho^+ \eta ^\prime$&III& $4.34$ &$3.81$ &$2.86$&$-0.21$&$ 0.0 $ &$ 0.56$   \\ \hline
$ B^0 \to \omega \pi^0$       &II & $0.43$ &$0.13$ &$0.15$&$ 46.8$&$ 70.6$ &$-1.97$ \\
$ B^+ \to \omega \pi^+ $      &III& $6.59$ &$3.85$ &$0.77$&$ 4.33$&$ 2.87$ &$-0.41$   \\
$ B^0 \to \omega \eta $       &II & $0.37$ &$0.05$ &$0.82$&$ 16.0$&$ 73.8$ &$-0.08$   \\
$ B^0 \to \omega \eta^\prime$ &II & $0.23$ &$0.006$ &$0.68$&$ 13.6$&$ 363 $ &$-0.89$   \\ \hline
$ B^0 \to \phi \pi^0$         &V  & $0.04$ &$0.002$ &$0.30$&$ 39.6$&$ 13.5$ &$ 32.2$  \\
$ B^+ \to \phi \pi^+ $        &V  & $0.09$ &$0.005$ &$0.64$&$ 39.6$&$ 13.5$ &$ 32.2$  \\
$ B^0 \to \phi \eta $         &V  & $0.02$ &$0.001$ &$0.11$&$ 39.6$&$ 13.5$ &$ 32.2$  \\
$ B^0 \to  \phi \eta^\prime $ &V  & $0.01$ &$0.001$ &$0.09$&$ 39.6$&$ 13.5$ &$ 32.2$  \\ \hline
$ B^+ \to  \bar K^{*0} K^+ $  &IV & $0.65$ &$0.79$ &$1.13$&$ 54.9$&$ 51.0$ &$ 45.2$  \\
$ B^0 \to \bar{K}^{*0} K^0$   &IV & $0.61$ &$0.75$ &$1.07$&$ 54.9$&$ 51.0$ &$ 45.2$  \\
$ B^+ \to K^{*+}  \bar K^0$   &V  & $0.001$&$0.001$&$0.005$&$-87.0$&$-68.2$ &$ 519 $   \\
$ B^0 \to K^{*0} \bar{ K}^0$  &IV & $0.001$&$0.001$&$0.005$&$-87.0$&$-68.2$ &$ 519 $   \\ \hline
$ B^0 \to \rho^0 K^0$         &IV & $0.34$ &$0.35$ &$0.52$&$-34.9$&$-34.9$ &$-27.6$  \\
$ B^+ \to \rho^0 K^+ $        &IV & $0.47$ &$0.41$ &$0.47$&$ 19.0$&$ 33.8$ &$ 51.5$  \\
$ B^0 \to  \rho^- K^{+} $     &I  & $0.41$ &$0.47$ &$0.58$&$-23.8$&$-21.5$ &$-17.8$  \\
$ B^+ \to \rho^+ K^0$         &IV & $0.02$&$0.005$ &$0.05$&$-84.4$&$-90.4$ &$ 907 $   \\ \hline
$ B^+ \to K^{*+} \eta$        &IV & $2.96$ &$2.95$ &$2.96$&$ 21.6$&$ 23.5$ &$ 27.3$  \\
$ B^+ \to K^{*+} \etap$       &III& $0.27$ &$0.34$ &$0.86$&$-59.3$&$-5.79$ &$ 260$   \\
$ B^0 \to K^{*0} \eta$        &IV & $2.84$ &$3.13$ &$3.78$&$ 22.6$&$ 23.0$ &$ 23.5$  \\
$ B^0 \to K^{*0} \etap$       &V  & $0.08$ &$0.27$ &$1.23$&$-75.7$&$ 188 $ &$ 292$   \\ \hline
$ B^0 \to K^{*+} \pi^-$       &IV & $13.1$ &$14.7$ &$18.1$&$ 52.5$&$ 51.7$ &$ 50.2$  \\
$ B^0 \to K^{*0} \pi^0$       &IV & $4.09$ &$4.93$ &$6.90$&$ 67.8$&$ 63.3$ &$ 56.1$  \\
$ B^+ \to K^{*+ }\pi^0$       &IV & $7.15$ &$8.02$ &$9.96$&$ 44.5$&$ 44.4$ &$ 44.1$  \\
$ B^+ \to  K^{*0} \pi^+$      &IV & $11.5$ &$14.0$ &$19.9$&$ 55.8$&$ 51.7$ &$ 45.8$  \\ \hline
$ B^+ \to \phi K^+ $          &V  & $33.4$ &$17.9$ &$1.31$&$ 51.2$&$ 55.9$ &$ 118$   \\
$ B^0 \to \phi K^0 $          &V  & $31.6$ &$16.9$ &$1.24$&$ 51.2$&$ 55.9$ &$ 118$   \\ \hline
$ B^0 \to \omega K^0 $        &V  & $5.07$ &$0.01$ &$17.4$&$ 53.1$&$ 489 $ &$ 31.0$  \\
$ B^+  \to \omega K^+ $       &V  & $5.31$ &$0.22$ &$21.2$&$ 50.3$&$-10.2$ &$ 28.7$  \\
\hline
\end{tabular}\end{center}
\end{table}

For considered thirty seven $B \to PV$ decays, three general features are as
follows:
\begin{itemize}
\item
The theoretical predictions in the SM and TC2 model as given in tables
(\ref{bpvsm}-\ref{bpv2}) are all consistent with the new experimental
measurements and upper limits.

\item
For most decay modes, the differences induced by using whether BSW or LQQSR form
factors in calculations are small, $\sim 15\% $.

\item
The new electroweak penguin play a more important role for $B \to PV$
decays than they do for $B \to PP$ decays.

\end{itemize}

For five $B \to \rho \pi$ and two $B \to \rho^+ \etapp$ decay modes, the NP
contributions are very small, $< 6\%$ for $N_c^{eff}=2-\infty$ as shown
in \tab{bpv1}, and can be neglected. For $B \to \rho^0 \eta$ decay, the NP
enhancement can be as large as $\sim 110\%$ for $N_c^{eff}=3$.

For $B \to \omega \pi $ decays, the NP contributions are small, $< 13\%$
for $N_c^{eff}=2-\infty$. For $B \to \omega \etapp $ decays, the NP
contributions can be large but show a strong $N_c^{eff}$ dependence.
The agreement between the theoretical prediction and CLEO measurement for
${\cal B}(B \to \omega \pi^+)$ remain unchanged in TC2 model.

For four $B \to \phi \pi, \phi \etapp$ and four $B \to K^{*} \overline{K}$
decay modes, the NP contributions can be as large as a factor of 4,
but strongly depend on $N_c^{eff}$.
For two $B \to \phi K$ decays, the NP enhancements are about $30\%$ and
insensitive to the variation of $N_c^{eff}$. It is clear that
the Belle data of $B \to K^+ \phi$ \cite{belle2000} prefer a small
effective number of colors, say $\sim N_c^{eff}2$.
For four Class-IV $B \to K^* \pi$ decays, the NP enhancements can be as
large as $90\%$, and are insensitive to the variation of $N_c^{eff}$.

For Class-I $B^0 \to \rho^- K^+$ decay, the NP correction is
about $-20\%$ and insensitive to $N_c^{eff}$. For $B^+ \to \rho^+ K^0$
decay, however, the NP correction can be large in size, a factor of 17
enhancement for $N_c^{eff}=\infty$,  but very sensitive to the variation
of $N_c^{eff}$. For the remaining two $B \to \rho^0 K$
decays, the NP enhancements are large in size and insensitive to the
value of $N_c^{eff}$.

\subsubsection{$B \to K^* \etapp$ decays}

Very recently, CLEO reported their first observation \cite{cleo9912} of
$B \to K^* \eta$ decays:
\beq
{\cal B}(B^+ \to K^{*+} \eta )&=&(26.4^{+9.6}_{-8.2} \pm 3.3)
\times 10^{-6},\\
{\cal B}(B^0 \to K^{*0} \eta )&=&(13.8^{+5.5}_{-4.6} \pm 1.6)
\times 10^{-6},
\eeq
while the theoretical predictions in the SM and TC2 model are
\beq
{\cal B}(B^+ \to K^{*+} \eta )&=&\left \{ \begin{array}{ll}
(1.5 - 3.8)\times 10^{-6} & {\rm in \ \ SM }, \\
(1.9 - 6.1)\times 10^{-6}  &  {\rm in \ \ TC2},\\  \end{array} \right., \\
{\cal B}(B^+ \to K^{*0} \eta )&=&\left \{ \begin{array}{ll}
(1.5 - 4.5)\times 10^{-6} & {\rm in \ \ SM }, \\
(2.3 - 7.2)\times 10^{-6}  &  {\rm in \ \ TC2},\\  \end{array} \right.,
\eeq
where the uncertainties induced by using the BSW or LQQSE form factors, and
setting $k^2=m_b^2/2 \pm 2 GeV^2$, $\eta=0.34 \pm 0.08$, $N_c^{eff}=2-\infty$,
and $\mpcc=200\pm 100$ GeV, have been taken into account.
Although the central values of the theoretical predictions for
${\cal B}(B \to K^* \eta)$ decays are much smaller than the central
values of the data, the theoretical predictions are still consistent with
the data since the experimental errors are still rather large.
Further refinement of the data will show whether there is a real
difference between the data and theoretical predictions.
The new physics enhancements to the decay rates are significant
( $\sim 60\%$ ) in size, insensitive to variation of $N_c^{eff}$ and
hence helpful to improve the agreement
between the theoretical predictions and the data, as illustrated in
Figs.(\ref{fig:fig8},\ref{fig:fig9}) where the upper dots band shows
the CLEO data \cite{cleo99,cleo9912}.

Fig.(\ref{fig:fig8}) and Fig.(\ref{fig:fig9}) show the mass and
$N_c^{eff}$-dependence of the decay rates ${\cal B}(B^+ \to K^{*+} \eta )$
and
${\cal B}(B^+ \to K^{*0} \eta )$, respectively. For Fig.\ref{fig:fig8}a and
Fig.\ref{fig:fig9}a, we set $N_c^{eff}=3$.  For Fig.\ref{fig:fig8}b and
Fig.\ref{fig:fig9}b, we set $\mpcc=200$ GeV. In these two figures, the
dot-dashed line refers to the SM prediction, while the short-dashed ( solid )curve
corresponds to the theoretical prediction with the inclusion of NP effects
from new gluonic ( both gluonic and electroweak ) penguins. It is clear that
the electroweak penguin play an important role for these two decay modes.

For other two $B \to K^* \etap$ decays,  the new physics
enhancement can be significant in size, from $-70\%$ to $\sim 200\%$,
but strongly depend on the variation of $N_c^{eff}$, as shown in \tab{bpv1}.
The theoretical predictions for these two decay modes are still far
below the current CLEO upper limits.

\section{CP-violating asymmetries in $B\to PP, PV$ decays} \label{sec:acp}

As is well-known, there are three possible manifestations of CP violation in
B system\cite{slac504,buras98,cpbook,waldi99}:
the direct CP violation or CP violation in decay, the indirect CP violation
or CP violation in mixing due to the interference between mixing amplitudes,
and finally the CP violation in interference between decays with and without
mixing. For the measurements of CP violation in B system, great progress has
been achieved recently\cite{cleoacp,cdfacp}.

In ref.\cite{ali9805}, Ali et al. estimated the CP-violating
asymmetries in charmless hadronic decays $B \to PP, PV, VV$,
based on the effective Hamiltonian with generalized factorization.
In another paper \cite{cct98}, Chen {\it et al}.  calculated the
CP-violating asymmetries in charmless hadronic decays of $B_s$ meson.
We here will follow the same procedure of \cite{ali9805} to estimate
the new physics effects on the CP-violating asymmetries of $B \to PP,
PV$ decays in the TC2 model.

In TC2 model, no new weak phase has been introduced through the interactions
involving new particles and hence the mechanism of CP violation in TC2 model
is the same as in the SM. But the CP-violating asymmetries ${\cal  A}_{CP}$
may be changed by the inclusion of
new physics contributions through the interference between the ordinary
tree/penguin amplitudes in the SM and the new strong and electroweak penguin
amplitudes in TC2 model. The real and imaginary part of effective
Wilson coefficients $C_i^{eff}$ and effective number $a_i$ will be modified by
new physics effects and hence the pattern of ${\cal  A}_{CP}$ for two-body
charmless hadronic B decays will be changed accordingly.

\subsection{Formalism}

For charged B decays the direct CP violation is defined as
\beq
{\cal  A}_{CP} = \frac{\Gamma(B^+ \to f) -\Gamma(B^- \to \bar{f})}{
\Gamma(B^+ \to f) + \Gamma(B^- \to \bar{f})}\label{eq:acpp}
\eeq
in terms of partial decay widths.

For neutral $B^0 (\bar{B}^0)$ decays, the situation becomes complicated
because of $B^0-\bar{B}^0$ mixing, and hence the time dependent CP
asymmetry for the decays of states that were tagged as pure $B^{0}$ or
$\bar{B}^0$ at production is defined as
\beq
{\cal  A}_{CP}(t) = \frac{\Gamma(B^0(t) \to f )
 -\Gamma(\bar{B}^0(t) \to \bar{f} )}{
\Gamma(B^0(t) \to f) + \Gamma(\bar{B}^0(t) \to \bar{f})}.\label{eq:acp0}
\eeq
According to the characteristics of the final states $f$, neutral B decays can be
classified into four cases as described in \cite{ali9805}.
For case-1, $f$ or $\bar{f}$ is not a common final state
of $B^0$ and $\bar{B}^0$, and the CP-violating asymmetry is independent of time. We
use Eq.(\ref{eq:acpp}) to calculate the CP-violating asymmetries for CP-class-1
decays: the charged B and case-1 neutral B decays.

For CP-class-2  (class-3) B decays where $\obar{B^0} \to (f = \bar{f})$ with
$f^{CP}=\pm f $ ( $f^{CP}\neq \pm f $ )\cite{ali9805}, the time-dependent and
time-integrated CP asymmetries are of the form
\beq
{\cal  A}_{CP}(t)&=&\aepa \cos(\Delta m\; t) + \aepb \sin(\Delta m\; t),
\label{eq:acptd}\\
{\cal  A}_{CP}&=&\frac{1}{1+x^2}\aepa + \frac{x }{1 + x^2} \aepb ,
\label{eq:acpti}
\eeq
where $\Delta m =m_H -m_L$ is the mass difference between mass eigenstates
$|B^0_H>$ and $|B^0_L>$, $x=\Delta m/\Gamma \approx 0.73$ for the case of
$B_d^0-\bar{B}_d^0$ mixing \cite{pdg98}, and
\beq
\aepa&=& \frac{1- |\lambda_{CP}|^2}{1 + |\lambda_{CP}|^2}, \ \
\aepb= \frac{-2 Im(\lambda_{CP})}{1 + |\lambda_{CP}|^2}, \label{eq:aep}\\
\lambda_{CP}&=&\frac{V^*_{tb}V_{td}}{V_{tb}V^*_{td}}
\frac{ <f|H_{eff}|\bar{B}^0> }{ <f|H_{eff}|B^0>} . \label{eq:lambda}
\eeq
For the formulae being used to calculate  ${\cal  A}_{CP}$ for the more
complicated CP-class-4 B decays, one can see Eqs.(36-40) of
ref.\cite{ali9805}. We also define the ratio
\beq
\delta {\cal A}_{CP}= \frac{{\cal  A}_{CP}^{TC2}
- {\cal  A}_{CP}^{SM}}{{\cal  A}_{CP}^{SM}}
\eeq
to measure the new physics effects on the SM predictions of ${\cal  A}_{CP}$
of B meson decays.

As an example, we present here the explicit calculation for the Class-III-1 decay
$B^{\pm} \to \pi^{\pm} \pi^0$. The decay amplitude $ M(B^- \to \pi^- \pi^0)$ has
been given in Eq.(\ref{eq:mpp3}) where all $a_i$ should be taken for transitions
$b \to d$. For the charged conjugate amplitude we have
\beq
M(B^+ \to \pi^+ \pi^0)&=& -i\, \frac{G_F}{2} f_{\pi} F^{B \to \pi}_0(m_{\pi}^2)
(m_B^2-m_{\pi}^2) \left \{ V_{ub}^*V_{ud} \left ( a_1  + a_2 \right ) \right. \non
 &&  \left. - V_{tb}^* V_{td} \times \frac{3}{2}
 \left ( - a_7 + a_9 + a_{10}  + a_8 R_2 \right ) \right\},
\label{eq:mpp3b}
\eeq
where the ratio $R_2$ has been given in Eq.(\ref{eq:r2}), and all $a_i$ are taken
for transitions $\bar{b} \to \bar{d}$. The CP
asymmetry for this decay mode is then defined as
\beq
{\cal  A}_{CP}(B^{\pm} \to \pi^{\pm} \pi^0) =
\frac{|M(B^+ \to \pi^+ \pi^0)|^2-|M(B^- \to \pi^- \pi^0)|^2}{
|M(B^+ \to \pi^+ \pi^0)|^2 +|M(B^- \to \pi^- \pi^0)|^2} .\label{eq:acp3}
\eeq

\subsection{Numerical results}

In  tables \ref{acpp1}-\ref{acpv2b}, we present  numerical results of
${\cal  A}_{CP}$ in $B \to P P $ and $B \to PV$ decays in the SM and
TC2 model. We show the numerical results for the
case of using BSW form factors only since the form factor dependence is weak.
In second column of Tables \ref{acpp1}-\ref{acpv2b}, the roman number and arabic
number denotes the classification of the decays $B \to PP, PV$ using
$N_c^{eff}$-dependence and the CP-class for each decay mode as defined in
\cite{ali9804,ali9805}, respectively. The first and second error of
the theoretical predictions  correspond to the uncertainties induced by setting
$\delta k^2 = \pm 2 GeV^2$ and $ \delta \eta =\pm 0.08$, respectively.

\begin{table}[tb]
\begin{center}
\caption{ CP-violating asymmetries ${\cal  A}_{CP}$ in $B \to P P $ decays
(in percent) in the SM using $\rho=0.12$ and $N_c^{eff}=2,3,\infty$
for   $k^2=m_b^2/2 \pm 2 GeV^2$ and $\eta=0.34\pm 0.8$.
The first and second error of the ratios corresponds to
$\delta k^2 =\pm 2 GeV^2$ and $\delta \eta =\pm 0.08$, respectively.}
\label{acpp1}
\vspace{0.2cm}
\begin{tabular}{l|l|c|c|c} \hline \hline
Channel & Class & $2$&$3$ & $\infty$ \\ \hline
$\obar{B^0} \to \pi^+ \pi^-$       & I-2  &  $ 23.7^{+0.3 +16.2}_{-1.3 -15.9}$&$ 23.4^{+0.3 +16.5}_{-1.2 -15.9}$&$  23.0^{+0.3 +16.6}_{-1.2 -15.9}$ \\
$\obar{B^0} \to \pi^0 \pi^0$       & II-2 &  $-54.9^{+9.5 +1.0}_{-3.9 -1.2}$&$ 15.3^{+6.1 +2.6}_{-3.4 -3.0}$&$  48.0^{+1.0 +7.8}_{-2.9 -12.7}$ \\
$ B^{\pm} \to \pi^{\pm} \pi^0$     & III-1&  $ 0.1^{+0.01 +0.02}_{-0.02 -0.01}$&$ 0.1^{+0.01 +0.02}_{-0.02 -0.01}$&$  0.$ \\
$\obar{B^0} \to \eta \eta$         & II-2 &  $ 57.5^{+1.5 +2.7 }_{-2.5 -7.2}$&$ 13.8^{+4.7 +2.3}_{-2.6 -2.8}$&$ -53.1^{+6.9 +0.7}_{-2.9 -1.5}$ \\
$ \obar{B^0} \to \eta \eta^\prime$ & II-2 &  $ 60.8^{+2.0 -4.2}_{-4.3 -0.7}$&$ 20.0^{+5.9 +3.5}_{-3.3 -4.0}$&$ -52.6^{+6.9 +0.4}_{-2.9 -1.5}$ \\
$ \obar{B^0} \to \etap\eta^\prime$ & II-2 &  $ 44.8^{+1.4 +14.4}_{-5.2 -16.3}$&$ 36.2^{+7.6 +6.0}_{-4.8 -7.2}$&$ -47.3_{-1.9 -1.4}^{+6.5 +0.7}$ \\
$ B^{\pm} \to \pi^{\pm} \eta$      & III-1&  $ 12.1^{+2.4 +0.3}_{-5.2 -0.2}$&$ 14.3^{+2.7 +0.3}_{-5.6 -1.2}$&$  18.1^{+2.7 +2.6}_{-5.0 -3.3}$  \\
$ B^{\pm} \to \pi^{\pm} \eta^\prime$&III-1&  $ 12.6^{+2.9 +0.9}_{-6.3 -1.2}$&$ 15.5^{+3.4 +0.3}_{-7.2 -0.9}$&$  22.4^{+4.2 +1.4}_{-8.0 -2.7}$ \\
$ \obar{B^0}\to \pi^0 \eta$        & V-2  &  $ 28.5^{+2.8 +5.2}_{-1.5 -5.8}$&$ 14.3^{+5.1 +2.4}_{-2.8 -2.9}$&$ -9.9_{-4.5 -1.3}^{+8.2 +1.7}$ \\
$ \obar{B^0} \to \pi^0 \eta^\prime$& V-2  &  $ 53.5^{+0.3 +8.4}_{-0.1 -10.3}$&$ 24.9^{+7.2 +4.2}_{-4.2 -5.0}$&$ -16.8_{-7.7 -1.9}^{+13 +2.6}$ \\
$B^{\pm} \to K^{\pm} \pi^0$        & IV-1 &  $-5.6_{-1.6 -1.2}^{+2.9 +1.2}$&$-5.0_{-1.3 -1.0}^{+2.5 +1.2}$&$ -3.8_{-1.0 -0.9}^{+1.7 +0.9}$ \\
$\obar{B^0} \to K^{\pm} \pi^{\mp}$ & IV-1 &  $-6.1_{-1.7 -1.3}^{+3.2 +1.4}$&$-6.2_{-1.8 -1.3}^{+3.2 +1.4}$&$ -6.4_{-1.8}^{+3.4} \pm 1.4$ \\
$B^{\pm} \to K_S^0 \pi^{\pm}$      & IV-1 &  $-1.3\pm 0.1 \pm 0.3$&$-1.2 \pm 0.1 \pm 0.3$&$ -1.2 \pm 0.1 \pm 0.3$ \\
$\obar{B^0} \to K_S^0 \pi^0$       & IV-2 &  $ 34.4^{+0.3 +5.0}_{-0.6 -6.4}$&$ 31.2\pm 0.0_{-5.9}^{+4.8}$&$  25.6^{+0.9 +4.1}_{-0.6 -5.0}$ \\
$ B^{\pm}\to  K^{\pm} \eta$        & IV-1 &  $ 4.0^{+1.7 +0.8}_{-3.3 -0.9}$&$ 2.9^{+1.4 +0.7}_{-2.5 -0.6}$&$  1.0^{+0.8}_{-1.3}\pm 0.2$ \\
$ B^{\pm} \to  K^{\pm} \eta^\prime$& IV-1 &  $-4.4_{-1.1 -1.1}^{+1.9 +1.0}$&$-3.6_{-0.8}^{+1.4}\pm 0.8$&$ -2.5_{-0.4}^{+0.7}\pm 0.5$ \\
$\obar{B^0} \to K_S^0 \eta$        & IV-2 &  $ 34.7^{+0.4 +5.0}_{-0.6 -6.4}$&$ 30.9\pm 0.0_{-5.9}^{+4.7}$&$  23.7^{+1.2 +3.9}_{-0.7 -4.6}$ \\
$\obar{B^0}\to K_S^0 \eta^\prime$  & IV-2 &  $ 29.7^{+0.3 +4.7}_{-0.2 -5.7}$&$ 31.2\pm 0.0_{-5.9}^{+4.8}$&$  33.2^{+0.2 +5.0}_{-0.3 -6.2}$  \\
$ B^{\pm} \to K^{\pm} \bar{K}_S^0$ & IV-1 &  $ 10.5^{+5.1 +1.9}_{-2.6 -2.2}$&$ 10.4^{+5.0 +1.8}_{-2.5 -2.2}$&$  10.2^{+5.0 +1.8}_{-2.6 -2.1}$ \\
$ \obar{B^0} \to K^0\bar{K}^0$     & IV-2 &  $ 13.5^{+5.0 +2.3}_{-2.6 -2.7}$&$ 13.4^{+4.9 +2.2}_{-2.7 -2.7}$&$  13.1^{+4.9 +2.3}_{-2.6 -2.6}$ \\ \hline
\end{tabular}\end{center}
\end{table}

\begin{table}[tb]
\begin{center}
\caption{ Same as in \tab{acpp1} but in the TC2 model and
assuming $\mpcc=200$ GeV. }
\label{acpp2}
\vspace{0.2cm}
\begin{tabular}{l|l|c|c|c|c|c|c} \hline \hline
 &&  \multicolumn{3}{c|}{ TC2 }&
\multicolumn{3}{c}{ $\delta {\cal A}_{CP} [\%]$} \\ \cline{3-8}
Channel & Class&$2$&$3$ & $\infty$ & $2$& $3  $ &$\infty$  \\ \hline
$\obar{B^0} \to \pi^+ \pi^-$        & I-2  & $ 27.3$ & $26.9$& $ 26.3$ & $ 15.3$ & $ 14.9$ & $ 14.3$\\
$\obar{B^0} \to \pi^0 \pi^0$        & II-2 & $-55.5$ & $14.9$& $ 49.8$ & $ 1.0$ & $-2.9$ & $ 3.7$\\
$ B^{\pm} \to \pi^{\pm} \pi^0$      & III-1& $ 0.07$ & $0.05$& $ 0.0 $ & $ 0.4$ & $ 0.4$ & $ -  $\\
$\obar{B^0} \to \eta \eta$          & II-2 & $ 51.9$ & $10.7$& $-50.9$ & $-9.9$ & $-22.4$ & $-4.2$\\
$ \obar{B^0} \to \eta \eta^\prime$  & II-2 & $ 60.4$ & $15.4$& $-53.7$ & $-0.7$ & $-23.2$ & $ 2.0$\\

$ \obar{B^0} \to \etap\eta^\prime$  & II-2 & $ 55.3$ & $27.5$& $-50.7$ & $ 23.3$ & $-24.0$ & $ 7.2$\\
$ B^{\pm} \to \pi^{\pm} \eta$       & III-1& $ 12.0$ & $13.7$& $ 15.7$ & $-0.7$ & $-4.0$ & $-13.5$\\
$ B^{\pm} \to \pi^{\pm} \eta^\prime$&III-1 & $ 13.1$ & $15.9$& $ 21.5$ & $ 4.3$ & $ 2.4$ & $-4.1$\\
$ \obar{B^0}\to \pi^0 \eta$         & V-2  & $ 23.8$ & $11.8$& $-7.9 $ & $-16.3$ & $-17.1$ & $-19.8$\\
$ \obar{B^0} \to \pi^0 \eta^\prime$ & V-2  & $ 46.2$ & $21.3$& $-14.8$ & $-13.7$ & $-14.2$ & $-12.3$\\

$B^{\pm} \to K^{\pm} \pi^0$         & IV-1 & $-3.8 $ & $-3.4 $& $-2.7 $ & $-31.8$ & $-30.8$ & $-28.9$\\
$\obar{B^0} \to K^{\pm} \pi^{\mp}$  & IV-1 & $-4.8 $ & $-5.0 $& $-5.2 $ & $-20.5$ & $-20.0$ & $-19.1$\\
$B^{\pm} \to K_S^0 \pi^{\pm}$       & IV-1 & $-1.2 $ & $-1.1 $& $-1.0 $ & $-15.0$ & $-15.6$ & $-16.4$\\
$\obar{B^0} \to K_S^0 \pi^0$        & IV-2 & $ 34.3$ & $31.3$& $ 26.1$ & $-0.2$ & $ 0.2$ & $ 2.1$\\
$ B^{\pm}\to  K^{\pm} \eta$         & IV-1 & $ 4.11$ & $2.7 $& $ 0.9 $ & $ 3.5$ & $-3.6$ & $-16.7$\\
$ B^{\pm} \to  K^{\pm} \eta^\prime$ & IV-1 & $-3.34$ & $-2.8 $& $-2.0 $ & $-24.6$ & $-22.4$ & $-18.8$\\
$\obar{B^0} \to K^0_S \eta$         & IV-2 & $ 35.2$ & $31.3$& $ 24.7$ & $ 1.2$ & $ 1.5$ & $ 4.1$\\
$\obar{B^0}\to K_S^0 \eta^\prime$   & IV-2 & $ 30.1$ & $31.2$& $ 32.9$ & $ 1.1$ & $ 0.04$ & $-1.0$\\
$ B^{\pm} \to K^{\pm} \bar{K}_S^0$  & IV-1 & $ 8.8 $ & $8.6 $& $ 8.3 $ & $-16.4$ & $-17.0$ & $-18.0$\\
$ \obar{B^0} \to K^0\bar{K}^0$      & IV-2 & $ 11.5$ & $11.3$& $ 10.9$ & $-15.2$ & $-15.8$ & $-16.7$\\ \hline
\end{tabular}\end{center}
\end{table}

\begin{table}
\begin{center}
\caption{ CP-violating asymmetries ${\cal A}_{CP}$ in $B \to P V $ decays
(in percent) with $b\to d$ transition in the SM using $\rho=0.12$ and
$N_c^{eff}=2,3,\infty$ for   $k^2=m_b^2/2 \pm 2$ and $\eta=0.34\pm 0.8$.
The first and second error of the ratios corresponds to
$\delta k^2 =\pm 2$ and $\delta \eta =\pm 0.08$, respectively. }
\label{acpv1}
\vspace{0.2cm}
\begin{tabular}{l|l|c|c|c} \hline \hline
Channel & Class & $2$&$3$ & $\infty$   \\ \hline
$ B^0/\bar{B}^0 \to \rho^+  \pi^-/\rho^- \pi^+ $    &I-4  &$ 3.2_{-0.7 -18}^{+1.2 +22.3}$ &$  3.2_{-0.7 -18}^{+1.2 +22.3}$ &$ 3.4_{-0.6 -18}^{+1.3 +22.3}$   \\
$ B^0 /\bar{B}^0 \to \rho^-  \pi^+/\rho^+ \pi^-  $  &I-4  &$ 5.8^{+0.7 +10.5}_{-1.8 -8.8}$ &$  5.8^{+0.7 +10.4}_{-1.9 -8.9}$ &$ 5.8^{+0.7 +10.4}_{-1.8 -8.9}$   \\
$ \obar{B^0} \to \rho^0 \pi^0$                      &II-2 &$-36.1_{-1.1 -4.8}^{+4.4 +5.7}$ &$  21.4^{+8.6 +3.5}_{-5.1}$ &$ 23.1^{+0.4 +16.9}_{-1.8 -16.4}$   \\
$ B^{\pm} \to \rho^0 \pi^{\pm} $                    &III-1&$-4.1^{+2.9 +0.8}_{-1.2 -1.0}$ &$ -5.4^{+3.9 +1.0}_{-1.8 -1.6}$ &$-10.7^{+10.2 +2.0}_{-4.6 -3.1}$   \\
$ B^{\pm} \to \rho^{\pm} \pi^0$                     &III-1&$ 2.7^{+0.7 +0.4}_{-1.5 -0.3}$ &$  3.0^{+0.8 +0.5}_{-1.7 -0.4}$ &$ 3.6^{+0.9 +0.5}_{-2.0 -0.4}$   \\
$ \obar{B^0}\to \rho^0 \eta $                       &II-2 &$-49.4_{-9.8 -0.5}^{+10.7 +2.3}$ &$  24.9^{+7.7 +4.2}_{-4.5 -5.0}$ &$ 63.8^{+2.3 +1.8}_{-4.6 -4.5}$   \\
$ \obar{B^0}\to \rho^0 \eta^\prime$                 &II-2 &$ 8.8^{+0.5 +2.7}_{+5.0 -5.7}$ &$ -26.9^{+12.5 +3.4}_{-39 -2.0}$ &$ 34.9^{+1.4 +19.8}_{-6.5 -17.4}$   \\
$ B^{\pm}\to \rho^{\pm} \eta $                      &III-1&$ 4.0^{+1.0 +0.6}_{-2.3 -0.5}$ &$  4.5^{+1.1 +0.7}_{-2.5 -0.6}$ &$ 5.9^{+1.4 +0.8}_{-3.2 -0.7}$   \\
$ B^{\pm}\to\rho^{\pm}\eta^\prime$                  &III-1&$ 3.9^{+1.1 +0.9}_{-2.4 -0.6}$ &$  4.5^{+1.2 +1.0}_{-1.7 -0.6}$ &$ 5.9^{+1.6 +1.1}_{-3.5 -0.9}$   \\
$ \obar{B^0} \to\omega \pi^0$                       &II-2 &$ 51.1^{+0.7 +7.7}_{-0.9 -10.7}$ &$  22.1^{+6.5 +3.8}_{-3.7 -4.5}$ &$ 33.0^{+0.1 +11.6}_{-0.8 -14.4}$   \\
$ B^{\pm}\to \omega \pi^{\pm}$                      &III-1&$ 10.2^{+2.3 +0.4}_{-4.9 -0.7}$ &$  8.5^{+2.0 +0.7}_{-4.4 -0.9}$ &$-2.1^{+1.6 +0.4}_{-0.6 -0.6}$   \\
$ \obar{B^0} \to\omega \eta $                       &II-2 &$ 52.2^{+1.2 +5.3}_{-3.3 -11.0}$ &$  22.4^{+5.4 +3.9}_{-3.0 -4.6}$ &$ 7.6^{+0. +17.4}_{-0.3 -14.6}$   \\
$ \obar{B^0}\to\omega \eta^\prime$                  &II-2 &$ 32.3^{+0.9 +17.0}_{-3.6 -16.6}$ &$  39.9_{-13.5 -7.7}^{+20.4 +5.1}$ &$ 21.3^{+0.1 +15.9}_{-0.2 -15.5}$ \\
$ \obar{B^0} \to\phi \pi^0$                         &V-2  &$ 16.0^{+5.6 +2.7}_{-4.1 -3.2}$ &$  1.4^{+0.6 +0.2}_{-0.4 -0.3}$ &$ 11.9^{+4.5 +2.0}_{-2.5 -2.4}$   \\
$ B^{\pm}\to \phi \pi^{\pm} $                       &V-1  &$ 12.7^{+5.6 +2.4}_{-3.0 -2.6}$ &$  1.0^{+0.8 +0.2}_{-0.3 -0.1}$ &$ 9.1 ^{+4.7 +1.6}_{-2.4 -1.9}$   \\
$ \obar{B^0} \to \phi \eta $                        &V-2  &$ 16.0^{+5.6 +2.7}_{-3.1 -3.2}$ &$  1.4^{+0.6 +0.2}_{-0.3 -0.3}$ &$ 11.9^{+4.5 +2.0}_{-2.5 -2.4}$   \\
$ \obar{B^0} \to \phi \eta^\prime$                  &V-2  &$ 16.0^{+5.6 +2.7}_{-3.1 -3.2}$ &$  1.4^{+0.6 +0.2}_{-0.3 -0.3}$ &$ 11.9^{+4.5 +2.0}_{-2.5 -2.4}$   \\
$ B^{\pm} \to \bar K^{*0} K^{\pm}$                  &IV-1 &$ 13.4^{+5.7 +2.4}_{-3.0 -2.8}$ &$  12.6^{+5.6 +2.4}_{-2.9 -2.6}$ &$ 11.6^{+5.3 +2.1}_{-2.8 -2.1}$   \\
$ B^0/\bar{B}^0 \to \bar{K}^{*0} K_S^0/K^{*0}K_S^0$ &IV-4 &$ 13.8^{+5.8 +2.5}_{-3.1 -2.9}$ &$  13.2^{+5.7 +2.4}_{-3.0 -2.7}$ &$ 12.3^{+5.5 +2.2}_{-2.9 -2.5}$   \\
$ B^{\pm} \to K^{*\pm}  K_S^0$                      &V-1  &$-7.9^{+5.0 +1.2}_{-17 -0.8}$ &$ -4.6^{+6.3 +0.5}_{-30.3 -0.2}$ &$ 75.8^{+0.6 +8.6}_{-12.2 -12.3}$   \\
$ B^0/\bar{B}^0 \to K^{*0} K_S^0/ \bar{K}^{*0}K_S^0$&IV-4 &$-11.6^{+3.0 +2.4}_{-5.8 -2.1}$ &$ -10.1^{+2.6 +1.9}_{-5.3 -2.1}$ &$-7.7^{+2.2 +0.7}_{-4.3 -1.5}$   \\ \hline
\end{tabular}\end{center}
\end{table}

\begin{table}
\begin{center}
\caption{ Same as in \tab{acpv1} but with $b\to s$ transition.}
\label{acpv1b}
\vspace{0.2cm}
\begin{tabular}{l|l|c|c|c} \hline \hline
Channel & Class & $2$&$3$ & $\infty$   \\ \hline
$ \obar{B^0} \to \rho^0 K_S^0$     &IV-1 &$ 15.1^{+0.5 +2.7}_{-1.1 -3.1}$ &$ 32.1\pm 0.0^{+5.1}_{-6.2}$ &$ 46.0^{+1.4 +1.2}_{-0.6 -3.6}$   \\
$ B^{\pm} \to \rho^0 K^{\pm} $     &IV-1 &$-17.9^{+9.5 +3.0}_{-4.3 -4.0}$ &$-18.9^{+10.1 +1.9}_{-4.9 -1.6}$ &$-9.7_{-2.7 -1.5}^{+5.1 +1.8}$   \\
$ \obar{B^0} \to  \rho^- K^{+} $   &I-1  &$-11.7^{+7.2 +1.0}_{-2.9 -0.7}$ &$-12.2^{+7.5 +1.1}_{-3.0 -0.9}$ &$-13.0^{+8.0 +1.4}_{-3.2 -1.3}$   \\
$ B^{\pm} \to \rho^{\pm} K_S^0$    &IV-1 &$ 1.7^{+0.2 +0.4}_{-0.1 -0.4}$ &$ 2.7^{+0.6 +0.7}_{-0.3 -0.6}$ &$-2.3^{+2.1 +0.5}_{-7.9 -0.6}$   \\
$ B^{\pm} \to K^{*\pm} \eta$       &IV-1 &$-7.3^{+4.1 -1.2}_{-2.1 +1.5}$ &$-7.3^{+4.1 -1.3}_{-2.2 +1.5}$ &$-7.2_{-2.1 -1.4}^{+4.0 +1.6}$   \\
$ B^{\pm} \to K^{*\pm} \etap$      &III-1&$-29.4_{-2.4 -1.9}^{+16 +3.6}$ &$-54.4^{+22.0 -0.7}_{-0.8 +2.4}$ &$-83.0_{-12.5 -2.3}^{+47.4 +2.4}$   \\
$ \obar{B^0} \to K^{*0} \eta$      &IV-1 &$-1.8_{-0.3 }^{+0.6 }\pm 0.4$ &$-1.0_{-0.0}^{+0.1}\pm 0.2$ &$ 0.6^{+0.5 +0.1}_{-1.0 -0.2}$   \\
$ \obar{B^0} \to K^{*0} \etap$     &V-1  &$-4.3_{-0.3}^{+4.2}\pm 1.0$ &$ 4.4^{+8.1 +1.0}_{-1.1 -1.0}$ &$ 15.3^{+7.5 +3.2}_{-13 -3.4}$   \\
$ \obar{B^0} \to K^{*+} \pi^-$     &IV-1 &$-13.8_{-4.4 -2.0}^{+8.0 +2.6}$ &$-13.9_{-4.5 -2.0}^{+8.2 +2.6}$ &$-14.0_{-4.6 -1.9}^{+8.2 +1.6}$   \\
$ \obar{B^0} \to K^{*0} \pi^0$     &IV-1 &$ 0.2^{+0.7 +0.1}_{-1.2 -0.0}$ &$-1.7\pm 0.0 \pm 0.4$ &$-4.3_{-1.0 -0.9}^{+1.9 +1.0}$   \\
$ B^{\pm} \to K^{*\pm }\pi^0$      &IV-1 &$-11.3_{-3.6 -1.8}^{+6.5 +2.2}$ &$-10.4_{-3.2 -1.7}^{+6.0 +2.1}$ &$-8.7_{-2.7 -1.7}^{+4.9 +1.9}$   \\
$ B^{\pm} \to  K^{*0} \pi^{\pm}$   &IV-1 &$-1.5\pm 0.1 _{-0.4}^{+0.3}$ &$-1.5_{-0.1 -0.3}^{+0.1 +0.4}$ &$-1.4_{-0.0 -0.3}^{ +0.1 +0.4}$   \\
$ B^{\pm} \to \phi K^{\pm} $       &V-1  &$-1.5\pm 0.1 _{-0.4}^{+0.3}$ &$-1.6\pm 0.1 \pm 0.4$ &$-2.5\pm 0.1 \pm 0.6$   \\
$ \obar{B^0} \to \phi K_S^0 $      &V-2  &$ 31.1\pm 0.0^{+4.8}_{-5.9}$ &$ 31.1\pm 0.0^{+5.7}_{-5.9}$ &$ 30.6 \pm 0.0^{+4.7}_{-5.8}$   \\
$ \obar{B^0} \to \omega K_S^0 $    &V-2  &$ 23.5^{+1.4 +3.8}_{-0.9 +4.6}$ &$ 31.4^{+1.2 +4.1}_{-11.4 -5.6}$ &$ 24.2^{+1.1 +4.0}_{-0.7 -4.8}$   \\
$ B^{\pm}  \to \omega K^{\pm} $    &V-1  &$-11.5_{-3.7 -2.0}^{+6.6 +2.3}$ &$-17.8^{+10.3 +2.7}_{-4.1 -3.6}$ &$ 0.2^{+0.4}_{-0.8}\pm 0.1$   \\ \hline
\end{tabular}\end{center}
\end{table}

\begin{table}
\begin{center}
\caption{CP-violating asymmetries ${\cal A}_{CP}$ in $B \to P V $ decays
(in percent) with $b\to d$ transitions in the TC2 model using
$\rho=0.12$, $\eta=0.34$, $k^2=m_b^2/2$, $\mpcc=200 $ GeV
and $N_c^{eff}=2,3,\infty$.}
\label{acpv2}
\vspace{0.2cm}
\begin{tabular} {l|l|c|c|c|c|c|c} \hline \hline
 & & \multicolumn{3}{c|}{TC2 }&
\multicolumn{3}{c}{ $\delta {\cal A}_{CP} [\%]$}  \\ \cline{3-8}
Channel &Class &$2$& $3$ & $\infty$ & $ 2$& $3$ & $\infty$ \\ \hline
$ B^0 \to \rho^+  \pi^- $                         &I-4  &$ 6.5$ &$ 6.1$ &$ 5.4$ &$  104$ &$ 88.1$ &$ 62.5$\\
$ B^0 \to \rho^-  \pi^+ $                         &I-4  &$ 7.5$ &$ 7.3$ &$ 7.0$ &$  30.9$ &$ 27.2$ &$ 21.1$\\
$ \obar{B^0} \to \rho^0 \pi^0$                    &II-2 &$-36.7$ &$ 20.8$ &$ 24.7$ &$  1.7$ &$-2.7$ &$ 6.7$\\
$ B^{\pm} \to \rho^0 \pi^{\pm} $                  &III-1&$-4.3$ &$-5.8$ &$-11.0$ &$  7.3$ &$ 7.0$ &$ 3.1$\\
$ B^{\pm} \to \rho^{\pm} \pi^0$                   &III-1&$ 2.9$ &$ 3.2$ &$ 3.9$ &$  6.6$ &$ 6.5$ &$ 6.4$\\
$ \obar{B^0} \to \rho^0 \eta $                    &II-2 &$-34.6$ &$ 16.8$ &$ 58.4$ &$ -29.9$ &$-32.5$ &$-8.4$\\
$ \obar{B^0} \to \rho^0 \eta^\prime$              &II-2 &$-1.5$ &$-16.6$ &$ 41.9$ &$ -117$ &$-38.2$ &$ 20.2$\\
$ B^{\pm} \to \rho^{\pm} \eta $                   &III-1&$ 4.2$ &$ 4.8$ &$ 6.3$ &$  6.9$ &$ 6.7$ &$ 6.0$\\
$ B^{\pm} \to \rho^{\pm} \eta ^\prime$            &III-1&$ 4.2$ &$ 4.8$ &$ 6.3$ &$  7.7$ &$ 7.5$ &$ 7.1$\\
$ \obar{B^0} \to \omega \pi^0$                    &II-2 &$ 48.8$ &$ 21.8$ &$ 41.0$ &$ -4.5$ &$-1.3$ &$ 24.3$\\
$ B^{\pm} \to \omega \pi^{\pm} $                  &III-1&$ 10.6$ &$ 8.9$ &$-2.2$ &$  3.7$ &$ 5.0$ &$ 8.0$\\
$ \obar{B^0} \to \omega \eta $                    &II-2 &$ 56.2$ &$ 15.3$ &$ 3.3$ &$  7.8$ &$-31.5$ &$-55.9$\\
$ \obar{B^0} \to \omega \eta^\prime$              &II-2 &$ 41.6$ &$ 66.0$ &$ 23.1$ &$  28.6$ &$ 65.6$ &$ 8.1$\\
$ \obar{B^0} \to \phi \pi^0$                      &V-2  &$ 16.8$ &$ 0.7$ &$ 9.3$ &$  5.4$ &$-53.2$ &$-22.1$\\
$ B^{\pm} \to \phi \pi^{\pm} $                    &V-1  &$ 13.6$ &$ 0.4$ &$ 6.9$ &$  7.4$ &$-54.1$ &$-23.8$\\
$ \obar{B^0} \to \phi \eta $                      &V-2  &$ 16.8$ &$ 0.7$ &$ 9.3$ &$  5.4$ &$-53.2$ &$-22.1$\\
$ \obar{B^0} \to  \phi \eta^\prime $              &V-2  &$ 16.8$ &$ 0.7$ &$ 9.3$ &$  5.4$ &$-53.2$ &$-22.1$\\
$ B^{\pm} \to  \bar K^{*0} K^{\pm} $              &IV-1 &$ 10.6$ &$ 9.9$ &$ 8.9$ &$ -21.2$ &$-21.8$ &$-22.6$\\
$ \obar{B^0} \to \bar{K}^{*0} K_S^0/K^{*0}K_S^0$  &IV-4 &$ 11.1$ &$ 10.5$ &$ 9.7$ &$ -19.6$ &$-20.3$ &$-21.2$\\
$ B^{\pm} \to K^{*\pm}  K_S^0$                    &V-1  &$-0.4$ &$ 84.4$ &$ 18.3$ &$ -94.5$ &$-1951$ &$-75.8$\\
$ \obar{B^0}\to K^{*0} K_S^0/ \bar{K}^{*0}K_S^0$  &IV-4 &$-8.2$ &$-6.8$ &$-4.6$ &$ -29.0$ &$-33.4$ &$-40.4$\\ \hline
\end{tabular}\end{center}
\end{table}

\begin{table}
\begin{center}
\caption{Same as in \tab{acpv2} but with $b\to s$ transitions. }
\label{acpv2b}
\vspace{0.2cm}
\begin{tabular} {l|l|c|c|c|c|c|c} \hline \hline
 & & \multicolumn{3}{c|}{TC2 }&
\multicolumn{3}{c}{ $\delta {\cal A}_{CP} [\%]$}  \\ \cline{3-8}
Channel &Class &$2$& $3$ & $\infty$ & $ 2$& $3$ & $\infty$ \\ \hline
$ \obar{B^0} \to \rho^0 K_S^0$                    &IV-1 &$ 19.2$ &$ 32.1$ &$ 45.0$ &$ 26.6$ &$ -0.09$ &$-2.1$\\
$ B^{\pm} \to \rho^0 K^{\pm} $                    &IV-1 &$-8.9$ &$-6.9$ &$-2.8$ &$-50.4$ &$ -63.3$ &$-71.2$\\
$ \obar{B^0} \to  \rho^- K^{+} $                  &I-1  &$-18.2$ &$-17.6$ &$-16.4$ &$ 55.2$ &$  44.1$ &$ 26.0$\\
$ B^{\pm} \to \rho^{\pm} K_S^0$                   &IV-1 &$ 2.8$ &$ 2.6$ &$-2.1$ &$ 65.1$ &$ -0.6$ &$-9.5$\\
$ B^{\pm} \to K^{*\pm} \eta$                      &IV-1 &$-4.8$ &$-5.1$ &$-5.4$ &$-33.6$ &$ -31.1$ &$-25.2$\\
$ B^{\pm} \to K^{*\pm} \etap$                     &III-1&$-66.7$ &$-90.3$ &$-44.4$ &$ 127$ &$  65.9$ &$-46.5$\\
$ \obar{B^0} \to K^{*0} \eta$                     &IV-1 &$-1.3$ &$-0.8$ &$ 0.3$ &$-27.2$ &$ -20.9$ &$-52.1$\\
$ \obar{B^0} \to K^{*0} \etap$                    &V-1  &$-24.2$ &$-4.5$ &$ 4.4$ &$ 469$ &$ -203$ &$-71.2$\\
$ \obar{B^0} \to K^{*+} \pi^-$                    &IV-1 &$-9.5$ &$-9.9$ &$-10.6$ &$-31.0$ &$ -28.5$ &$-24.0$\\
$ \obar{B^0} \to K^{*0} \pi^0$                    &IV-1 &$ 0.2$ &$-1.6$ &$-3.7$ &$ 0.5$ &$ -8.0$ &$-14.3$\\
$ B^{\pm} \to K^{*\pm }\pi^0$                     &IV-1 &$-6.6$ &$-6.2$ &$-5.5$ &$-41.6$ &$ -39.8$ &$-36.4$\\
$ B^{\pm} \to  K^{*0} \pi^{\pm}$                  &IV-1 &$-1.2$ &$-1.2$ &$-1.1$ &$-18.9$ &$ -19.5$ &$-20.3$\\
$ B^{\pm} \to \phi K^{\pm} $                      &V-1  &$-1.3$ &$-1.4$ &$-2.2$ &$-13.2$ &$ -13.2$ &$-13.5$\\
$ \obar{B^0} \to \phi K_S^0 $                     &V-2  &$ 31.2$ &$ 31.2$ &$ 30.8$ &$ 0.4$ &$  0.4$ &$ 0.6$\\
$ \obar{B^0} \to \omega K_S^0 $                   &V-2  &$ 24.8$ &$ 31.5$ &$ 25.4$ &$ 5.8$ &$  0.4$ &$ 5.1$\\
$ B^{\pm}  \to \omega K^{\pm} $                   &V-1  &$-8.5$ &$-19.9$ &$ 0.06$ &$-25.9$ &$  11.6$
&$-66.9$\\\hline
\end{tabular}\end{center}
\end{table}

The SM predictions for the CP-violating asymmetries of fifty seven B meson decay
modes investigated here as given in tables \ref{acpp1}-\ref{acpv2b} are well
consistent with those given in \cite{ali9805}. For details of the parametric
dependence of ${\cal  A}_{CP}$ in the SM, one can see \cite{ali9805}.
We here focus on the new physics effects on ${\cal  A}_{CP}$ of B
meson decays.

Very recently, CLEO reported their first measurements of CP-violating
asymmetries  for five decay modes\cite{cleoacp},  $B^{\pm} \to K^{\pm}
\pi^0,\; K_S^0 \pi^{\pm}, \; \omega \pi^{\pm} $ and $\obar{B^0}
\to K^{\pm} \pi^{\mp}$. They conclude that the measured asymmetries
are consistent with zero in all five decay modes studied
\cite{cleoacp}.

Using the same input parameters as
in \tab{acpp1}, we find the theoretical predictions in TC2 model
for those five decay modes
\beq
{\acp}(B \to K^{\pm} \pi^0)&=& (-3.4 ^{+1.6}_{-0.9}
\pm 0.8 ^{+0.7\; +0.4}_{-0.4\; -0.3} )\times 10^{-2}~, \label{eq:acppp11}\\
{\acp}(B \to K^{\pm} \pi^{\mp})&=& (-5.0 ^{+2.5}_{-1.5}
\pm 1.1 ^{+0.1\; +0.3}_{-0.2\; -0.2} )\times 10^{-2}~, \label{eq:acppp12}\\
{\acp}(B \to K_S^0 \pi^{\mp})&=& (-1.1 \pm 0.1 ^{+0.2}_{-0.1}
\pm 0.1 \pm 0.1 )\times 10^{-2}~, \label{eq:acppp13}\\
{\acp}(B \to K^{\pm} \etap )&=& (-2.8 ^{+1.0}_{-0.6}
\pm 0.7 ^{+0.8}_{-0.5} \pm 0.1 )\times 10^{-2}~, \label{eq:acppp16}\\
{\acp}(B \to \omega \pi^{\pm})&=& (8.9 \pm 0.1 ^{+0.4\; +1.7}_{-0.7\; -11.1}
\pm 0.1 )\times 10^{-2}~,\label{eq:acppv11}
\eeq
where the central values correspond to setting $k^2=m_b^2/2$, $\eta=0.34$
and $ N_c^{eff} =3$, while the first to fourth error is induced by
considering the uncertainty $\delta k^2= \pm 2 GeV^2$, $\delta \eta=\pm 0.8$,
$ 2\leq N_c^{eff} \leq \infty $ and $\delta \mpcc =\pm 100$ GeV, respectively.
For first four $B \to PP$ decay modes, the uncertainty induced by varying
$N_c^{eff}$ is smaller or in comparable size with other three
uncertainties. For $B \to \omega \pi^{\pm}$ decay mode, however,
the uncertainty induced by varying $N_c^{eff}$ dominate over other
three uncertainties.

The CLEO measurements, the $90\% CL$ region and the theoretical predictions in the
SM and TC2 model are  as given in \tab{acpexp}. The theoretical predictions
are taken from tables 9-14 and eqs.(\ref{eq:acppp11}-\ref{eq:acppv11}).
One also should note that the sign
convention as being used here in Eq.(\ref{eq:acpp},\ref{eq:acp0}) to define
${\cal A}_{CP}$ is opposite to that used in \cite{cleoacp}, we therefore changed
the sign of the theoretical predictions of ${\cal A}_{CP}$ in \tab{acpexp}
in order to be consistent with those reported results by CLEO.

\begin{table}
\begin{center}
\caption{ CLEO measurements for ${\cal A}_{CP}$ in $B \to K \pi, K \etap,
\omega \pi$ decays [22], and the corresponding theoretical predictions
in the SM and TC2 model.} \label{acpexp}
\begin{tabular}{l|c|c|c|c} \hline \hline
Channel&  ${\cal A}_{CP}^{exp}$ & $ 90\%\; CL \; Region$& ${\cal  A}_{CP}^{SM}$ & ${\cal  A}_{CP}^{TC2}$ \\ \hline
$B^{\pm} \to K^{\pm} \pi^0$        &$-0.29 \pm 0.23$ &$[-0.67, 0.09] $ &$[-0.001, 0.079] $&$[0.009, 0.058] $  \\
$\obar{B^0} \to K^{\pm} \pi^{\mp}$ &$-0.04 \pm 0.16$ &$[-0.30, 0.22] $ &$[0.015, 0.096] $&$[0.010, 0.080] $  \\
$B^{\pm} \to K_S^0 \pi^{\pm}$      &$+0.18 \pm 0.24$ &$[-0.22, 0.56] $ &$[0.007, 0.017] $&$[0.006, 0.015] $  \\
$ B^{\pm} \to  K^{\pm} \eta^\prime$&$-0.03 \pm 0.12$ &$[-0.17, 0.23] $ &$[0.003, 0.062] $&$[0.002, 0.047] $  \\
$ B^{\pm} \to \omega \pi^{\pm}$    &$-0.34 \pm 0.25$ &$[-0.75, 0.07] $ &$[-0.129,0.007] $&$[-0.102, 0.031] $
\\ \hline
\end{tabular}\end{center}
\end{table}

It is easy to see that the CP-violating asymmetries of all five decay modes
studied are small in size in both the SM and TC2 model, and well consistent
with the CLEO data. For all five decay
modes, the new physics corrections are also small: which will change the SM
predictions by about $20\%$.  \fig{fig:fig10} and \fig{fig:fig11} show
the mass and $N_c^{eff}$ dependence for $B^{\pm} \to K^{\pm} \etap $ and
$B^{\pm} \to \omega \pi^{\pm}$ in the SM (the dots lines or curves) and
TC2 model (the solid curves) \footnote{ In these two figures we use the
same sign convention as CLEO Collaboration \cite{cleoacp}. }

From the theoretical predictions and the CLEO measurements as given in
tables \ref{acpp1}-\ref{acpexp},  the following general features about
the CP-violating asymmetry of charmless hadronic B meson decays under
study in this paper can be understood:

\begin{itemize}

\item
The  CP-violating asymmetries depend weakly on whether we use the BSW or
LQQSR form factors. The inclusion of NP contributions does not change
this feature.

\item
The $\mpcc$ dependence of ${\cal  A}_{CP}$ is weak:  $\delta \acp$ is
about $\pm 15\%$ as one varies $\mpcc$ in the range $ 100 {\rm GeV}
\leq \mpcc \leq 300 {\rm  GeV} $.

\item
For twenty $B \to PP$ decays, the new physics corrections to ${\cal  A}_{CP}$
are generally small or moderate in  magnitude. The largest correction is
about $-30\%$ for the decay $B^{\pm} \to K^{\pm} \pi^0 $, and about
$\pm 20\%$ for decay modes $\obar{B^0} \to \pi^+ \pi^-, \pi^0 \eta,
K^+ \pi^-, K^0 \bar{K}^0$ and $B^+ \to K^0 \pi^+, K^0 \bar{K}^0$.
For four class-II $B \to \eta \etapp, \etap \etap $ and $\pi^0 \pi^0$ decays,
they have large CP violating asymmetries ( around $\pm 50\%$), but
unfortunately also
have strong $N_c^{eff}$-dependence in both the SM and TC2 model.

\item
For $B \to PV$ decays, however, the NP corrections to ${\cal  A}_{CP}$ can
be rather large for many decay modes, as illustrated in \tab{acpv2}. For
class-I-4 decay $B^0/\bar{B}^0
\to \rho^+ \pi^-/\rho^- \pi^+$, the new physics correction is
$(60 \sim 100)\%$ for $N_c^{eff}=2 - \infty$. For decay $B^+
\to K^{*+} \bar{K}^0$ the correction  can even reaches a factor of 20
for $N_c^{eff}=2$ due to strong interference between the
contributions from the $W-$penguin and new charged-scalar penguins.

\item
For most class-I, III and IV decays, the $N_c^{eff}$-dependence
and $k^2$-dependence of $\delta {\cal A}_{CP}$ are weak. For most
class-V decays, however, the $N_c^{eff}$-dependence of $\delta
{\cal A}_{CP}$ is strong.

\item
For most decay modes considered here, the new physics corrections on
${\cal A}_{CP}$ in the TC2 model are still much smaller than existed
theoretical uncertainties, and therefore will be masked by the later.
Low experimental statistics and large theoretical uncertainties
together prevent us from testing the TC2 model through the studies
of CP-violating asymmetries at present.

\end{itemize}

According to relevant studies \cite{he98} for these decay modes, we
know that the FSI may
provide a new strong phase and therefore enhance ${\cal A}_{CP}$ to a level
$20\% -40\%$, new physics with new large phases may also increase the
${\cal A}_{CP}$ to a level $ 40\% -60\%$. Although there is still no
evidence for direct CP violation in B system, the CLEO measurements
ruled out large part of the parameter space for ${\cal A}_{CP}$.
The key problem is that the measurements are currently statistics
limited.

\section{Summary and discussions} \label{sec:sum}

In this paper, we calculated the new physics contributions to the branching ratios
and CP-violating asymmetries of two-body charmless hadronic B meson decays
$B \to PP, PV$ in the TC2 model by employing the NLO effective Hamiltonian
with generalized factorization. We will present the calculation for the new
physics effects on ${\cal B}(B \to VV)$ and $\acp(B\to VV)$  in a forthcoming
paper\cite{xiao20b}

We firstly evaluate analytically all strong and electroweak charged-scalar
penguin diagrams in the quark level processes $b \to s V^*$ with
$V=\gamma, gluon,\; Z$, extract out the corresponding Inami-Lim functions
$C_0^{TC2}, D_0^{TC2}, E_0^{TC2}$  and  $C_{g}^{TC2}$ which describe the
NP contributions to the Wilson coefficients $C_i(M_W)$ ($i=3-10$) and
$C_{g}(M_W)$, combine these new functions with their SM counterparts,
run all Wilson coefficients  from the high energy scale
$\mu\approx O(M_W)$ down to the lower energy scale $\mu =O(m_b)$ by using
the QCD renormalization equations, find the effective Wilson coefficients
$C_i^{eff}$, and finally calculate the branching ratios and CP-violating
asymmetries after inclusion of NP contributions in the TC2 model.

\begin{figure}[t] 
\vspace{-60pt}
\begin{minipage}[t]{0.95\textwidth}
\centerline{\epsfxsize=\textwidth \epsffile{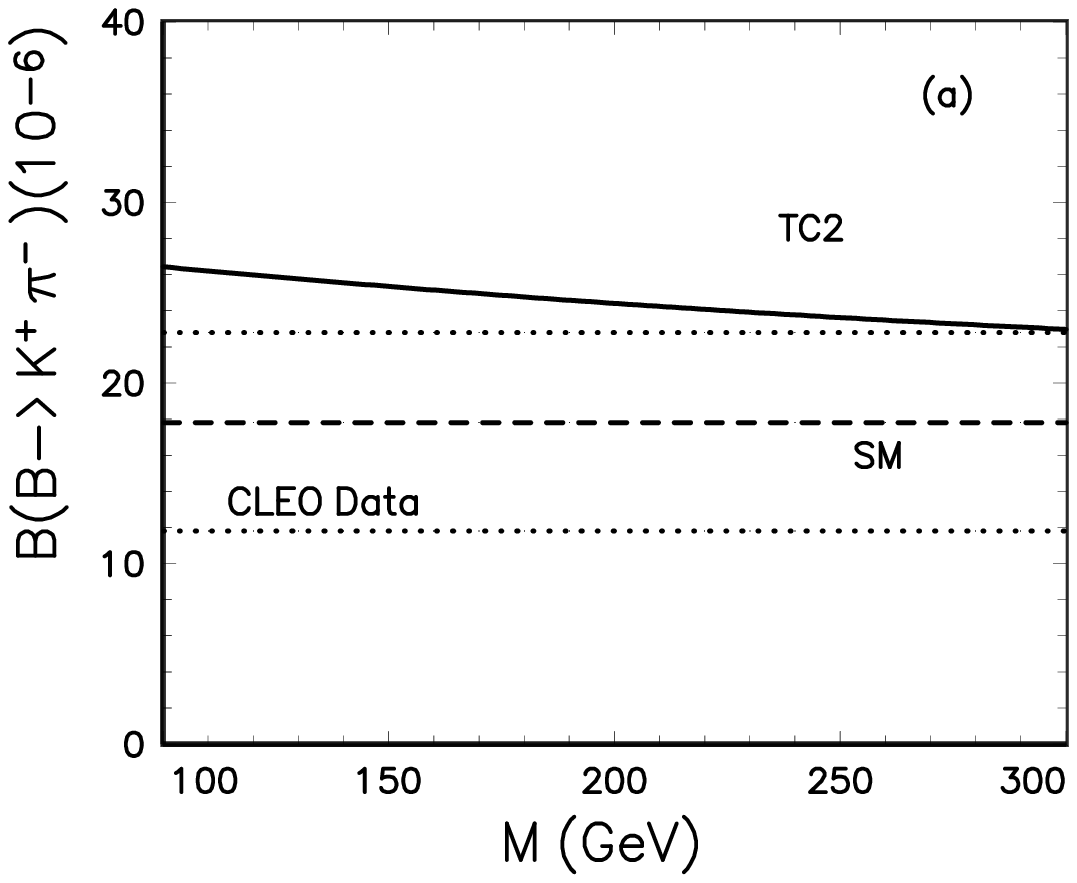}}
\vspace{-60pt}
\centerline{\epsfxsize=\textwidth \epsffile{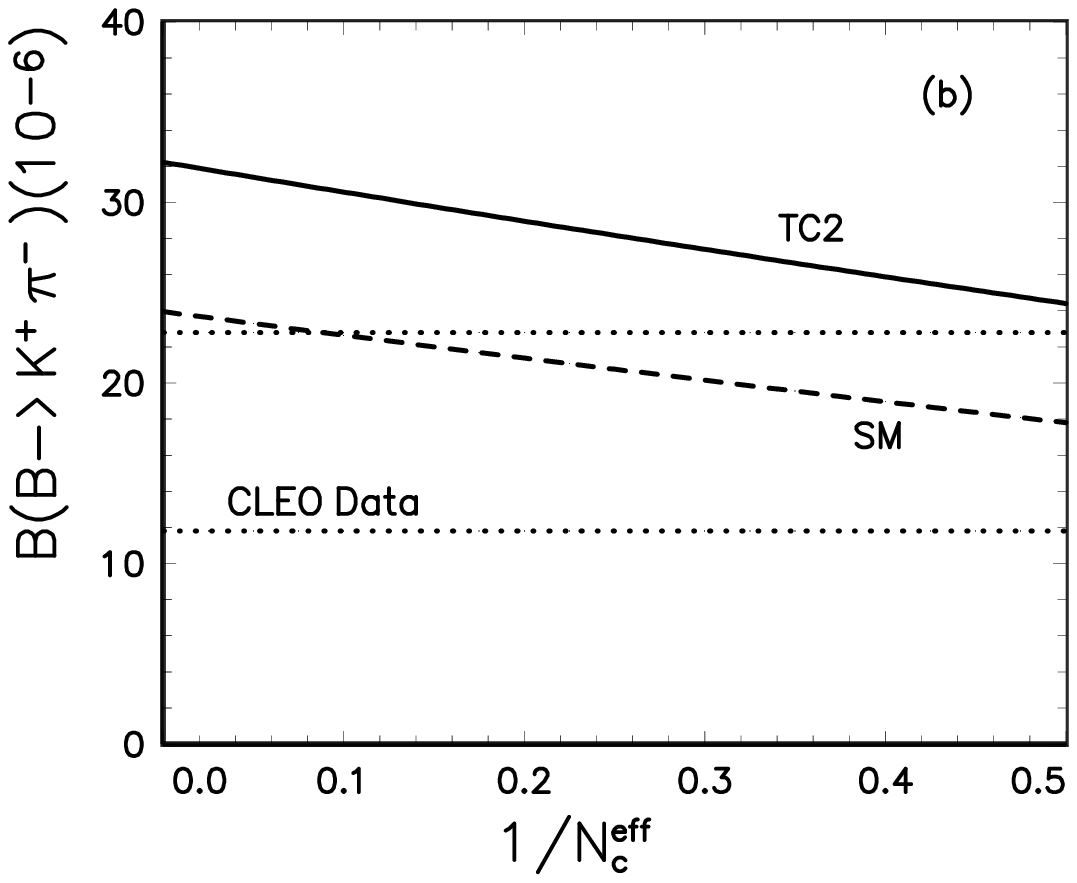}}
\caption{Same as \fig{fig:fig2} but for the case of $B \to K^+ \pi^-$
decay mode. The dots band corresponds to
the CLEO data with $2\sigma$ errors: ${\cal B}(B \to K^+ \pi^-)=
(17.2 ^{+5.6}_{-5.4})\times 10^{-6}$. }
\label{fig:fig3}
\end{minipage}
\end{figure}

\begin{figure}[t] 
\vspace{-60pt}
\begin{minipage}[t]{0.95\textwidth}
\centerline{\epsfxsize=\textwidth \epsffile{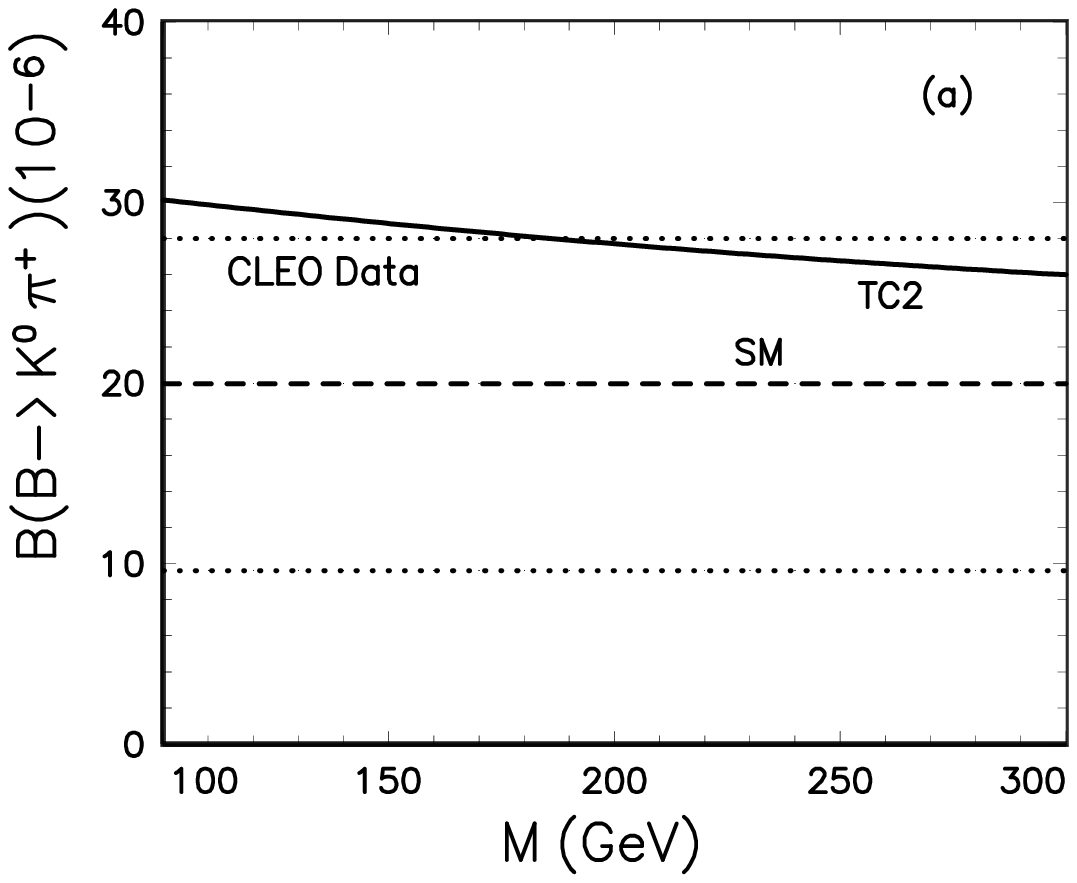}}
\vspace{-60pt}
\centerline{\epsfxsize=\textwidth \epsffile{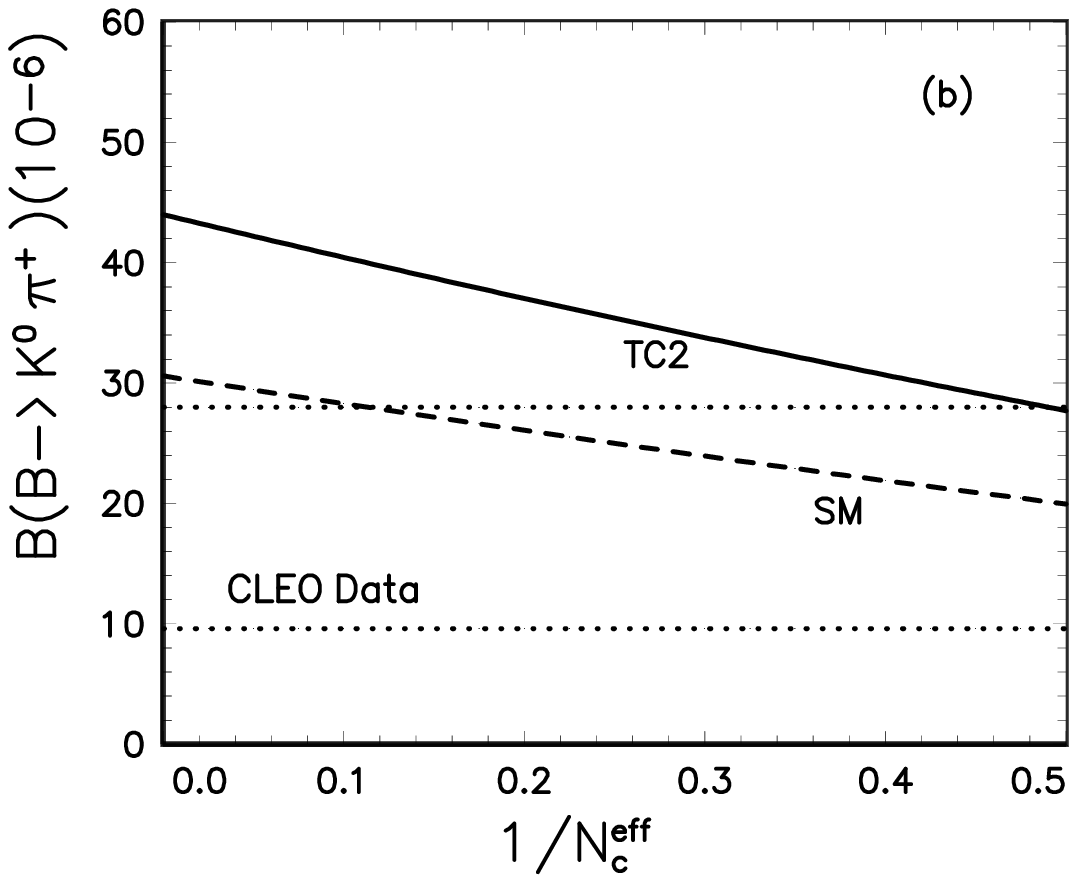}}
\caption{Same as \fig{fig:fig2} but for the case of $B \to K^0 \pi^+$
decay mode. The dots band corresponds to
the CLEO data with $2\sigma$ errors: ${\cal B}(B \to K^0 \pi^+)=
(18.2 ^{+9.8}_{-8.6})\times 10^{-6}$. }
\label{fig:fig4}
\end{minipage}
\end{figure}

\begin{figure}[t] 
\vspace{-60pt}
\begin{minipage}[t]{0.95\textwidth}
\centerline{\epsfxsize=\textwidth \epsffile{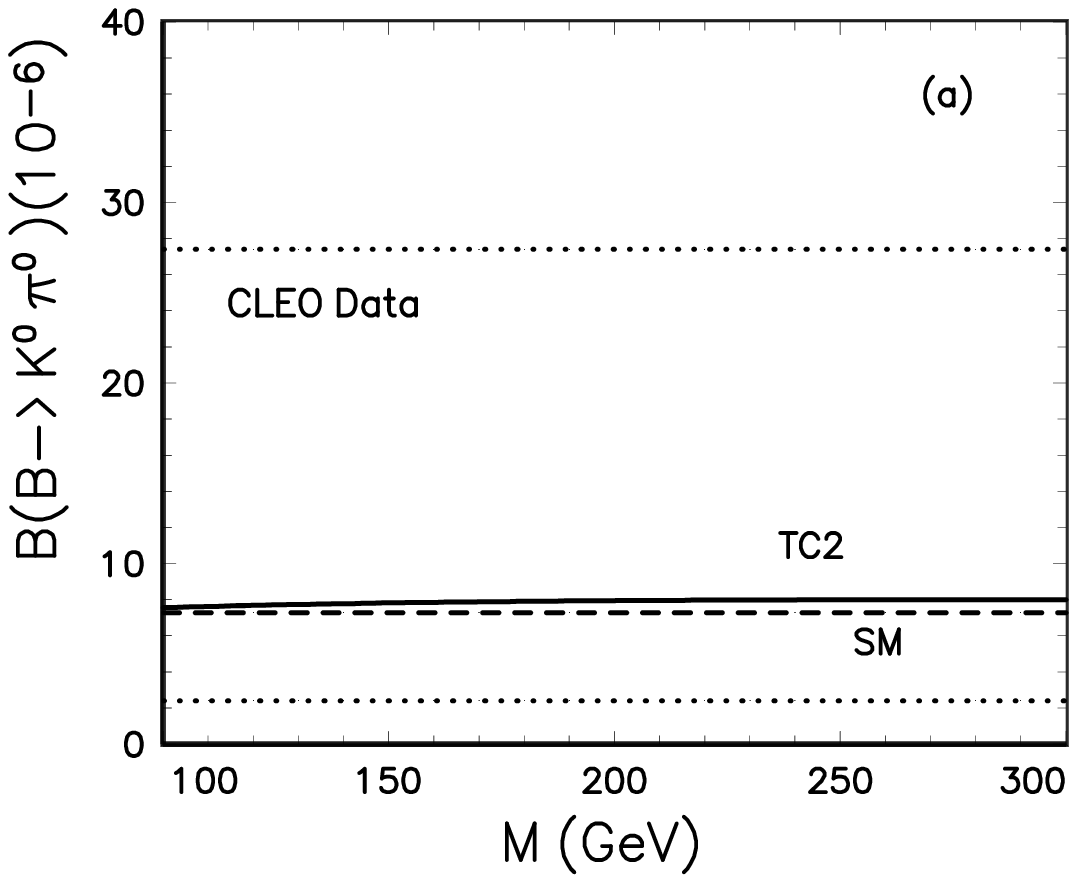}}
\vspace{-60pt}
\centerline{\epsfxsize=\textwidth \epsffile{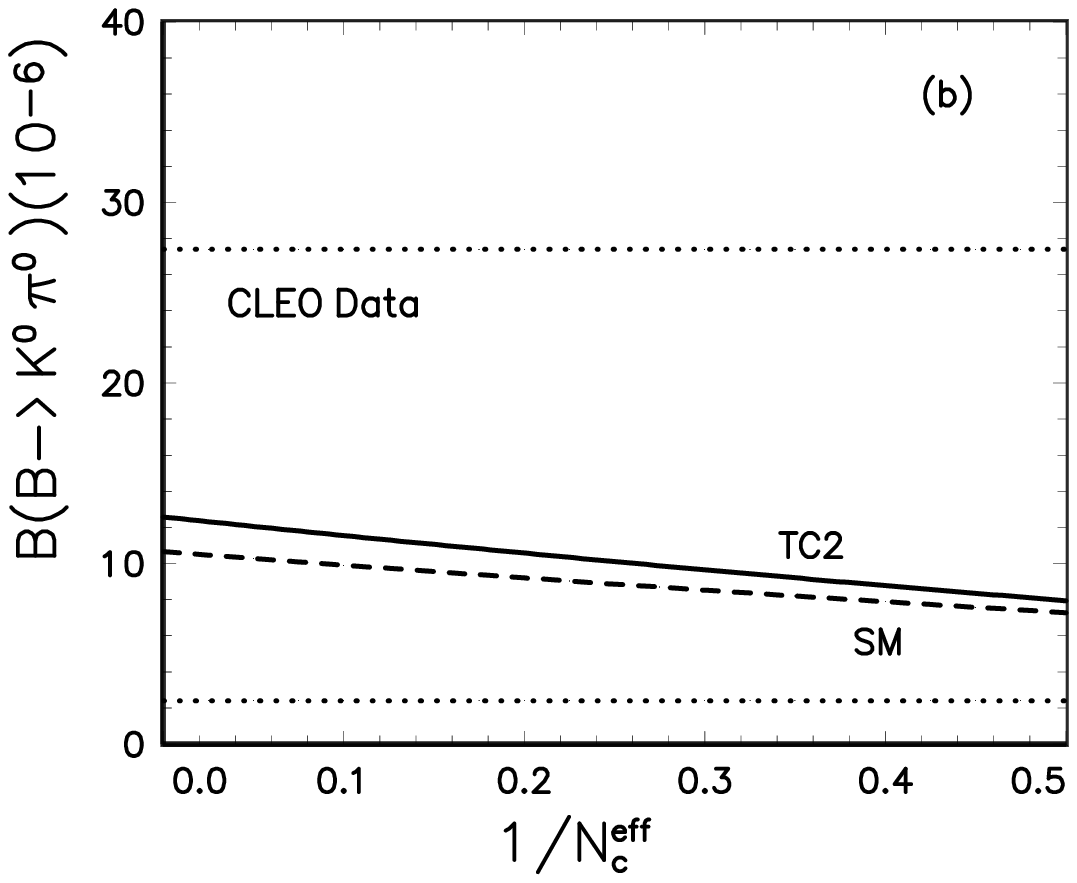}}
\caption{Same as \fig{fig:fig2} but for the case of $B \to K^0 \pi^0$
decay mode. The dots band corresponds to
the CLEO data with $1\sigma$ error: ${\cal B}(B \to K^0 \pi^0)=
(14.6 ^{+6.4}_{-6.1})\times 10^{-6}$. }
\label{fig:fig5}
\end{minipage}
\end{figure}

\begin{figure}[t] 
\vspace{-60pt}
\begin{minipage}[t]{0.95\textwidth}
\centerline{\epsfxsize=\textwidth \epsffile{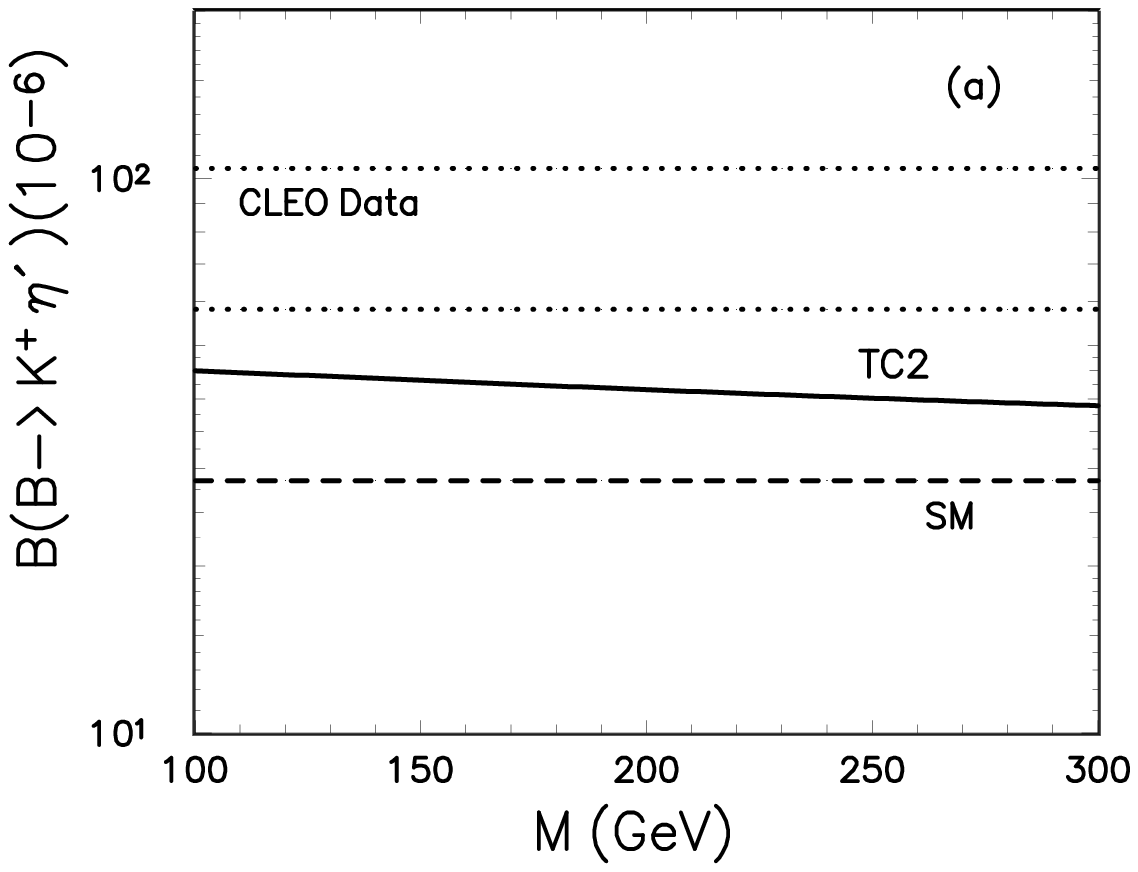}}
\vspace{-60pt}
\centerline{\epsfxsize=\textwidth \epsffile{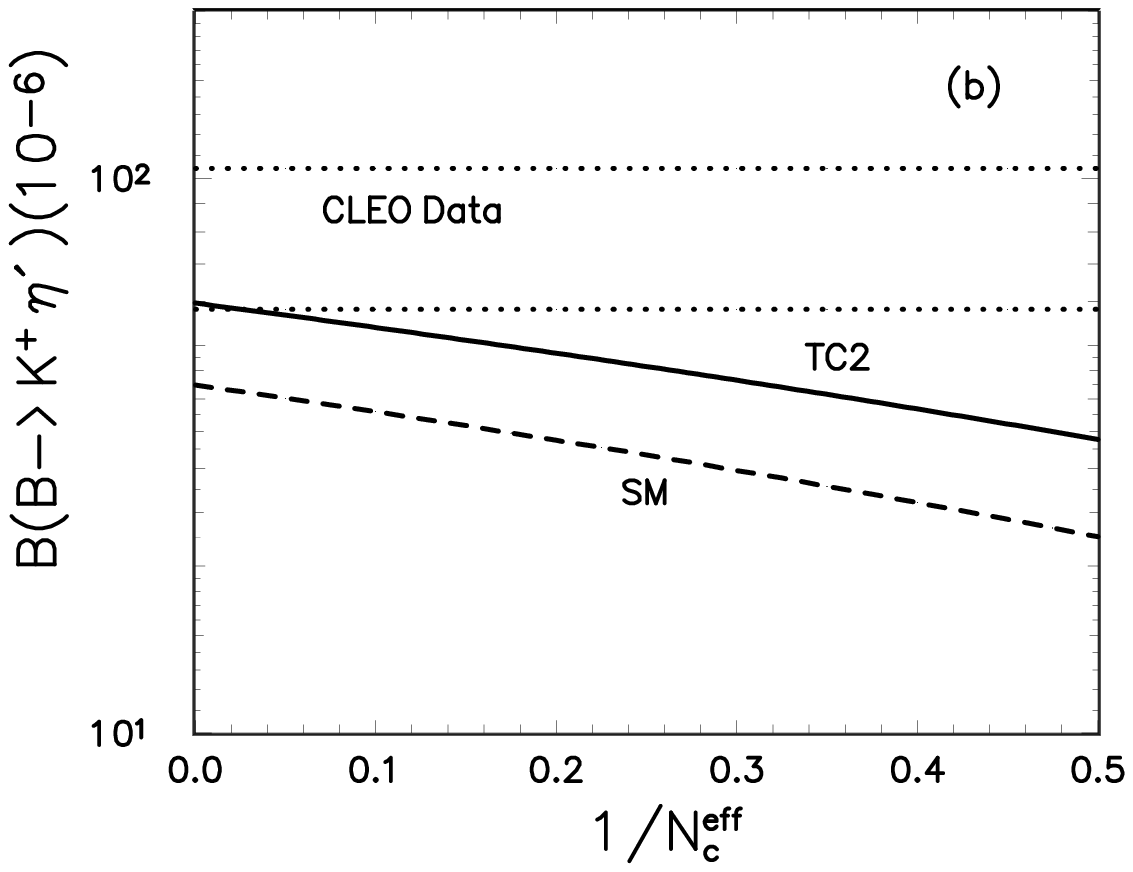}}
\caption{Plots of branching ratios of decays $B^+ \to K^+ \etap$ versus $\mpcc$
and $1/N_c^{eff}$ in the SM and TC2 model. For (a) and (b), we set $N_c^{eff}=3$ and
$\mpcc=200$GeV, respectively. The short-dashed line ( solid curve )
shows ${\cal  B}(B^+ \to  K^+ \etap)$ in the SM ( TC2 model ).
The dots band corresponds to the CLEO data with $2\sigma$ errors:
${\cal  B}(B^+ \to  K^+ \etap)=(80^{+24}_{-22})\times 10^{-6}$. }
\label{fig:fig6}
\end{minipage}
\end{figure}

\begin{figure}[t] 
\vspace{-60pt}
\begin{minipage}[t]{0.95\textwidth}
\centerline{\epsfxsize=\textwidth \epsffile{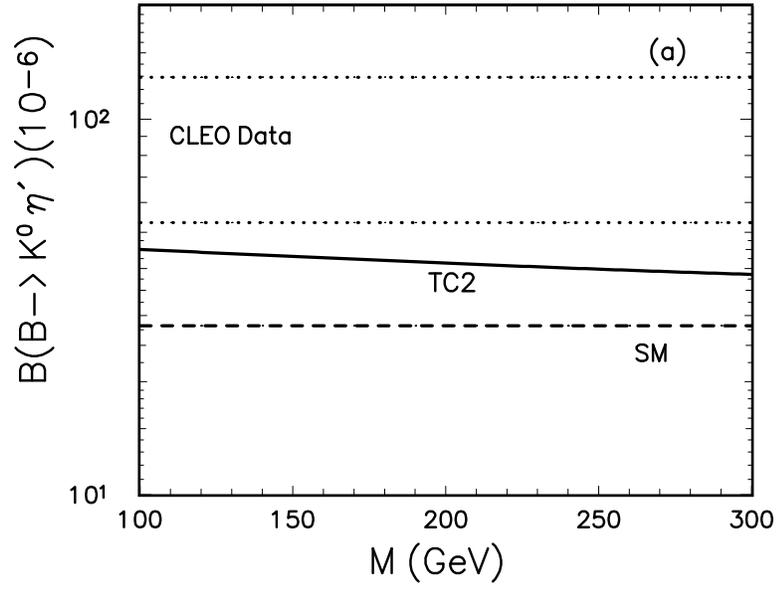}}
\vspace{-60pt}
\centerline{\epsfxsize=\textwidth \epsffile{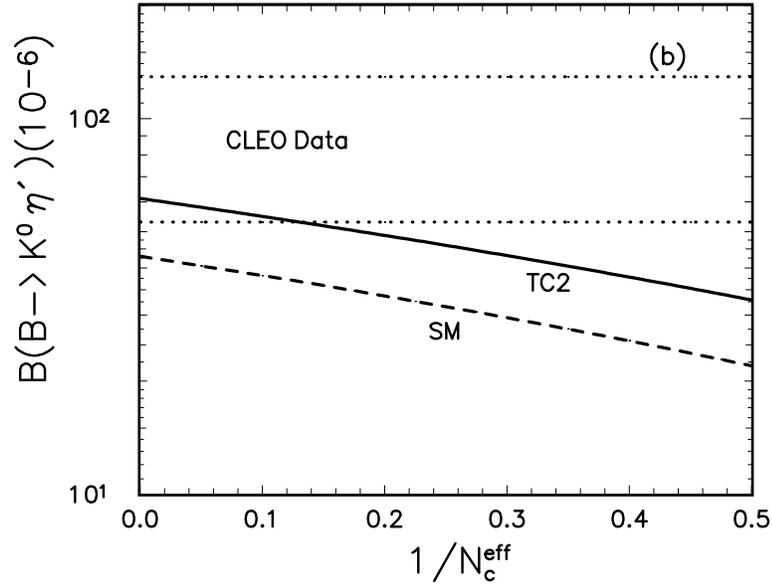}}
\caption{Same as \fig{fig:fig6} but for $B^0 \to K^0 \etap$ decay.
The dots band corresponds to the CLEO data with $2\sigma$ errors:
${\cal  B}(B^0 \to  K^0 \etap)=(89^{+40}_{-36})\times 10^{-6}$. }
\label{fig:fig7}
\end{minipage}
\end{figure}

\begin{figure}[t]
\vspace{-40pt}
\begin{minipage}[t]{0.95\textwidth}
\centerline{\epsfxsize=\textwidth \epsffile{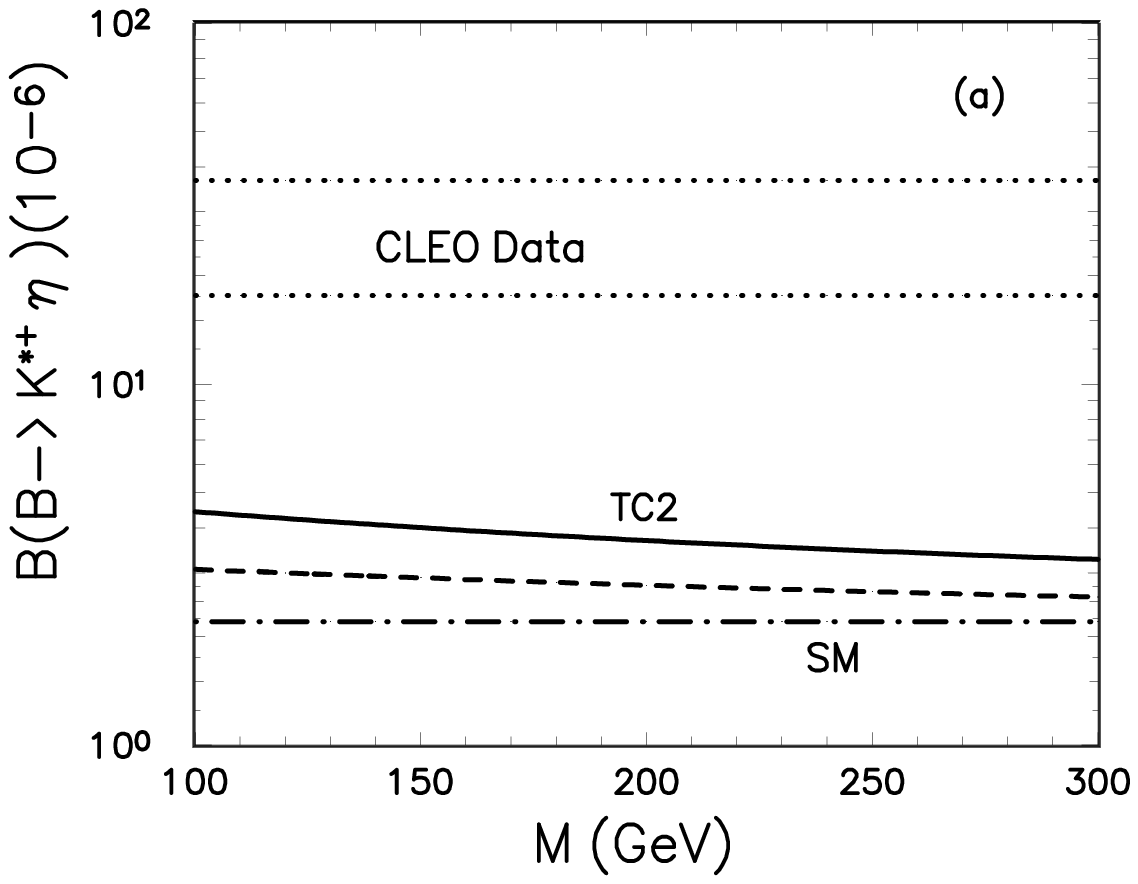}}
\vspace{-60pt}
\centerline{\epsfxsize=\textwidth \epsffile{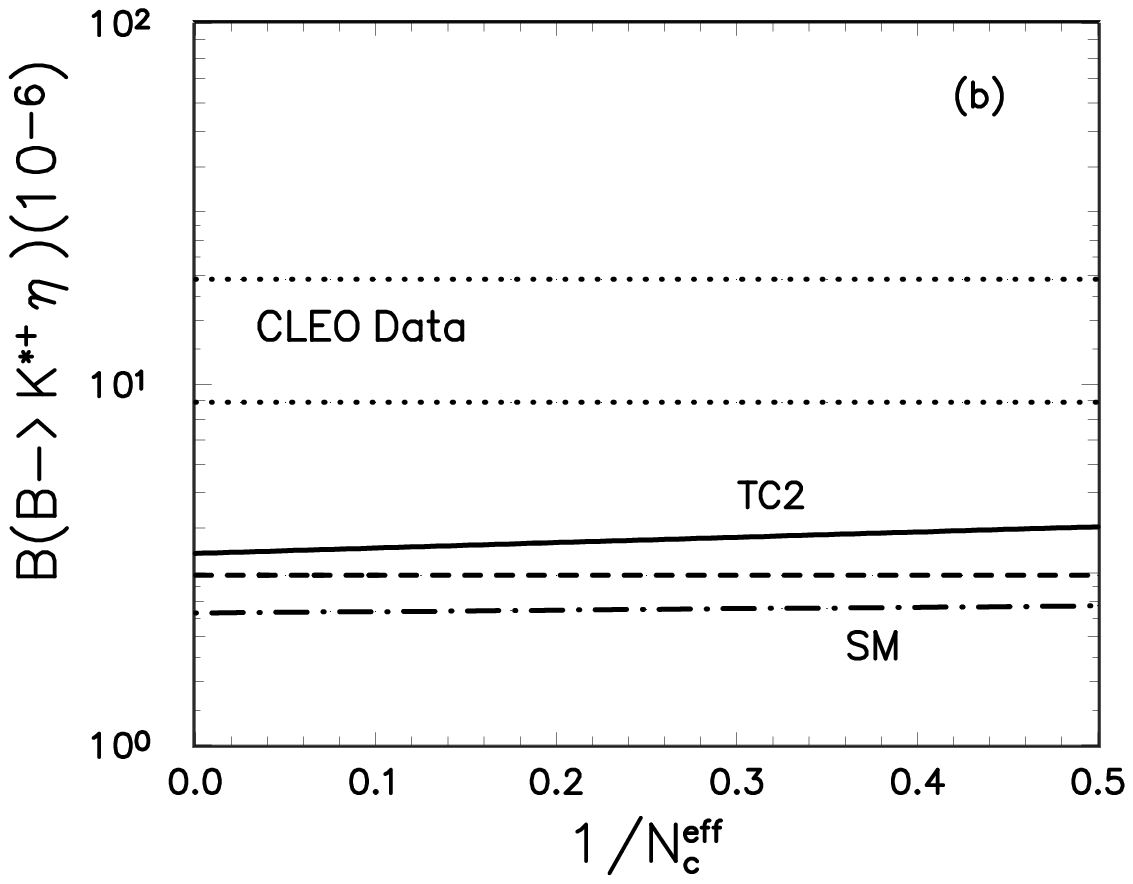}}
\caption{ Plots of ${\cal  B}(B^+ \to K^{*+} \eta)$ {\it versus} $\mpcc$
and $1/N_c^{eff}$ in the SM and TC2 model. For (a) and (b),
we set $N_c^{eff}=3$ and $\mpcc=200$GeV, respectively.
The dot-dashed line shows the SM prediction,
while the short-dashed and solid curve refer to the ratios with
the inclusion of contributions induced by new gluonic penguins and
both new gluonic  and electroweak penguins, respectively.
The upper band corresponds to the CLEO data with $1\sigma$ error:
${\cal  B}(B^+ \to  K^{*+} \eta) = ( 26.4^{+10.2}_{-8.8})\times 10^{-6}$. }
\label{fig:fig8}
\end{minipage}
\end{figure}

\begin{figure}[t]
\vspace{-40pt}
\begin{minipage}[t]{0.95\textwidth}
\centerline{\epsfxsize=\textwidth \epsffile{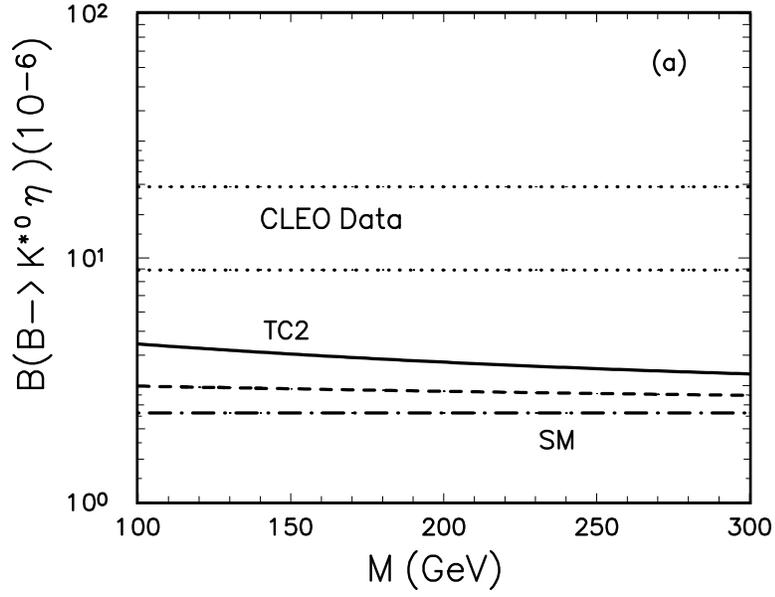}}
\vspace{-60pt}
\centerline{\epsfxsize=\textwidth \epsffile{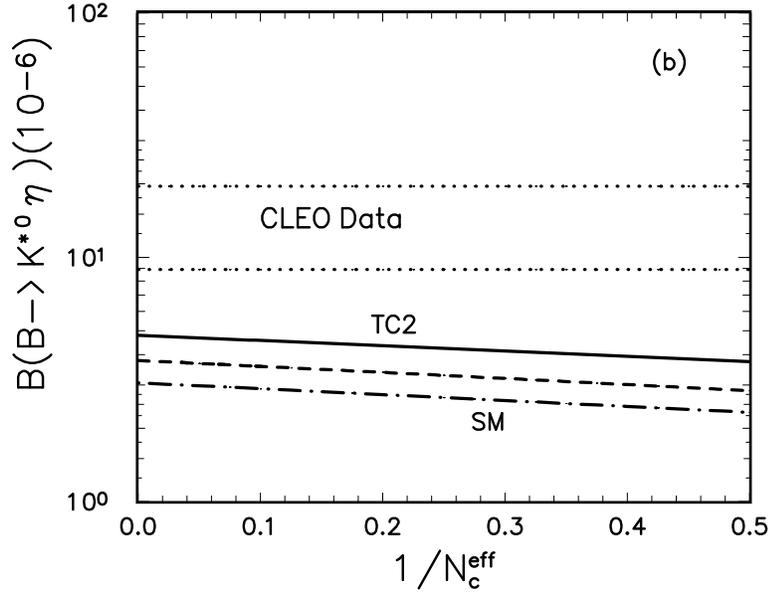}}
\caption{Same as Fig.(\ref{fig:fig8}), but for decay ${\cal  B}(B^0 \to
K^{*0} \eta)$. The upper band corresponds to the CLEO data with $1\sigma$
error: ${\cal  B}(B^0 \to  K^{*0} \eta) = ( 13.8^{+5.7}_{-4.9})
\times 10^{-6}$. }
\label{fig:fig9}
\end{minipage}
\end{figure}

\begin{figure}[t]
\vspace{-40pt}
\begin{minipage}[t]{0.95\textwidth}
\centerline{\epsfxsize=\textwidth \epsffile{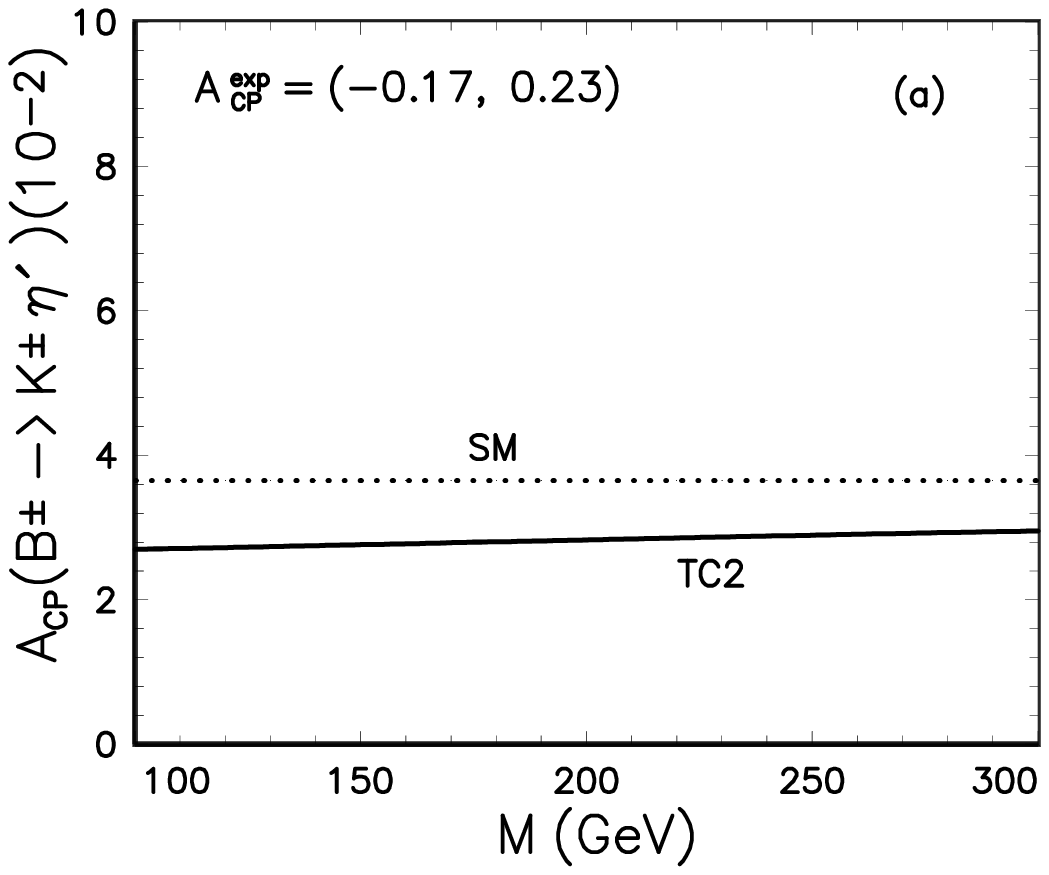}}
\vspace{-60pt}
\centerline{\epsfxsize=\textwidth \epsffile{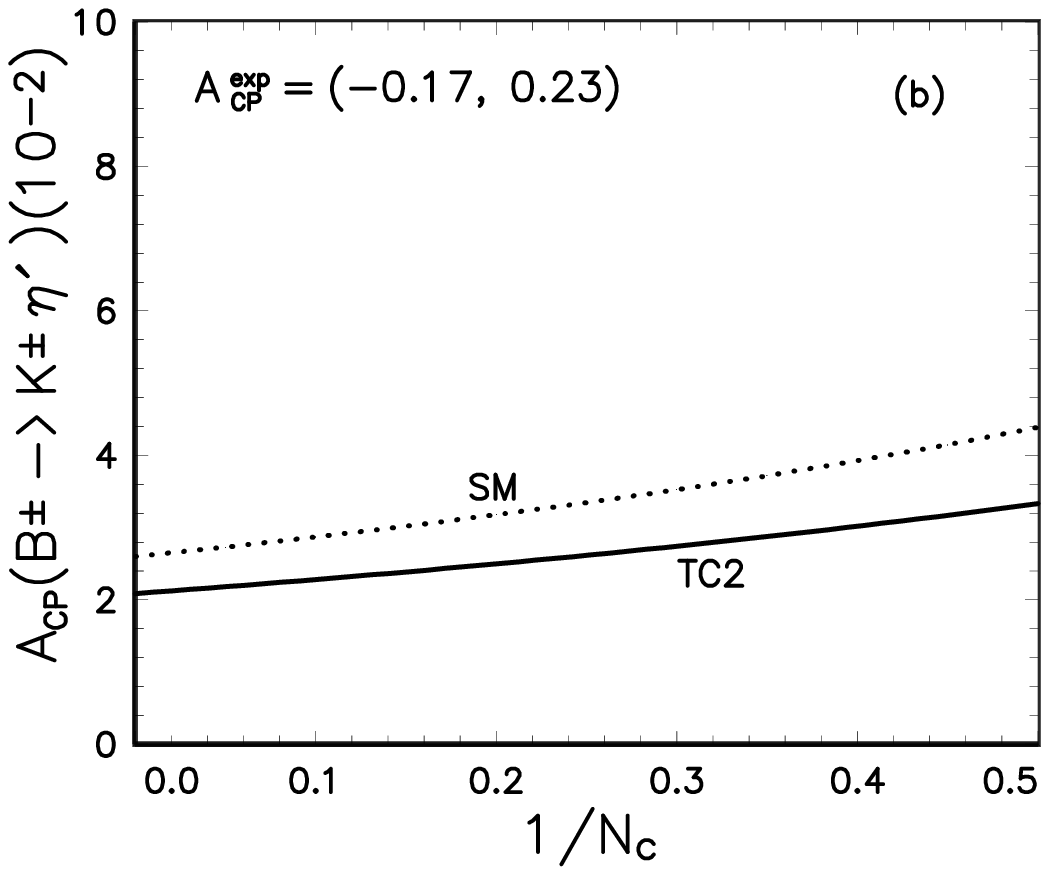}}
\caption{Plots of CP-violating asymmetries $\acp$ $vs$ $\mpcc$ and $1/N_c^{eff}$ for
decay $(B^{\pm} \to K^{\pm} \etap)$. For (a) and (b) we set $N_c^{eff}=3$ and
$\mpcc=200$ GeV, respectively. The $90\% C.L.$ allowed region from CLEO is
$\acp =[-0.17, 0.23]$. }
\label{fig:fig10}
\end{minipage}
\end{figure}

\begin{figure}[t]
\vspace{-40pt}
\begin{minipage}[t]{0.95\textwidth}
\centerline{\epsfxsize=\textwidth \epsffile{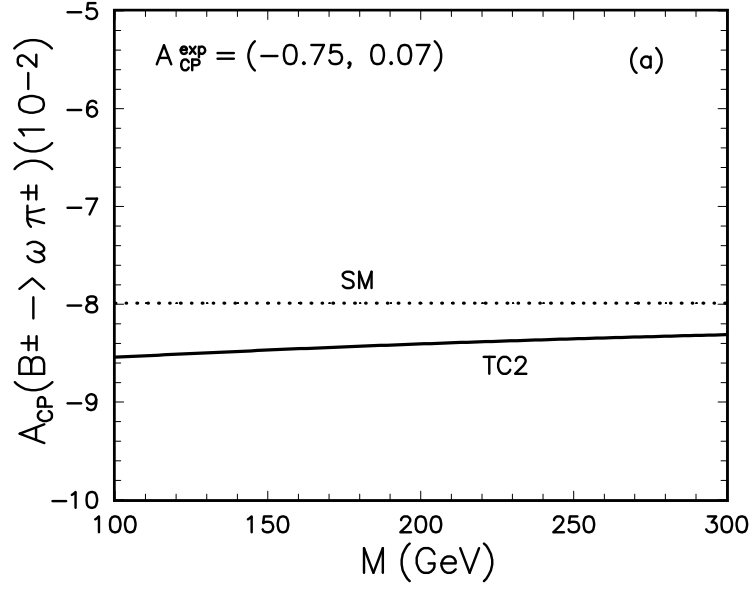}}
\vspace{-60pt}
\centerline{\epsfxsize=\textwidth \epsffile{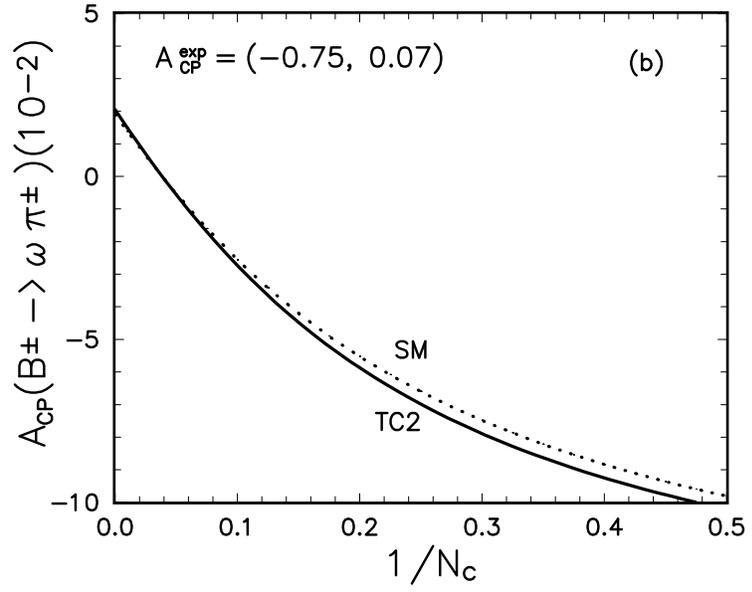}}
\caption{Same as \fig{fig:fig10} but for decay $(B^{\pm} \to \omega
\pi^{\pm})$. The $90\% C.L.$ allowed region from CLEO is $\acp
=[-0.75, 0.07]$. }
\label{fig:fig11}
\end{minipage}
\end{figure}

In section \ref{sec:br}, we calculated the branching ratios for fifty seven
$B \to PP, PV$ decays in the SM and TC2 model, presented the numerical
results
in tables (\ref{bpp1}-\ref{bpv2}) and displayed the $\mpcc$ and $N_c^{eff}$-dependence
of several interesting decay modes in Figs.(\ref{fig:fig2}-\ref{fig:fig9}).
From these tables and figures, the following conclusions can be reached:
\begin{itemize}

\item
The theoretical predictions in the TC2 model for all fifty seven decay modes
under study are well consistent with the experimental measurements and upper
limits within one or two standard deviations.

\item
The theoretical uncertainties induced by varying $k^2$, $\eta$ and $\mpcc$
are moderate within the range of $k^2 = m_b^2/2 \pm 2 GeV^2$, $\eta=0.34 \pm 0.08$
and $\mpcc=200\pm 100$ GeV. The dependence on whether we use
the BSW or LQSSR form factors are also weak. The $N_c^{eff}-$dependence vary
greatly for different decay modes.

\item
For most $B \to PP $ decay channels, the NP effects $\delta {\cal B}$ are
large in size and insensitive to the variations of the effective number
of colors $N_c^{eff}$. For many $B \to PV$ decays, however,
$\delta {\cal B}$ are sensitive to the variations of $N_c^{eff}$.
It seems that the $B\to K \pi$ and $B \to K\etap$
decay channels are good places to test the TC2 model.

\item
For most class-II, IV and V decay channels, such as $B \to \eta \etapp$,
$B \to K \pi$,$B \to K \etap$, $ etc,$ the NP enhancements to the decay
rates can be rather large,  from $~30\%$ to $~100\%$ $w.r.t$ the SM
predictions. So large enhancements will be measurable when enough
B decay events accumulated at B factories in the forthcoming years.

\item
For most decay modes, both new gluonic and electroweak  penguins  contribute
effectively.

\item
For  $B \to  K \etap$ decays, the new physics enhancements are significant,
$\sim 50\%$, and insensitive to the variations of $k^2$, $\eta$, $\mpcc$ and
$N_c^{eff}$ within considered parameter space. The theoretical predictions for
${\cal B}(B \to K \etap)$  become now consistent with the CLEO data
due to the inclusion of new physics effects in the TC2 model.

\end{itemize}

In section \ref{sec:acp}, we calculated the CP-violating asymmetries $\acp$
for  $B \to PP, PV$ decays in the SM and TC2 model, presented the
numerical results in tables (\ref{acpp1}-\ref{acpv2b}) and displayed the
$\mpcc$ and $N_c^{eff}$-dependence of $\acp$ for decays $B^{\pm} \to K^{\pm} \etap,
\omega \pi^{\pm}$ in Figs.(\ref{fig:fig10},\ref{fig:fig11}). In this paper,
the possible effects of FSI on $\acp$ are neglected. From those
tables and figures, the following conclusions can be drawn:

\begin{itemize}

\item
Although there is no new weak phase introduced in TC2 model,
the CP-violating asymmetries ${\cal  A}_{CP}$ can still be
changed through the interference
between the ordinary tree/penguin amplitudes in the SM and the new
strong and electroweak penguin amplitudes in TC2 model.

\item
The  CP-violating asymmetries depend weakly on whether we use the BSW or
LQQSR form factors in calculations in both the SM and TC2 model.

\item
For most $B \to PP$ decays,  $\delta {\cal  A}_{CP}$ are generally
small or moderate in magnitude ( $10\% -30\% $ ), and
insensitive to the variation of $\mpcc$ and $N_c^{eff}$.
But the four class-II decay modes $\obar{B^0} \to \pi^0 \pi^0,
\etapp \etapp$ have strong $N_c^{eff}$-dependence in both the
SM and TC2 model.

\item
For $B \to PV$ decays, however, $\delta {\cal  A}_{CP}$  can be rather
large for many decay modes. For decay $B^0/\bar{B}^0 \to
\rho^+ \pi^-/\rho^- \pi^+$, the new physics correction is
$(60 - 100)\%$ for $N_c^{eff}=2 - \infty$. For decay $B^+ \to K^{*+}
\bar{K}^0$ the correction can even reaches a factor of 20 for $N_c^{eff}=2$.
For most class-I, III and IV decays, the $N_c^{eff}$-dependence and
$k^2$-dependence
of $\delta {\cal A}_{CP}$ are weak. For most class-V decays, however, the
$N_c^{eff}$-dependence of $\delta {\cal A}_{CP}$ is strong.

\item
For the measured five decay modes $B \to K \pi, K \etap, \omega \pi$,
the new physics effects is only about $-20\%$ {\it w.r.t} the SM
predictions.  The theoretical
predictions for these five decay modes in the SM and TC2 model
are well consistent with the CLEO measurements.

\end{itemize}

\section*{ACKNOWLEDGMENTS}

Authors are very grateful to D.S. Du, K.T. Chao, C.S.Li, C.D. L\"u,
Y.D. Yang and M.Z.Yang for helpful discussions.
Z.J. Xiao acknowledges the support by the National
Natural Science Foundation of China under the Grant No.19575015 and 10075013,
the Excellent Young Teachers Program of MOE, P.R.China and the
Natural Science Foundation of Henan Province under the Grant No. 994050500.

\newpage

\section*{Appendix A: Input parameters} \label{app:a}

In this appendix we present relevant input parameters. We use the same set
of input parameters for the quark masses, decay constants, Wolfenstein
parameters and form factors as \cite{ali9804}.

\begin{itemize}

\item
Input parameters of electroweak and strong coupling constant, gauge boson masses,
B meson masses, light meson masses,
$\cdots$,  are as follows (all masses in unit of GeV )\cite{ali9804,pdg98}
\beq
\alpha_{em}&=&1/128, \;  \alpha_s(M_Z)=0.118,\;  \sin^2\theta_W=0.23,\;
G_F=1.16639\times 10^{-5} (GeV)^{-2}, \non
M_Z&=&91.187, \;   M_W=80.41,\;
m_{B_d^0}=m_{B_u^\pm}=5.279,\;   m_{\pi^\pm}=0.140,\;\non
m_{\pi^0}&=&0.135,\;   m_{\eta}=0.547,\; m_{\etap}=0.958,\;
m_{\rho}=0.770,\;  m_{\omega}=0.782,\non
m_{\phi}&=&1.019,\; m_{K^\pm}=0.494,\;  m_{K^0}=0.498,\;  m_{K^{*\pm}}=0.892,\;
m_{K^{*0}}=0.896,\non
\tau(B_u^\pm)&=& 1.64 ps,\; \tau(B_d^0)= 1.56 ps, \label{masses}
\eeq

\item
For the elements of CKM matrix, we use Wolfenstein parametrization, and fix
the parameters $A, \lambda, \rho$ to their central values,
$A=0.81,\; \lambda=0.2205, \; \rho=0.12$ and varying $\eta$ in the range of
$\eta=0.34 \pm 0.08$.

\item
We firstly treat the internal quark masses in the loops in connection with
the function $G(x_i,z)$ as constituent masses,
\beq
m_b=4.88 GeV, \; m_c=1.5 GeV,\; m_s=0.5 GeV, \; m_u=m_d=0.2 GeV.
\label{con-mass}
\eeq
Secondly, we will use the current quark masses for $m_i$ ($i=u,d,s,c,b$)
which appear through the equation of motion when working out the hadronic
matrix elements. For $\mu=2.5 GeV$, one finds\cite{ali9804}
\beq
m_b=4.88 GeV, \; m_c=1.5 GeV, m_s=0.122 GeV, \; m_d=7.6 MeV,\; m_u=4.2 MeV.
\label{cur-mass}
\eeq
For the mass of heavy top quark we also use $m_t=\overline{m_t}(m_t)=168 GeV$.

\item
For the decay constants of light mesons, the following values will be used
in the numerical calculations (in the units of MeV):
\beq
&&f_{\pi}=133,\; f_{K}=158,\; f_{K^*}=214, \; f_{\rho}=210, \;
f_{\omega}=195, \; f_{\phi}=233, \; \non
&&f^u_{\eta}=f^d_{\eta}=78,\; f^u_{\etap}=f^d_{\etap}=68,\;
f^c_{\eta}=-0.9,\; f^c_{\etap}=-0.23,\non
&&f^s_{\eta}=-113,\; f^c_{\etap}=141,
\label{fpis}
\eeq
where $f^u_{\etapp}$ and $f^s_{\etapp}$ have been defined in the two-angle-mixing
formalism with $\theta_0=-9.1^\circ$ and $\theta_8 =-22.2^\circ$\cite{fk97}
For more details about the mixings between $\eta$ and $\etap$,
one can see \cite{fk97,ali98}.

\end{itemize}

\section*{Appendix B: Form factors } \label{app:b}

\begin{itemize}

\item
The form factors at the zero momentum transfer in the Baner, Stech and Wirbel (BSW)
\cite{bsw87} model have been collected in Table 2 of ref.\cite{ali9804}. For the
convenience of the reader we list them here:
\beq
&&F_0^{B\pi}(0)=0.33,\; F_0^{BK}(0)=0.38,\;
F_0^{B\eta}(0)=0.145,\; F_0^{B\etap}(0)=0.135,\non
&&A_{0,1,2}^{B\rho}(0)=A_{0,1,2}^{B\omega}(0)=0.28,\;
A_{0}^{BK^*}(0)=0.32, \;A_{1,2}^{BK^*}(0)=0.33, \non
&&V^{B\rho}(0)=V^{B\omega}(0)=0.33, \; V^{BK^*}(0)=0.37.
\label{eq:bsw-f}
\eeq

\item
In the LQQSR approach, the form factors at zero momentum transfer being used in our
numerical calculations are \cite{ali9804},
\beq
&&F_0^{B\pi}(0)=0.36,\; F_0^{BK}(0)=0.41,\;
F_0^{B\eta}(0)=0.16,\; F_0^{B\etap}(0)=0.145,\non
&&\{ A_0,A_1,A_2,V \}( B \to \rho ) =\{0.30,0.27,0.26,0.35 \},\non
&&\{ A_0,A_1,A_2,V \}( B \to K^* )  =\{0.39,0.35,0.34,0.48 \},\non
&&\{ A_0,A_1,A_2,V \}( B \to \omega)=\{0.30,0.27,0.26,0.35 \}.
\label{eq:lqqsr-f}
\eeq

\item
The form factors $F_{0,1}(k^2),$ $A_{0,1,2}(k^2)$ and $V(k^2)$ were defined
in ref.\cite{bsw87} as
\beq
F_0(k^2)&=& \frac{F_0(0)}{1-k^2/m^2(0^+)},\ \
F_1(k^2)=   \frac{F_1(0)}{1-k^2/m^2(1^-)}, \non
A_0(k^2)&=& \frac{A_0(0)}{1-k^2/m^2(0^-)}, \ \
A_1(k^2)=   \frac{A_1(0)}{1-k^2/m^2(1^+)},  \non
A_2(k^2)&=& \frac{A_2(0)}{1-k^2/m^2(1^+)}, \ \
V(k^2) =    \frac{V(0)}{1-k^2/m^2(1^-)}.
\eeq

\item
The pole masses being used to evaluate the $k^2$-dependence of form factors
are,
\beq
\{ m(0^-),m(1^-),m(1^+),m(0^+) \}&=& \{ 5.2789, 5.3248,5.37,5.73 \}
\eeq
for $\bar{u}b$ and $ \bar{d}b$ currents. And
\beq
\{ m(0^-),m(1^-),m(1^+),m(0^+)\}&=& \{5.3693, 5.41,5.82,5.89\}
\eeq
for $\bar{s}b $ currents.

\end{itemize}

\newpage
\listoffigures

\newpage

\newpage

\newpage

\newpage

\newpage

\newpage

\newpage

\newpage

\newpage

\newpage

\newpage

\end{document}